\documentclass[%
 reprint,
 amsmath,amssymb,
 aps,
]{revtex4-2}

\usepackage{graphicx}% Include figure files
\usepackage{dcolumn}% Align table columns on decimal point
\usepackage{bm}% bold math
\usepackage{amsmath}
\usepackage{graphicx}
\usepackage{booktabs}
\usepackage{tabularx}
\usepackage{dcolumn}% Align table columns on decimal point
\usepackage{bm}% bold math
\usepackage{xspace} % to add space after command
\usepackage{rotating}
\usepackage{array}
\begin{document}

%\linenumbers

%\title{Uniform Distribution Technique for Neutrino Beam Scan Simulation}
\title{Uniform Distribution Technique for Neutrino Beam Simulation Studies}
 \author{D. A.\, Wickremasinghe}
 \email{athula@fnal.gov}
 \author{S.\, Ganguly}
 \author{K.\, Yonehara}
 \author{R.\, Zwaska}
 \address{Fermi National Lab, Batavia, IL, USA}
 \author{P.\, Snopok}
  \author{Y.\, Yu}
 \address{Illinois Institute of Technology, Chicago, IL, USA}
\date{\today}

\begin{abstract}
In Fermilab's neutrino facilities such as the Neutrinos at the Main Injector (NuMI) and the upcoming Long Baseline Neutrino Facility (LBNF), a proton beam strikes high-power target, producing positively and negatively charged pions and kaons. 
This paper introduces an efficient approach to generate multiple simulation samples, enabling the study of beam scan effects on target interactions for specific beam configurations. It includes factors like beam position, spot size, and magnetic horn current scan using simulated data. The method is valuable not only for neutrino experiments but could also be valuable for muon physics. By simplifying the generation of extensive simulation data, this technique facilitates the use of machine learning models to predict rare events and anomalies, making it a crucial tool for diverse research applications.

\end{abstract}
 
\maketitle

\section{Introduction}
%Muons from the Neutrinos from the Main Injector (NuMI) neutrino beam
The neutrino mass hierarchy and CP asymmetry in the lepton sector can be tested using long-baseline neutrino oscillation experiments that provide precision measurements \cite{NOvA:2016kwd,Abe:2011ts,T2K:2015sqm,DUNE:2015lol}. Neutrino flux at the Near Detector strongly correlates with proton beam misalignment. The alignment of the primary proton beam, target, and focusing horn greatly impacts the neutrino energy spectrum.\par
\begin{figure}%[htpb!]
  \begin{center}%\setlength{\unitlength}{1.0cm}
   \includegraphics[width=\linewidth]{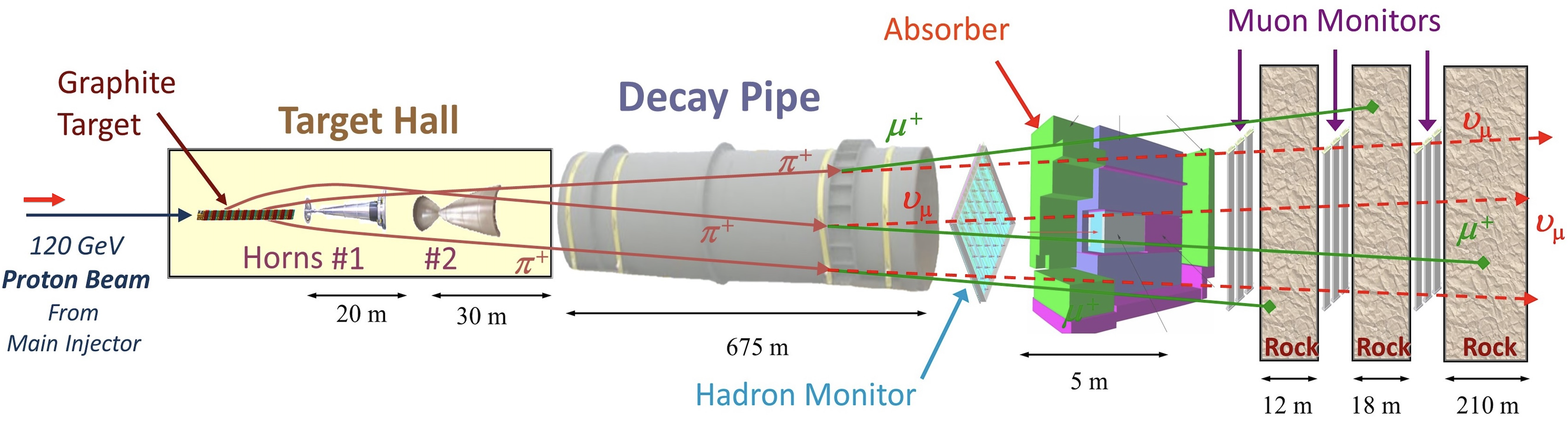}
\caption{Beamline layout for NuMI.}\label{fig:numibeamline}
    \end{center}
  \end{figure}

Neutrinos at the Main Injector (NuMI) \cite{Barenboim:2002mf} neutrino beam at Fermilab is generated by aiming high energy protons into a carbon target.
The NuMI beamline as depicted in Fig.~\ref{fig:numibeamline} produces an intense muon neutrino beam for NuMI neutrino experiments. 
A 120-GeV proton beam from the main injector collides with a fixed graphite target to provide neutrinos.
Using two focusing horns operating at 200\,kA horn current, charged particles produced from proton interactions with target nuclei are focused into a 675-m long, 2-m diameter cylindrical decay pipe.
As the mesons decay, they can produce neutrinos and muons before being absorbed by the hadron absorber. 
High-energy muons produced by meson decay could pass through muon monitors downstream of the hadron absorber.\par
\begin{figure}%[htpb!]
  \begin{center}%\setlength{\unitlength}{1.0cm}
   \includegraphics[width=\linewidth]{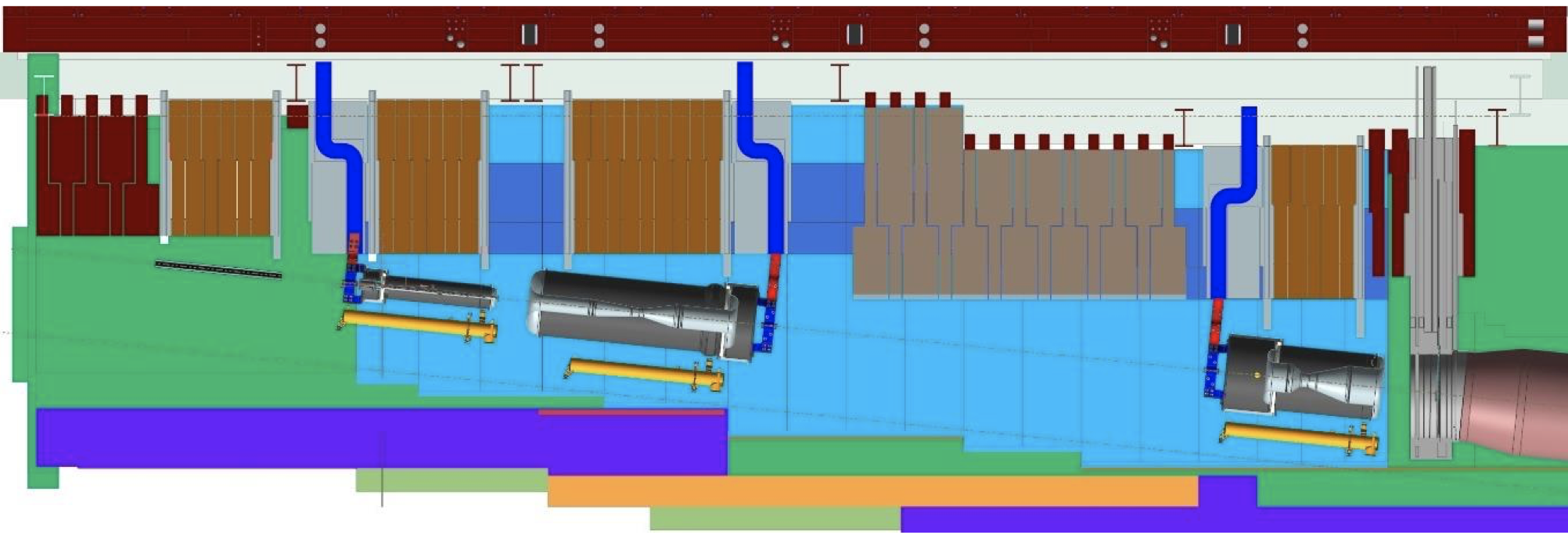}
\caption{An illustration of the upstream portion of the LBNF neutrino beamline. A horn-protection baffle, three focusing horns, and the decay pipe are shown inside the target chase respectively from left to right (the beam direction).}\label{fig:lbnfbeamline}
    \end{center}
  \end{figure}

A world-leading neutrino beam will be produced by the upcoming Long Baseline Neutrino Facility (LBNF)~\cite{lbnfbeamline} as shown in Fig.~\ref{fig:lbnfbeamline}. The baseline beamline design consists of a 1.2-MW, 120-GeV primary proton beam impinging on a cylindrical graphite target that measures 1.8\,m in length and 1.6\,cm in diameter. Using 300\,kA currents, three magnetic horns focus hadrons produced in the target, which is supported inside horn 1. In addition to the target chase, there is a 194-m long helium-filled decay pipe and a hadron absorber.\par
In both NuMI and LBNF, neutrinos and anti-neutrinos can be created by operating the focusing horns in forward or reverse current configurations.
We can estimate the effects of changes in the beam parameters by scanning the primary proton beam across the target. Typical beam scan processes \cite{Zwaska:2006px} involve systematically changing proton beam position on target, beam spot size and horn currents. In NuMI, these changes are visible in the response of each of the three muon monitors (MM1--MM3) that are identical to each other. Due to intervening shielding between two consecutive detectors, each detector sees a different range of muon momenta. Note that these initial ranges are subsets rather than distinct ranges.\par
\begin{figure}%[htpb!]
  \begin{center}%\setlength{\unitlength}{1.0cm}
   \includegraphics[width=\linewidth]{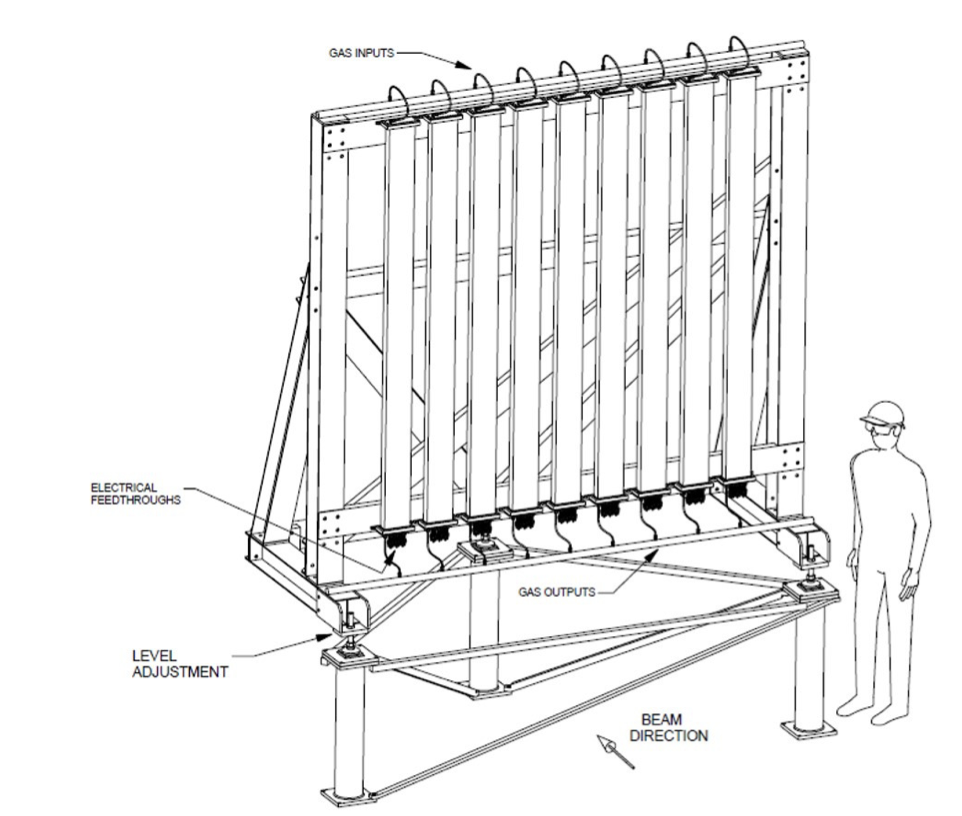}
\caption{Schematics of the NuMI muon monitors.}\label{fig:MM}
    \end{center}
  \end{figure}

Each muon monitor in the NuMI facility consists of a $9\times9$ array of ionization chambers as shown in Fig.~\ref{fig:MM}. There are two parallel-plate electrodes separated by a 0.3-cm gap in each ionization chamber filled with helium gas. LBNF will monitor its beam quality through several detectors, including the Muon Monitor System (MuMS). A NuMI-based approach will be used in the conceptual design of MuMS.\par
For the purpose of establishing a correlation between data and simulations, beam scan data must be compared with simulation results.
%In order to correlate the muon flux at the muon monitors to the neutrino flux at the neutrino detector, it is crucial to perform simulation studies on beam variations, as this helps with the understanding of the pion phase space variations. 
%Simulation studies on beam variations are crucial to understand the pion phase space variations. %This is in order to correlate the muon flux at the muon monitors to the neutrino flux at the neutrino detectors. 
Combining data and simulation also provides valuable beam diagnostics.
High-statistics simulations with ``realistic" Gaussian distributions for various beam and horn current settings generally require substantial computing resources and long simulation times. A Gaussian distribution characterizes the intensity profile of a Gaussian beam. In a Gaussian beam, the intensity is highest at the center and gradually decreases while moving away from the center in a transverse direction (perpendicular to the beam axis). 
Simulation time and computing resources can be significantly reduced by simulating a single high-statistics sample with a uniform distribution for each beam parameter where the uniform beam's intensity remains approximately the same from the center to the edges.
After that, a Gaussian weight can be calculated and applied to the sample in post-processing. 
To understand neutrino beam variations from muon monitor data and simulations, Monte Carlo samples with high statistics are essential.
Using a single high-statistics simulation sample with a uniform distribution, many Gaussian distributions can be created that are similar to those found in real beam scan studies.
This proposed technique can generate a significant number of Monte Carlo (MC) data samples by varying the incident beam parameters and horn current settings. 
Section~\ref{sec:simtools} provides a brief description of the simulation tools used in the studies.
Section~\ref{sec:tech} presents the details of the simulation technique we have proposed. 
Section~\ref{sec:simdata} contains information about the simulation data that have been generated.
In Sec.~\ref{sec:val} we compare uniform beam simulation results with the nominal beam simulation results and validate the uniform beam simulation technique. 
Section~\ref{sec:resource} gives details on the computer resources needed for these simulations. 
Section~\ref{sec:app} illustrates the application of the technique with a few beam scan examples with NuMI simulations. In Sec.~\ref{sec:dunesim} examples of uniform beam simulation technique are illustrated for the LBNF simulation.
Sec.~\ref{sec:machinelearning} highlights the usefulness of the technique for machine learning applications.
Sec.~\ref{sec:summary} briefly summarizes the details of the current state and the future applications of the proposed technique.
\section{Simulation Tools}\label{sec:simtools}
For NuMI and LBNF neutrino studies with simulation, g4numi and g4lbnf use end-to-end Monte Carlo simulation based on Geant4 \cite{GEANT4:2002zbu,Allison:2006ve}.
%The simulations account for the geometry of each beamline, including the targets, focusing horns, and decay volumes. %They consider the full evolution of particles produced in the first collisions between protons and targets, which includes both secondary meson production and subsequent interactions with beamline components.
The simulations account for beamline geometry, including the targets, focusing horns, and decay volumes. Both secondary meson production and subsequent interactions with beamline components are considered as part of the full sequence of particles produced in the first collisions between protons and targets.
When a neutrino is produced in a particle decay, information about its momentum, position, and ancestor is stored in the ``dk2nu" format. A dk2nu ntuple is essentially a list of neutrinos generated by a beam simulation. It is a flux ntuple format and library with methods for analyzing flux ntuples (e.g., location weights). DUNE, NOvA, and MINERvA are some of the Fermilab experiments that are using this ntuple format. 
Neutrinos are associated with dk2nu objects that contain detailed information about their kinematics and their hadron ancestors.\par
Based on the ancestry information provided in the g4numi/g4lbnf output, we extract the momentum and position of each pion/kaon that decayed to a neutrino. 
In the NuMI beamline, the intersection of Horn 1's front face and the primary proton beamline's trajectory marks the origin of the coordinate system used in this simulation which begins with a 120-GeV primary proton beam. The Z-axis points towards the beamline and the Y-axis points vertically up.
Geometrical features such as the target hall, decay pipe, absorber and muon monitors are included in the simulation that provides the location and kinematics of each decay into a neutrino.
%The g4numi package produces a ROOT-based output file containing n-tuples which contain neutrino production information. 
%The g4numi simulation used for the studies shown here applies a 0.02 MeV energy threshold 
All particles in g4numi have a default energy threshold of 0.02\,GeV. The simulation cuts off protons outside the beamline. In the muon monitor simulation, muons with energies less than 1\,GeV, as well as other particles, are omitted. Using a 1\,GeV muon cut for the second stage of the simulation (i.e., the muon monitor simulation) saves computing time and space. The threshold of the hadron absorber in front of the MM1 is about 5\,GeV. Therefore, cutting muons below 1\,GeV is safe in the decay pipe.
g4lbnf simulation uses the same tracking cuts as g4numi. 

\section{Simulation Technique}\label{sec:tech}
%The NuMI beam simulations for neutrino studies are performed using a GEANT4-based \cite{GEANT4:2002zbu,Allison:2006ve} Monte Carlo (MC) simulation package called g4numi. 
In this section, we first describe a technique to reproduce NuMI beam scans with simulated data. The pions generated from g4numi simulations are over-sampled to decay into multiple muons in this study. Then each muon is simulated through the absorber and the muon monitors. In order to predict the muon flux at muon monitors for selected beam and horn current settings, the output of the g4numi simulation is combined with the output of the muon monitor simulation. Using multiple muons as decay products is intended to decrease statistical uncertainty in the analysis of each muon monitor pixel. 
The simulation samples used here have all been generated in forward horn current (FHC) or neutrino mode, during which the horn current flows in the inner conductor in the same direction as the beam. 
As the first step of the proposed technique, a uniformly distributed single simulation data sample is generated for the selected beam scan variable. 
With the real beam width and beam positions, the uniformly distributed sample is then used to generate Gaussian samples for the selected Gaussian functions.
\begin{figure}%[htpb!]
  \begin{center}%\setlength{\unitlength}{1.0cm}
   \includegraphics[width=\linewidth]{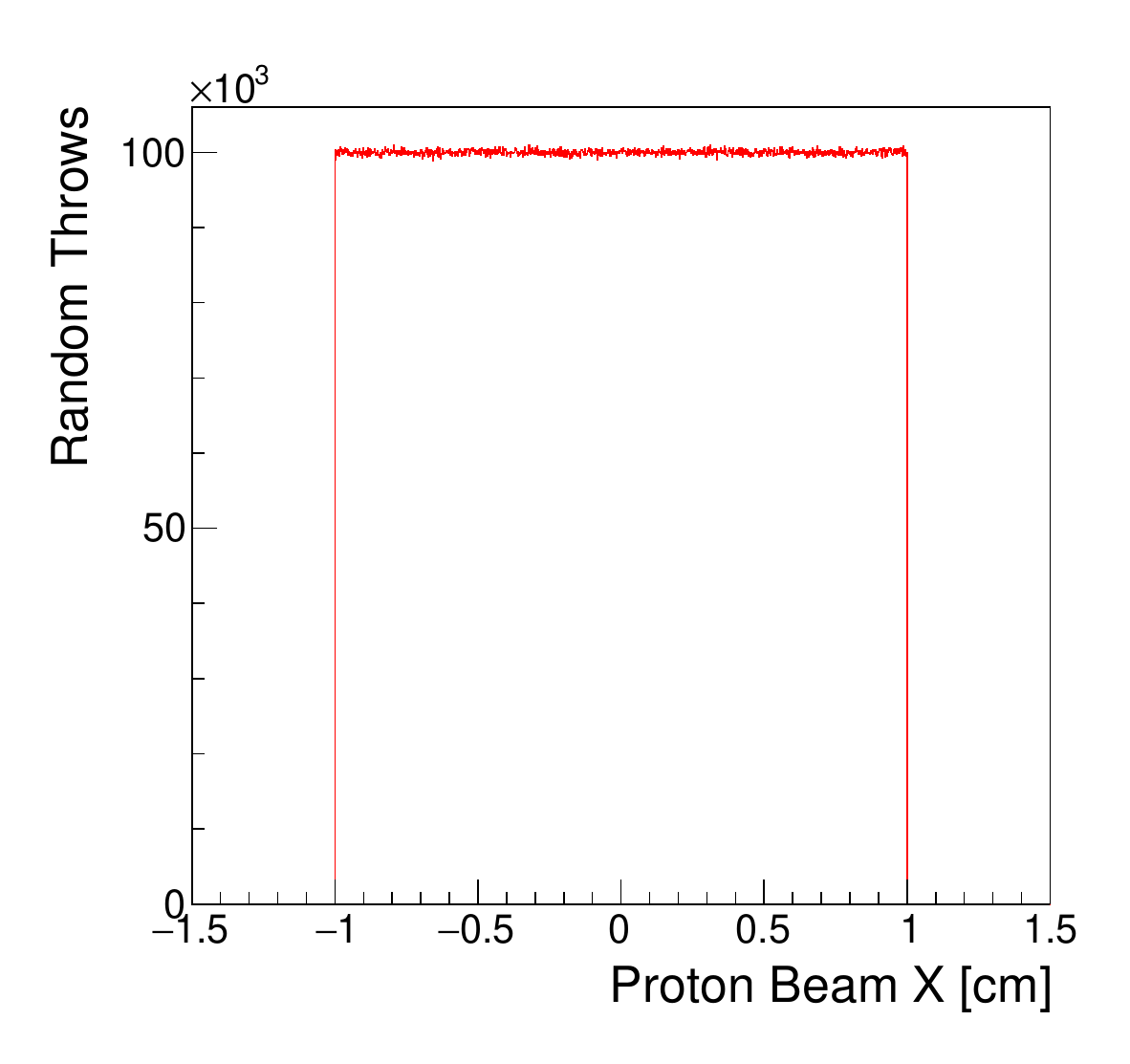}
\caption{Random throws along the horizontal beam positions to generate a uniform proton beam sample.}\label{fig:throws}
    \end{center}
  \end{figure}
Using this method, we can run the simulation scan through a set of beam variable values comparable to the actual beam scans. 
In this process, the weights ($w_i$) are calculated based on selected two-dimensional normalized Gaussian functions as shown in Eq.~\eqref{eq:gauss}:
\begin{equation}
w_i =  \frac{1}{2\pi \sigma_x \sigma_y}\cdot \exp\left \{-\frac{(x_i - \mu_x)^2}{2\sigma_x^2} -\frac{(y_i - \mu_y)^2}{2\sigma_y^2} \right\}\label{eq:gauss}
\end{equation}
The $w_i$ is the calculated weight at the $i^{th}$ beam position ($x_i,y_i$). The beam spot size ($\sigma_x$ and $\sigma_y$) and the beam centroid position at the target ($\mu_x,\mu_y$) have been fixed for selected beam settings. 
These weights are applied during the process of filling the histograms on the observable variables.
Steps involved in the proposed simulation technique are as follows:
\begin{itemize}
\item Prepare uniformly distributed simulation samples on a selected beam scan variable for a selected variable range.
\item Calculate the event weights according to a selected Gaussian function. 
\item Apply the calculated weights on the observable variables. 
\item Repeat the weight calculation algorithm by varying the Gaussian mean through the selected beam scan range. 
\end{itemize}
Every event from the uniformly distributed sample is contributing to each selected Gaussian sample based on their weight distribution. 
%The calculated weights are unique for each Gaussian profile.
%The selected Gaussian profiles may share the same weights at the overlapping points. 
%Events at the overlapping point provide equal contribution for selected Gaussian samples. 
%\textbf{\textcolor{red}{Add an explanation on how each event will get a different weight under different Gaussians even if they are created from the same uniform distribution. Add a pictorial description.}}
%This technique is demonstrated by combining the g4numi simulation with the muon monitor simulation packages. 
%To demonstrate the technique, a uniformly distributed sample of proton throws onto the target is shown in Fig.~\ref{fig:throws} where a Uniform distribution sample is generated along the horizontal proton beam positions with a 0.15 cm beam width and zero centroids. 

To demonstrate the technique, a uniformly distributed sample of protons on the target as shown in Fig.~\ref{fig:throws} is used to generate possible interactions with matter.
\begin{figure}%[htpb!]
  \begin{center}%\setlength{\unitlength}{1.0cm}
   \includegraphics[width=\linewidth]{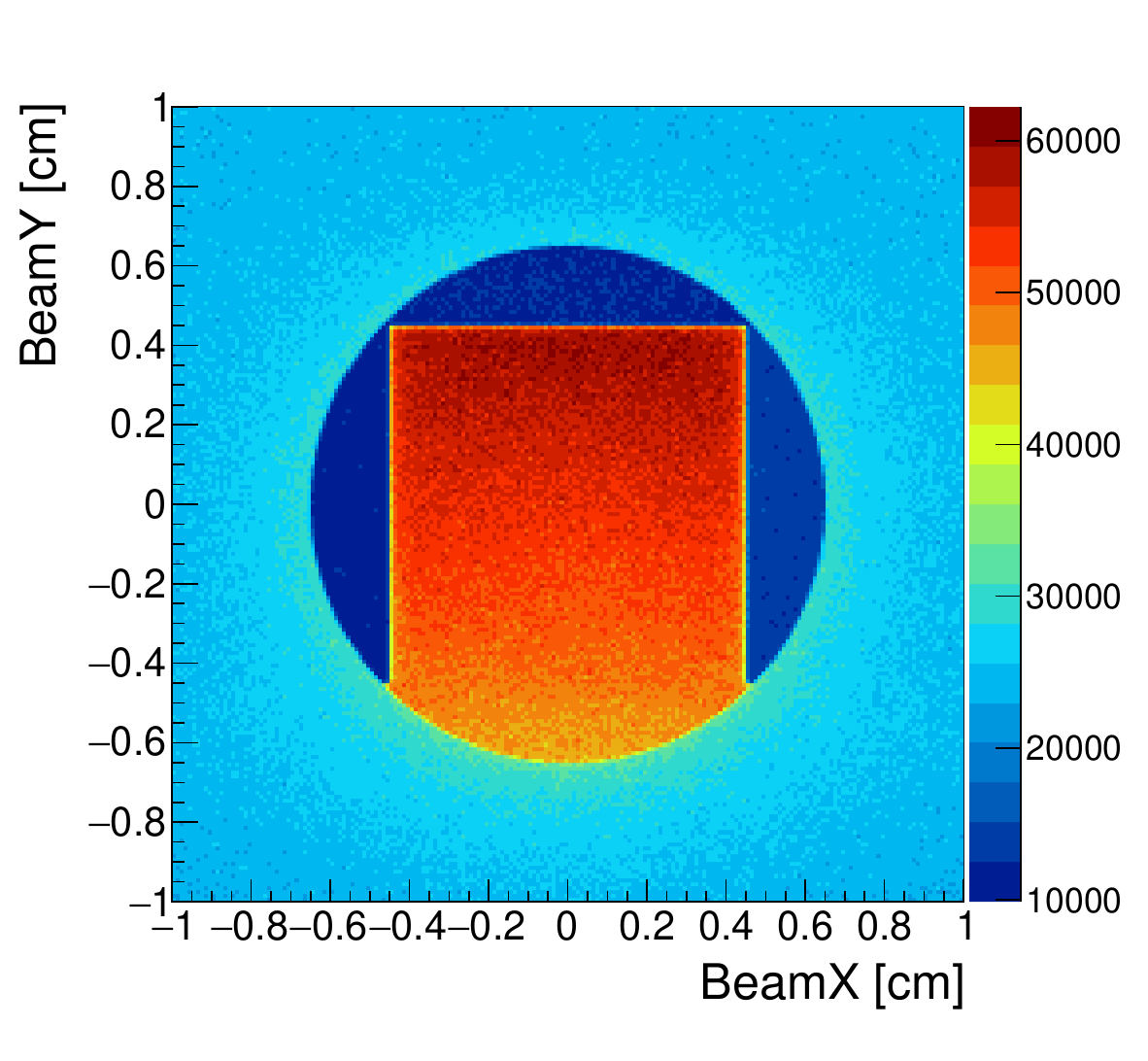}
\caption{Two-dimensional horizontal vs.\ vertical proton beam positions with uniform distribution sample for recorded beam interactions on the NuMI target, that has a neutrino candidate at the downstream neutrino detectors.}\label{fig:beamXY}
    \end{center}
  \end{figure}
From this uniformly distributed sample, a two-dimensional distribution of horizontal and vertical proton beam positions on the NuMI target is drawn as shown in Fig.~\ref{fig:beamXY}, for each recorded beam interaction that has a neutrino candidate at the downstream near detector.
The recorded events include beam interactions with the carbon target, aluminum baffle and downstream matter. 
\begin{figure}%[htp!]
  \begin{center}%\setlength{\unitlength}{1.0cm}
   \includegraphics[width=\linewidth]{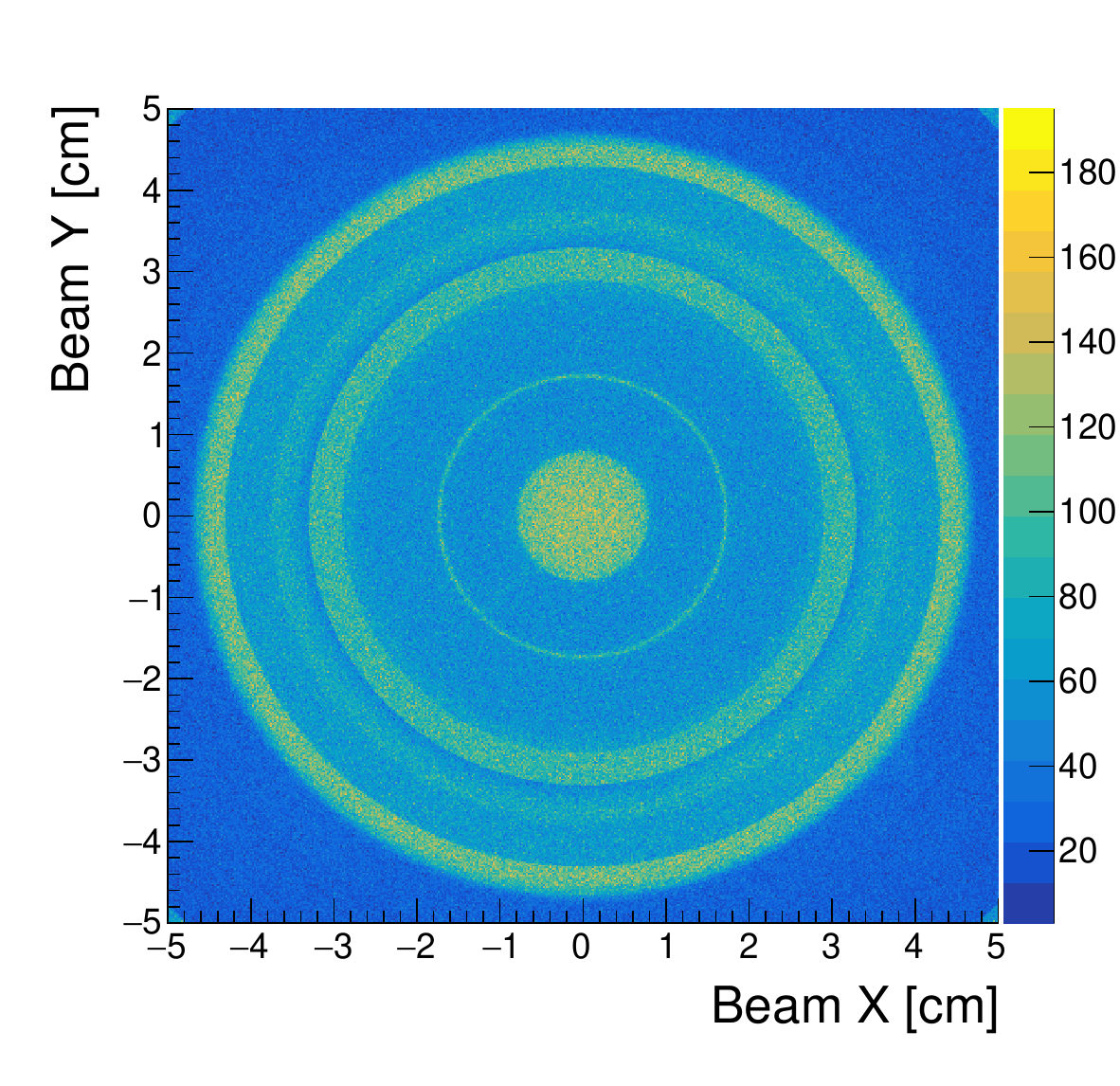} 
\caption{Two-dimensional horizontal vs.\ vertical proton beam positions with uniform distribution sample for recorded beam interactions on the LBNF target that has a neutrino candidate at the downstream neutrino detector.}\label{fig:beamXYLBNF}
    \end{center}
  \end{figure}
The same exercise has been performed with the g4lbnf simulation in the forward horn current mode to draw two-dimensional proton beam horizontal vs.\ vertical positions on the LBNF target as shown in Fig.~\ref{fig:beamXYLBNF}. A detailed description of the LBNF optimized beam design, performed using the g4lbnf simulation has been documented in the conceptual design report for the optimized LBNF beamline~\cite{lbnfbeamline}.
As shown in the Table.~\ref{tab:beam_widths} below, g4numi and g4lbnf use different nominal beam spot sizes. As per NuMI and LBNF specifications, these values have been chosen in all our studies since they represent the recommended beam spot sizes for high-power beam operations at the respective beamlines.
\begin{table}[bth]
\centering
\begin{tabular}{ c| >{\centering\arraybackslash}m{3.5cm}  }
\hline
 & Beam spot size (cm) $\sigma$ \\
\hline
  g4numi & 0.15  \\ 
\hline
 g4lbnf & 0.27 \\ 
\hline
\end{tabular}
\caption{Nominal beam spot size values for the g4numi and g4lbnf simulations.\label{tab:beam_widths}}
\end{table}

\begin{figure}%[htp!]
  \begin{center}%\setlength{\unitlength}{1.0cm}
    \includegraphics[width=\linewidth]{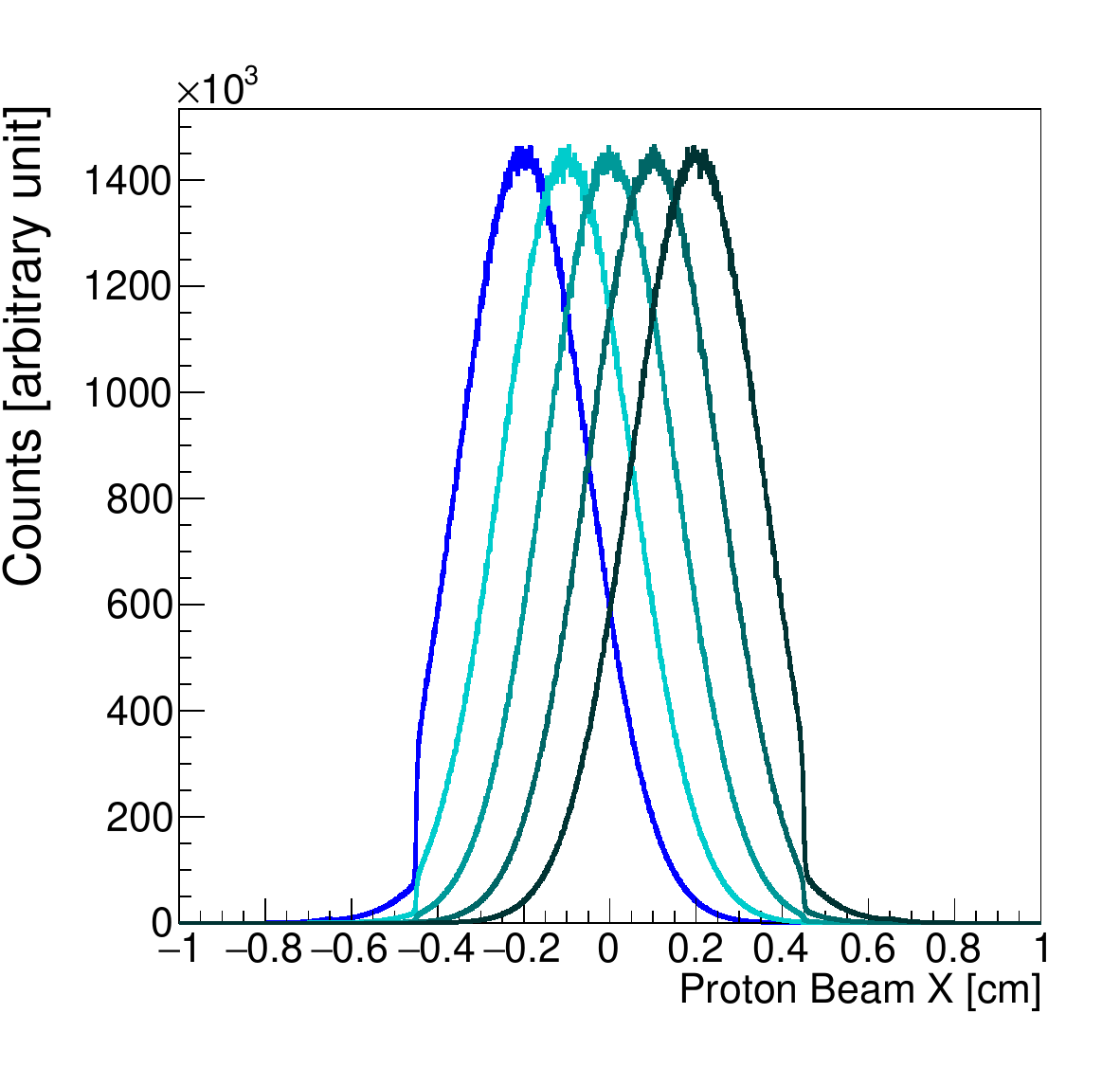}
    \includegraphics[width=\linewidth]{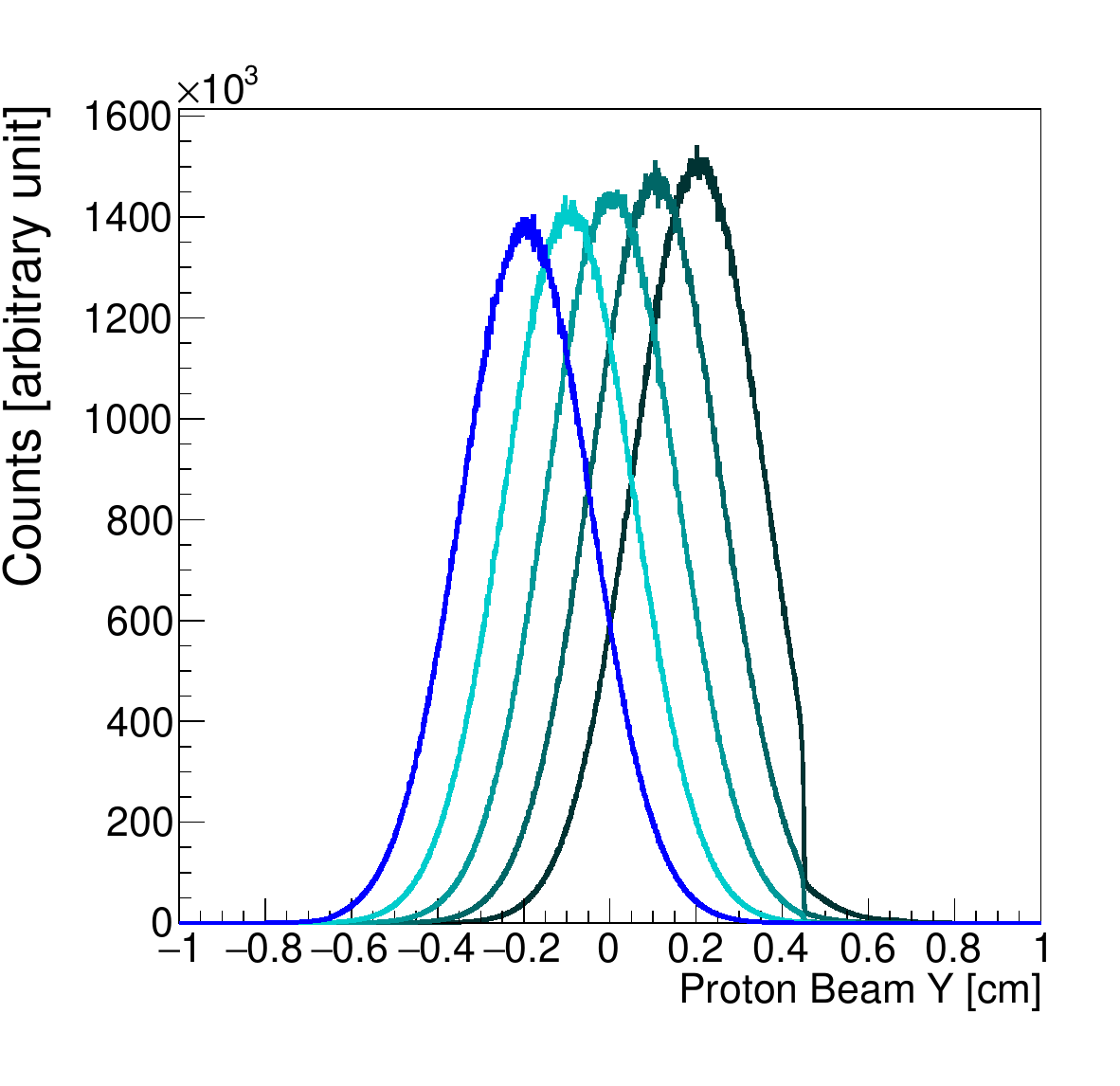} 
\caption{The horizontal (top) and vertical (bottom) proton beam positions for recorded beam interactions that have a neutrino candidate at the downstream neutrino detectors.}\label{fig:beamXnY}
    \end{center}
  \end{figure}

\begin{figure}%[htp!]
  \begin{center}%\setlength{\unitlength}{1.0cm}
  %\begin{tabular}{@{}cc@{}}
   %\includegraphics[width=.23\textwidth]{beamX_wi} &
    \includegraphics[width=\linewidth]{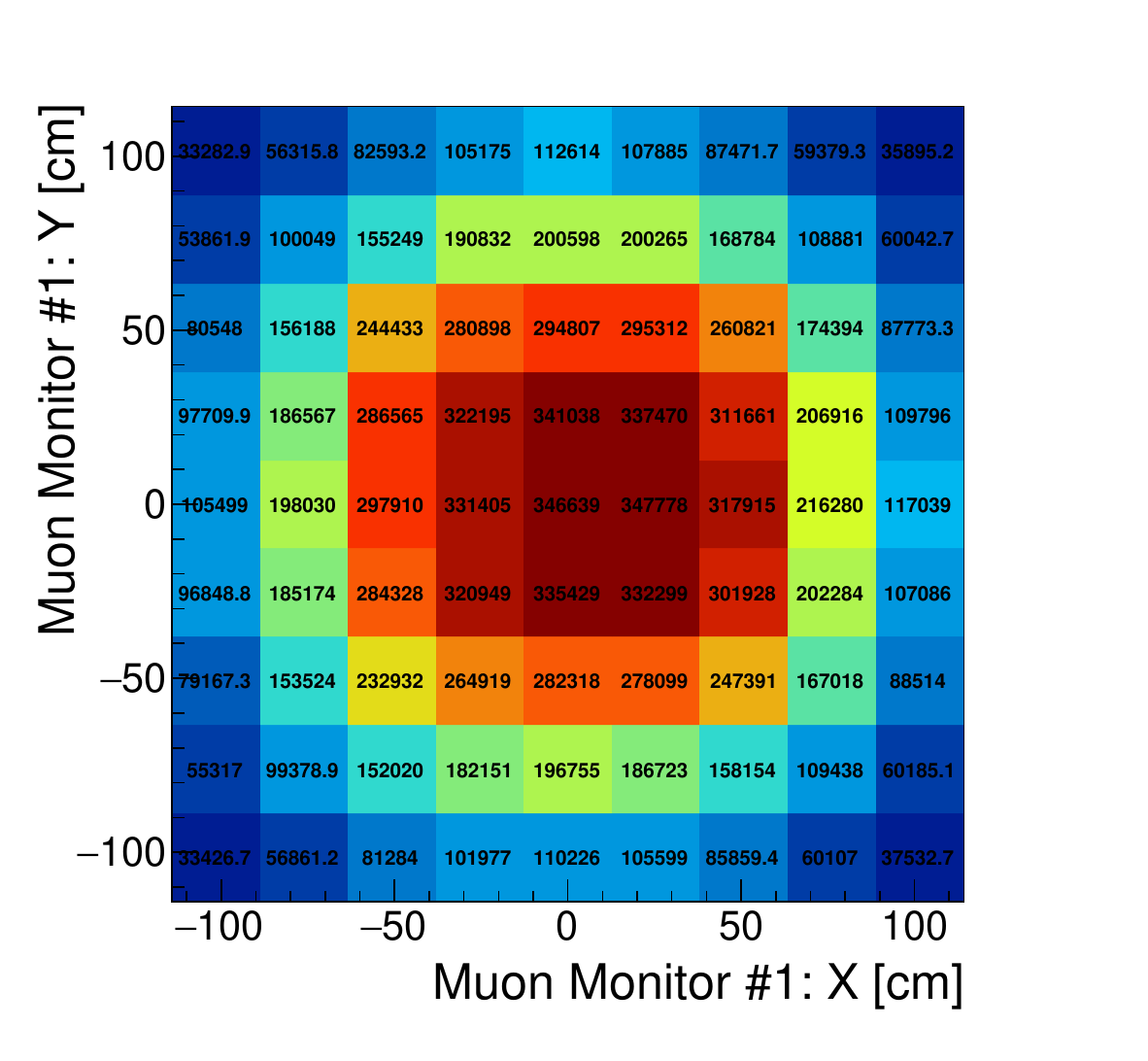} 
    %\end{tabular}
\caption{The observed two-dimensional muon distribution on NuMI muon monitor 1 for a given beam $\mu=0.0$\,cm and $\sigma = 0.15$\,cm.}\label{fig:wi_mm}
    \end{center}
  \end{figure}

%where a Uniform distribution sample is generated along the horizontal proton beam positions with a 0.15 cm beam width and zero centroids. 
%From the uniformly distributed proton throws, the weights for different proton beam settings are calculated using Eqn.\ref{eq:gauss}. 
Next, the Gaussian weights for different proton beam configurations are calculated using Eq.~\eqref{eq:gauss}. 
\begin{figure}%[htp!]
  \begin{center}%\setlength{\unitlength}{1.0cm}
   \includegraphics[width=\linewidth]{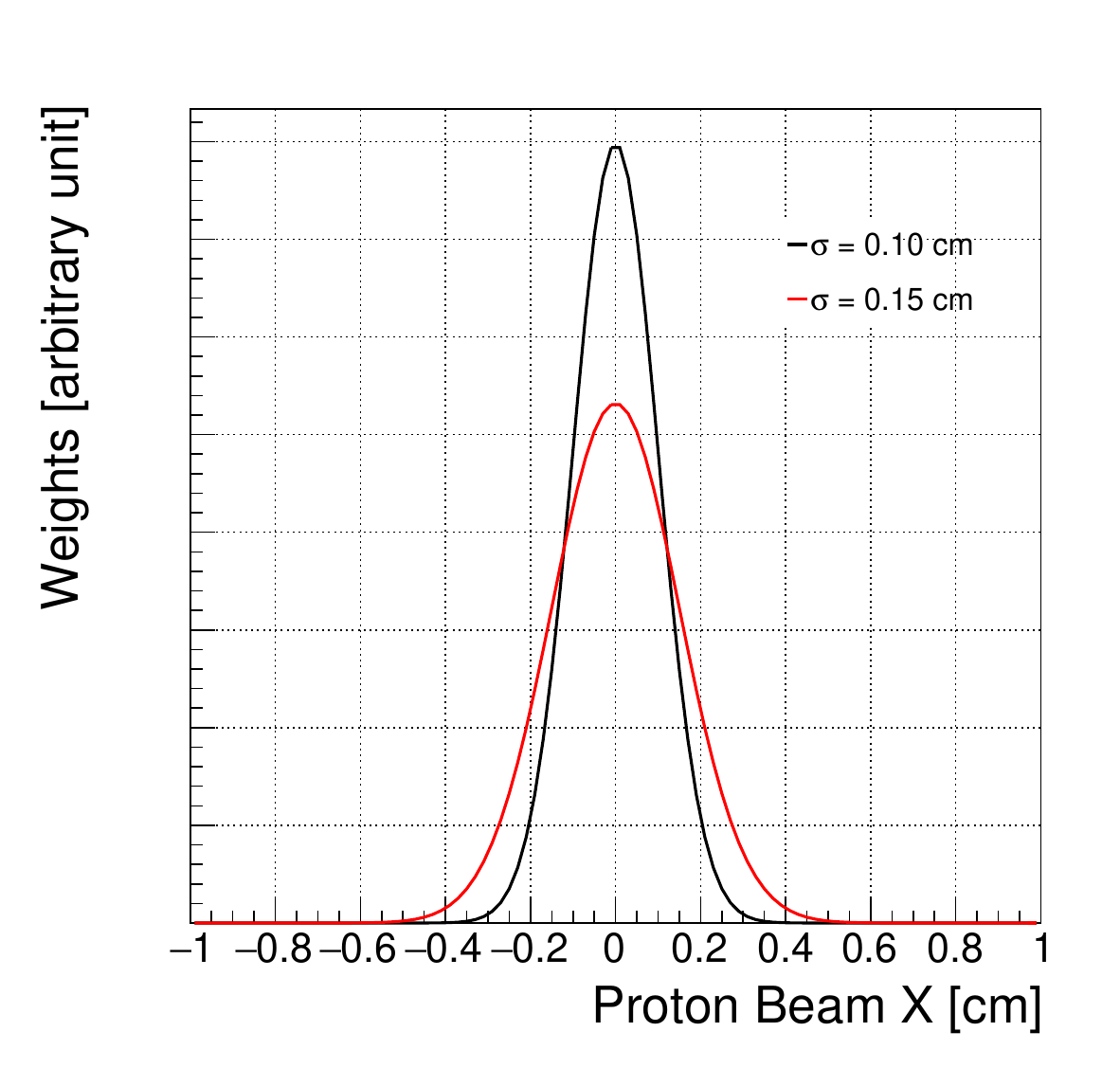}
   \includegraphics[width=\linewidth]{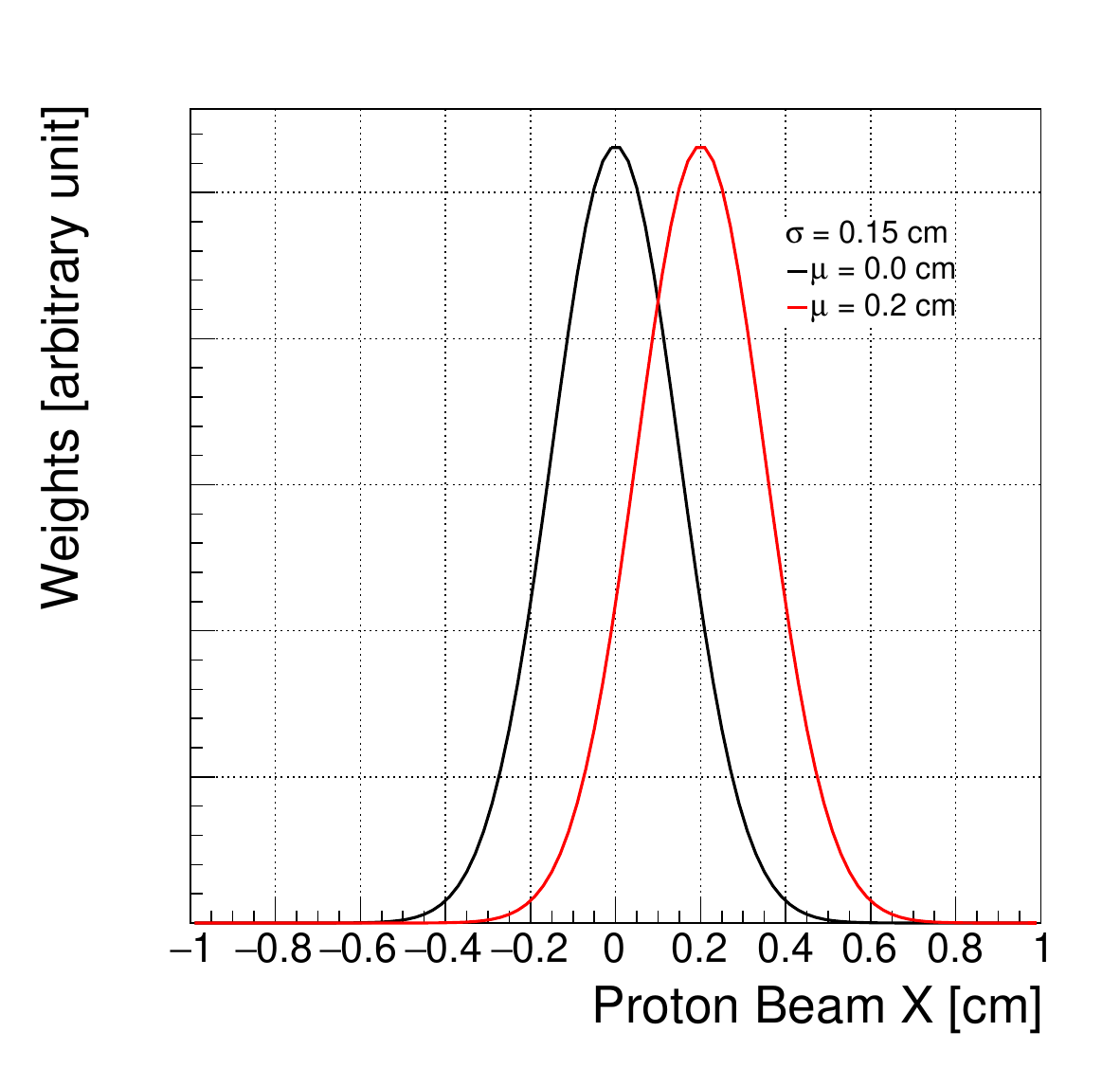}
\caption{Example of calculated weights for different beam settings with varying beam width (top) and mean (bottom).}\label{fig:wi}
    \end{center}
  \end{figure}
The top plot in Fig.~\ref{fig:wi} shows two examples of the calculated weights for a given beam setting with mean $\mu$ = 0.0\,cm and widths $\sigma$ = 0.10\,cm and $\sigma$ = 0.15\,cm. The bottom plot in Fig.~\ref{fig:wi} shows two examples of the calculated weights with beam $\sigma$ fixed at 0.15\,cm and varying $\mu$ between 0.0\,cm and 0.2\,cm. 
The third step is to weight each event in the distribution to select corresponding proton beam interactions horizontally and vertically as shown in Fig.~\ref{fig:beamXnY}. Five Gaussian distributions as shown in Fig.~\ref{fig:beamXnY} for the proton beam in the horizontal direction are drawn by changing the proton beam mean $\mu$ from $-0.2$\,cm to $0.2$\,cm while keeping the Gaussian width $\sigma$ fixed at 0.15\,cm. 
The vertical proton beam distribution is asymmetric since the target geometry is asymmetrical in y. In each Gaussian profile, the weights are distinct.
%By increasing the proton beam mean $\mu$ from $-0.2$ cm to $0.2$ cm and keeping the Gaussian width $\sigma$ at $ 0.15$ cm, five Gaussian distributions are drawn in the horizontal and the vertical direction respectively.
  %----------------------
 %\begin{figure}[htp!]
  %\begin{center}\setlength{\unitlength}{1.0cm}
  %\begin{tabular}{@{}cc@{}}
   %\includegraphics[width=.23\textwidth]{beamX} &
    %\includegraphics[width=.23\textwidth]{beamY} 
    %\end{tabular}
%\caption{The horizontal (left) and vertical (right) proton beam positions for recorded beam interactions which has a neutrino candidate at the downstream neutrino detectors.}\label{fig:beam_xy}
    %\end{center}
  %\end{figure}
This demonstrates that starting with a single uniformly distributed sample of proton beam in both horizontal and vertical directions, for which there is a recorded beam interaction, as many Gaussian distributions in either horizontal or vertical proton beam direction can be selected. 
In the simulation, pions decay into multiple muons, which are propagated through the absorber and the muon monitors. The corresponding weights have been applied to the observed muon flux from the Gaussian proton beam slice with $\mu=0.0$\,cm to obtain the two dimensional event distributions on the muon monitors as shown in Fig.~\ref{fig:wi_mm}. 
%The corresponding weighted distributions of the proton beam positions along the horizontal axis are shown in Fig.~\ref{fig:wi_mm} (left).
%Each individual event in the distribution in Fig.~\ref{fig:beamXY} in the beam horizontal direction has been weighted to select all corresponding proton beam interactions as shown in Fig.~\ref{fig:wi_mm} (left). 
  %------------

\begin{figure}%[htp!]
  \begin{center}%\setlength{\unitlength}{1.0cm}
   \includegraphics[width=\linewidth]{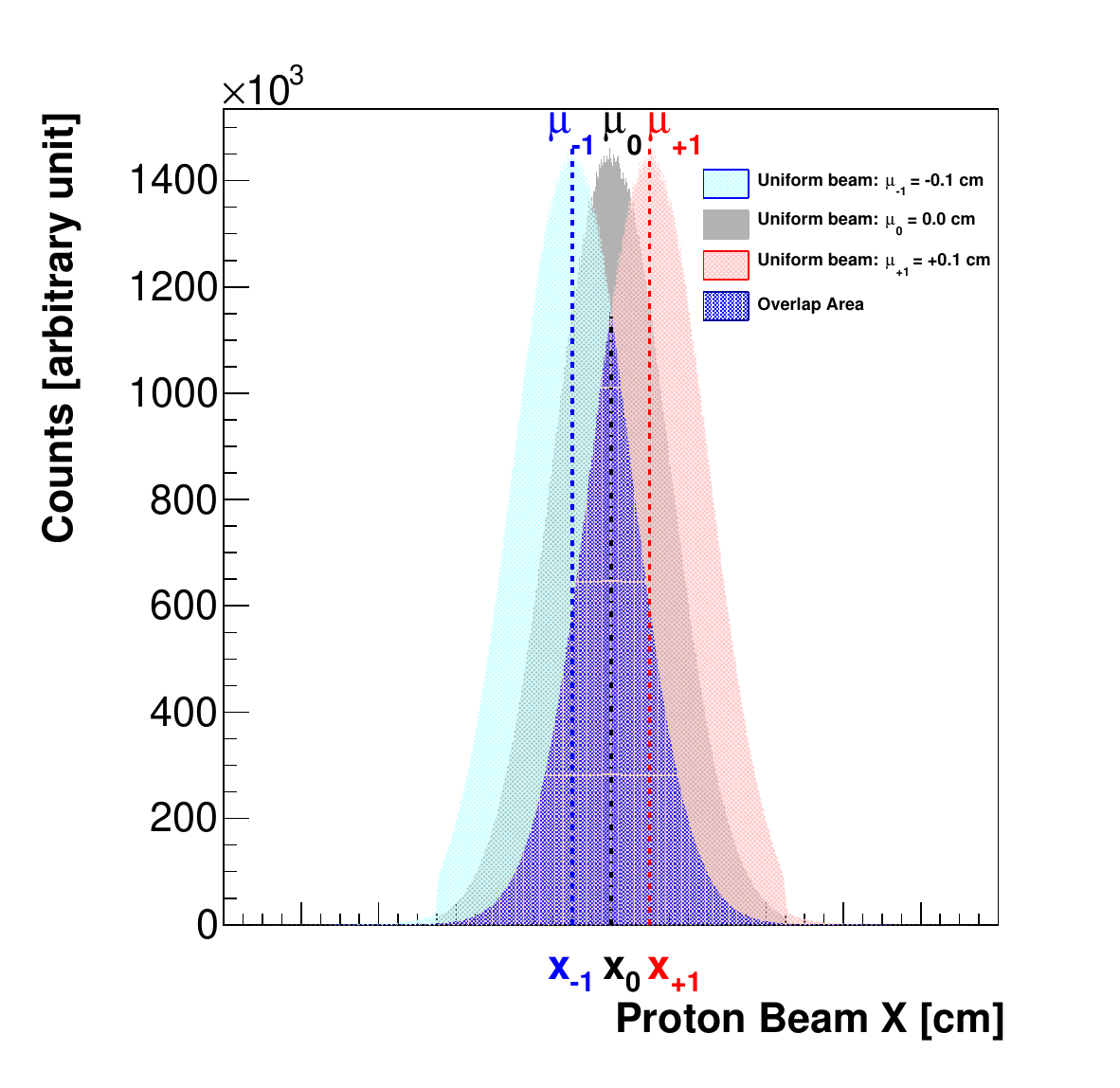}
\caption{Two selected Gaussian profiles from the uniform simulation, showing the volume of overlap between them.}\label{fig:overlap_gaussian}
    \end{center}
  \end{figure}

The uniform beam simulation allows us to generate as many Gaussian samples as we require. 
It is possible to draw samples with means $\mu_{i}$ with i ranging from $-\infty$ to $+\infty$. Figure \ref{fig:overlap_gaussian} illustrates an example case with three Gaussian samples with means of $\mu_{-1}$, $\mu_0$, and $\mu_{+1}$. The $\mu_{-1}$ value in this example is $-0.1$\,cm, the $\mu_0$ value is 0.0\,cm, and the $\mu_{+1}$ value is +0.1\,cm. For all three samples, the width $\sigma$ has been fixed at 0.15\,cm. 
Our reference Gaussian is the Gaussian sample with mean $\mu_0$.
By using the uniform beam method, we start with a static pool of randomly thrown events interacting with matter to produce neutrinos. 
Gaussian samples are generated based on weights by selecting events from the same static pool. However, this is not a concern, because we give different statistical weights to each event. These statistical weights determine with what probability these events will contribute to the Gaussian samples. 
Within one Gaussian sample, no two events will have the same statistical weight, as the weight is calculated using Eq.~\eqref{eq:gauss}. 
In the case of two Gaussian samples, there might be a scenario in which the same event could be shared with the same statistical weight between these two samples as shown in Fig.~\ref{fig:overlap_gaussian}. 
%In the uniform beam simulation, no two events will have the same proton beam positions on target. 

\begin{figure}%[htp!]
  \begin{center}%\setlength{\unitlength}{1.0cm}
   \includegraphics[width=\linewidth]{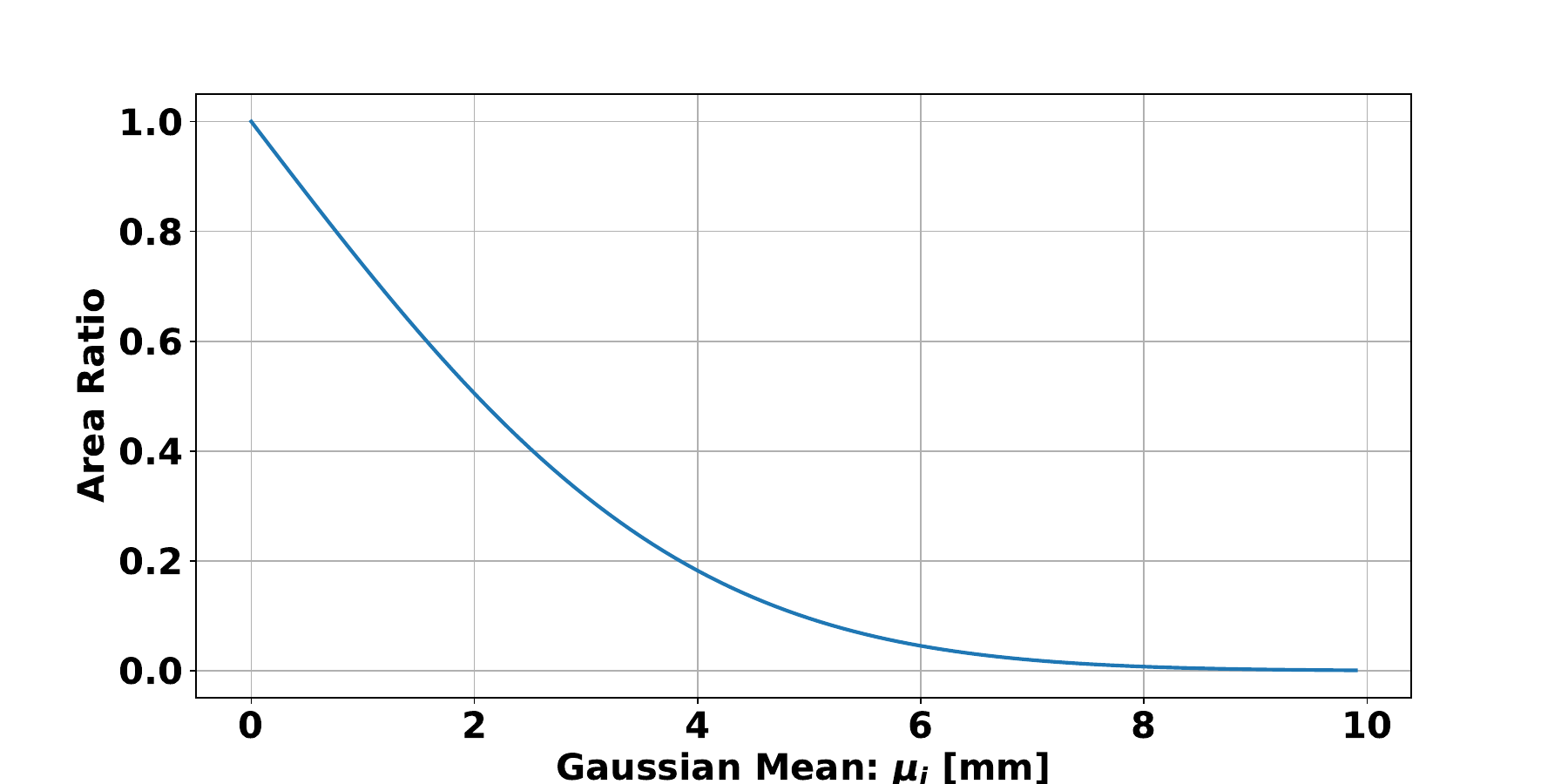}
\caption{Ratio of the overlapping volume of Gaussian beam profiles for different $|\mu - \mu_0|$.}\label{fig:area_r}
    \end{center}
  \end{figure}

In Fig.~\ref{fig:overlap_gaussian}, we can see that by plotting two Gaussian samples with $\mu_{-1}=-0.1$\,cm and $\mu_{+1}=+0.1$\,cm, an overlapping volume is created between these two Gaussian samples, as shown by the shaded blue region. 
%Two events belonging to two different Gaussian samples can have the same weight in this overlap region, but they are truly two different events as each will have a unique beam position on the target and hence unique interactions with matter.
It is possible for two events belonging to the above two Gaussian samples to have the same weight in this overlap region, but they are really two different events since they each have a unique beam position and will interact with matter differently.
Only the event at proton beam position $x_0$, shown by the black dotted line will have the same interactions. So, it is only the event shown by the black dotted line which will be shared by the Gaussian with $\mu_{-1} = -0.1$\,cm as well as the Gaussian with $\mu_{+1} = +0.1$\,cm.
%\textcolor{red}{However, the events within this overlap, each has unique interactions with matter except for 
%the event shown by the black dotted line at $\mu$ = 0.0 cm.
%As a result, the event along the black dotted line will be part of the Gaussian shown in blue and also the Gaussian shown in red, but the beam configurations are distinct in each case.}
Within one generated Gaussian sample (for example the Gaussian with mean $\mu_{-1}$), we are using the same event only once. 
Equation~\eqref{eq:gauss_overlap} estimates the ratio of the overlapping volume $R_i$ between a Gaussian sample with mean $\mu_i$ and a given beam width $\sigma$ as compared to the reference Gaussian sample with mean $\mu_0$.
%\begin{figure}[htp!]
 % \begin{center}\setlength{\unitlength}{1.0cm}
  % \includegraphics[width=.31\textwidth]{AllNewPlots/overlap_gaussians.pdf}
%\caption{Two selected Gaussian profiles from the uniform simulation, showing area of overlap between them.}\label{fig:overlap_gaussian}
 %   \end{center}
  %\end{figure}
\begin{equation}
R_i =  \frac{2\int_{-\infty}^{x_0}\mathcal{N}(\mu_i,\,\sigma^{2})dx}{\int_{-\infty}^{+\infty}\mathcal{N}(\mu_0,\,\sigma^{2})dx}\label{eq:gauss_overlap}
\end{equation}
Figure~\ref{fig:area_r} plots $R_i$ as a function of $\mu_i$ indicating that the overlapping volume will be maximum when the mean of the $i$-th Gaussian aligns with the mean of the central Gaussian slice (with $\mu_0$).
%$|\mu_i-\mu_0|$ = 0. 
%In this example, we have set $\mu_0$ to be 0.0 cm.
As the mean of the Gaussian $\mu_i$ moves away from $\mu_0$, the overlapping volume between two consecutive Gaussian samples will decrease according to Eq.~\ref{eq:gauss_overlap}. 
%The further apart the Gaussian samples are, or in other words 
The larger $|\mu_i - \mu_0|$ is, the smaller the overlap will be, reducing the weight of the shared event between two consecutive Gaussian samples. Consequently, the statistical contribution of this shared event will decrease according to Eq.~\ref{eq:gauss_overlap}.
Using Figure~\ref{fig:area_r}, we can determine the overlap between two Gaussian samples at a given separation between their means. By separating Gaussian means by 2$\sigma$, for example, the overlap volume is reduced from 100\% to 30\%, compared to the case when the samples have the same mean. 
\section{Simulation Data}\label{sec:simdata}
Validation and demonstration of the proposed technique has been performed by generating several data samples. 
We have generated a uniformly simulated beam sample of 1 billion protons on target (POT). Using FermiGrid, this sample was the largest we were able to produce under the limit of available storage per user. FermiGrid is a collection of computing resources that Fermilab Computing Division makes available through grid protocols. It would be possible to create multiple 1 billion uniform beam samples by utilizing supercomputers.\par
While real measurements utilize the three muon monitors to measure integrated flux of muon beams that exceed 5\,GeV and two other energies separated by at least 7\,GeV, simulated data samples can only be generated using muon monitors 1 and 2 as computing resources are limited. Simulating muon monitor 3 responses will require supercomputer resources in order to generate large statistics.\par
In each Gaussian beam sample, 250 million POTs were generated with different beam positions while maintaining a large beam spot size of 0.15\,cm. Based on the beam spot size ($\sigma$ = 0.15\,cm) and the cross sectional area of the target and the baffle, the vertical and horizontal limits of the uniform beam simulation data are chosen to be 1.0 $\times$ 1.0\,cm perpendicular to the beam direction.  %The horizontal and the vertical limitations of the uniform beam simulation data, 1.0 cm $\times$ 1.0 cm perpendicular to the beam direction, is chosen based on the beam spot size ($\sigma = 0.15$) cm and the cross sectional area of the target and the baffle. 
%We have selected the uniform beam window to cover approximately 7 $\sigma$ of the Gaussian beam which allowed us to take account the interaction of the protons with the matter in the Gaussian beam tale. This limitations covers the target and the significant amount of the baffle. 
The uniform beam window was chosen to cover approximately $7 \sigma$ of the Gaussian beam, which allowed us to incorporate the interaction of the protons with matter in the Gaussian beam tail. The target and a significant portion of the baffle are covered by this limit.
%Fig.~\ref{fig:hadAngle} shows a two dimensional distribution of total hadron momentum vs.\ hadron angle at decay for a Gaussian slice of 250 million POT with a beam mean $\mu$ = 0.0 cm. %By weighting the uniform simulation sample of 1 billion POT, this Gaussian slice was created.
In order to prepare the horizontal and vertical beam scan data samples, Gaussian beams with mean values in the range of $[-0.2,0.2]$\,cm are generated from uniform beam simulation data. These beams cover a minimum of $5~\sigma$ beam profiles for each beam configuration.
%This uniform beam data sample is used to prepare the horizontal and the vertical beam scan data samples by generating Gaussian beams in the rage of $\pm$ 0.2 cm covering minimum of 5 $\sigma$ beam profiles for each beam configurations.
%This uniform beam data covers 5$\sigma$ of the Gaussian beam from $\pm 0.2$ cm beam centroid locations on the target. 
The randomly thrown protons in the uniform beam simulation interact with matter to produce secondary particles and are only recorded if there is a neutrino candidate at the near detector from the hadron decay.
For each proton beam position along the horizontal direction, Gaussian samples were created by applying weights with a beam width of 0.15\,cm. As a result, every Gaussian sample contains $\sim$250 million POTs. This is estimated based on the generated 1 billion POTs in the uniform beam sample.\\
%\textbf{\textcolor{red}{Add explanation on Gaussian slice overlap, estimate N(x), W(x) in the overlapping region for both uniform and nominal simulation techniques.
%List of plots to add:\\
%1. Plot two Gaussian slices from uniform, showing the overlapping area highlighted in a different color.\\
%2. Calculate N(x) in the overlapping area.\\
%3. Propose to add uniform samples with multiple seeds, create one final uniform sample by combining those. Then apply the whole technique.\\
%4. Highlight the importance of random seeds.\\
%Overall, add arguments about The appropriateness of re-sampling techniques. How many times, can we pull events from the same uniform sample, before we have over-sampled?
%}}
%setting up the beam width as 0.15 cm for each beam position configuration. 
%Each sample has 250 M equivalent protons on the target by covering the weighted Gaussian distribution as single Gaussian beam. 
By applying the Gaussian weight to random throws of $N$ POTs, we calculated statistically that we would select $\sim$ $\frac{1}{4}$ of the starting N POTs in one Gaussian beam. In order to verify the calculation, we generated one uniform beam profile with 1 billion POTs from which we drew Gaussian beam profiles by applying weights. Our calculation shows that each of these Gaussian beam profiles will contain 250 million POTs. Following this, we compared these Gaussian beam profiles with Gaussian beams from nominal beam simulations each containing 250 million POTs with the same beam width of 0.15\,cm. Therefore, we would generate a uniform simulation sample just once with four times as many POTs as a nominal simulation sample. On the other hand, we would have to generate a new Gaussian with 250 million POTs every time if we were to generate a new reference simulation sample with different beam parameter settings.\par
\begin{figure}%[htpb!]
\begin{center}%\setlength{\unitlength}{0.6cm}
\includegraphics[width=\linewidth]{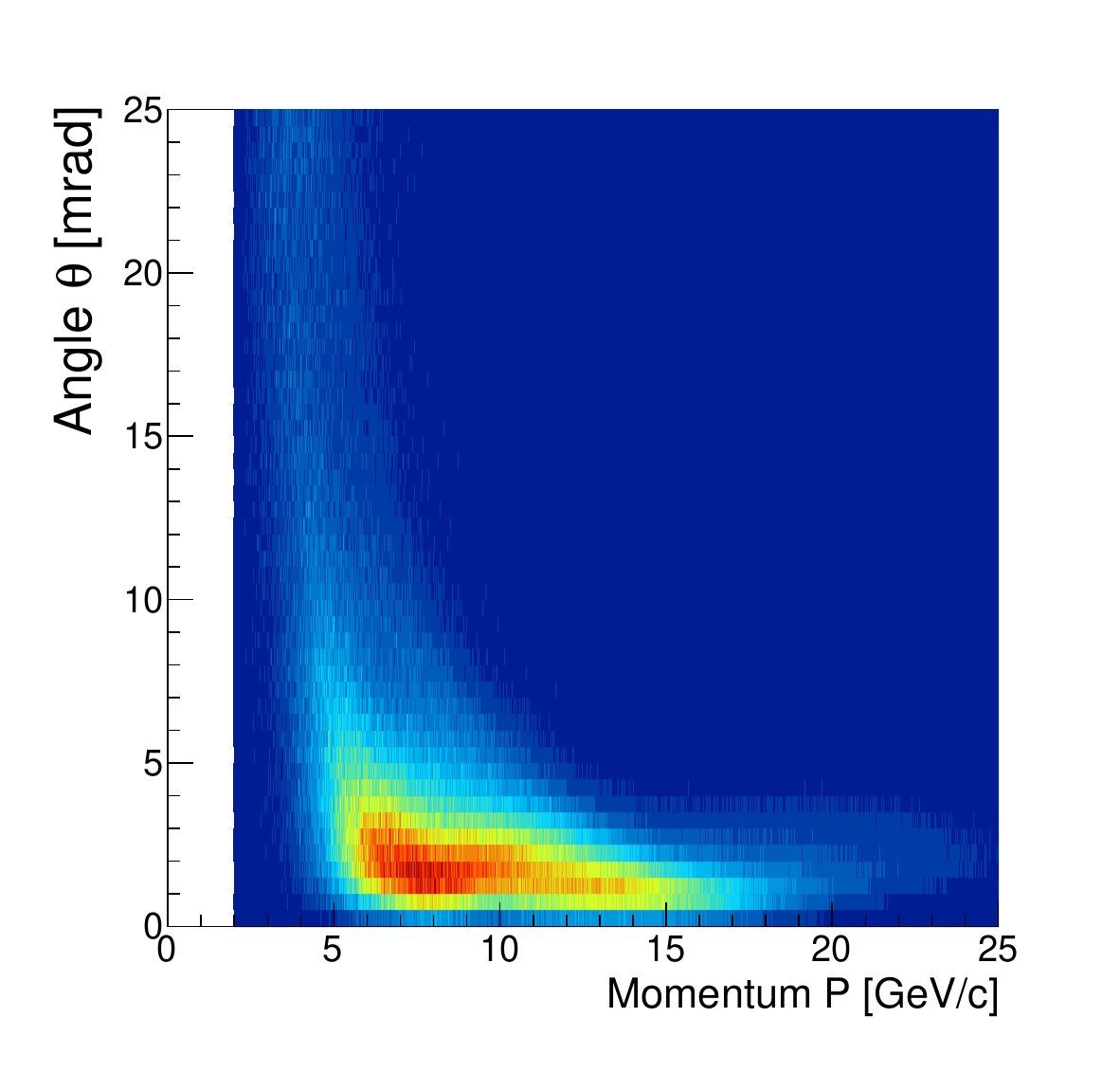}
\caption{Distributions of hadron angle and hadron momentum in the longitudinal direction at decay with uniform simulation for a Gaussian beam with $\mu$ = 0.0\,cm.}\label{fig:hadAngle}
\end{center}
\end{figure}
\begin{figure}%[htpb!]
\begin{center}%\setlength{\unitlength}{0.6cm}
\includegraphics[width=\linewidth]{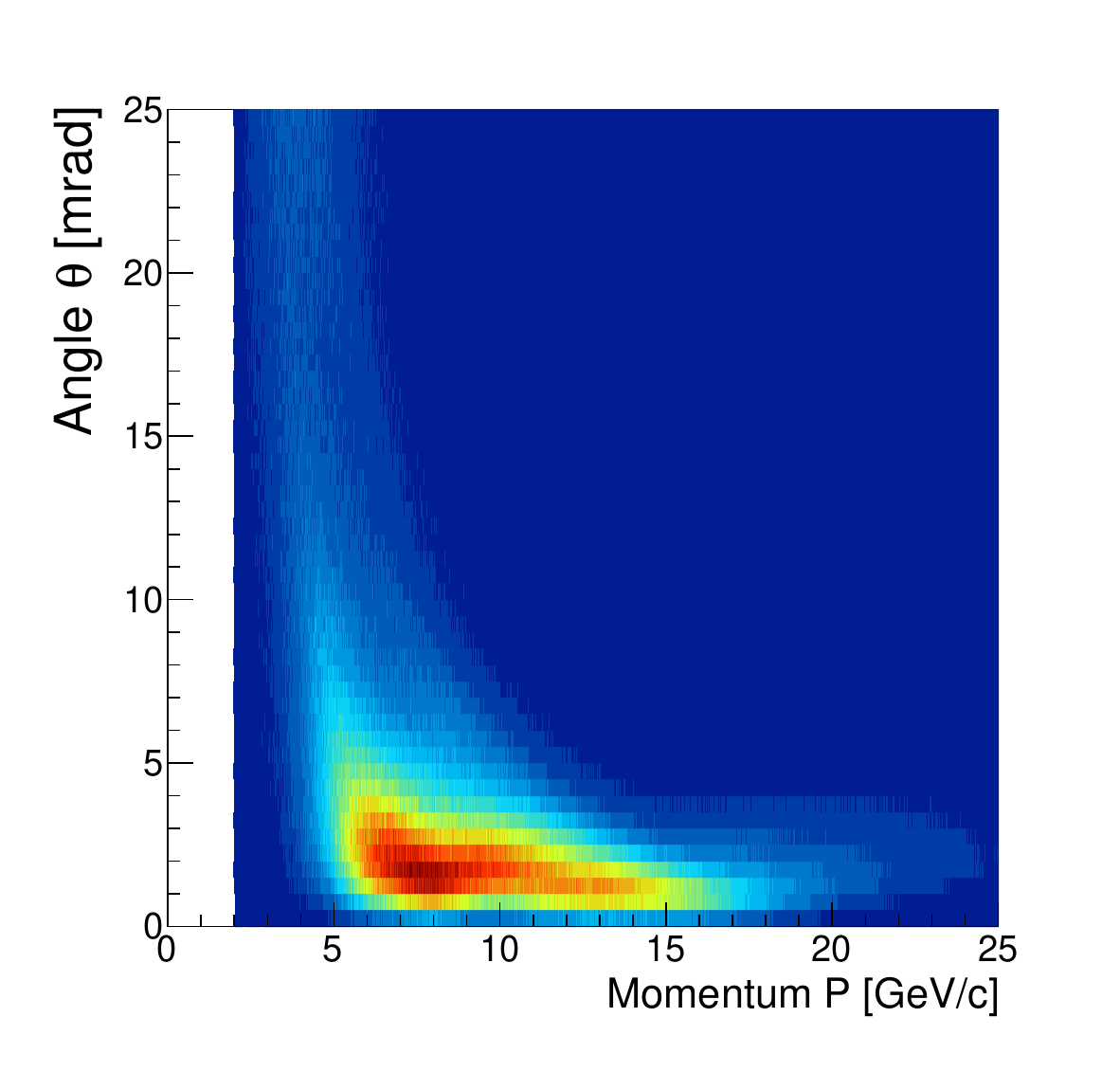}
\caption{Distributions of hadron angle and total hadron momentum  at decay with nominal simulation for a Gaussian beam of $\mu$ = 0.0\,cm.}\label{fig:hadAngle_nominal}
\end{center}
\end{figure}

Figure~\ref{fig:hadAngle} shows a two-dimensional distribution of total hadron momentum vs.\ hadron angle at decay for a Gaussian slice of 250 million POT with a beam mean $\mu$ = 0.0\,cm. By weighting the uniform simulation sample of 1 billion POT, this Gaussian slice was created. The same two-dimensional distribution of total hadron momentum vs.\ hadron angle at decay is plotted for a Gaussian slice of 250 million particles with a beam mean $\mu$ = 0.0\,cm using the nominal simulation, as shown in Fig.~\ref{fig:hadAngle_nominal}. 
These two plots only show events for which muon candidates with energies greater than 5\,GeV are present in muon monitor 1. 
%The advantage of doing so is that, since the POTs are increased by the factor of four only once, we can generate as many Gaussian beams as we need.
%Using this 1 billion POT, we generated a Gaussian beam in the horizontal direction with a mean of 0 cm. We selected 540 million events in the Gaussian beam as a result of beam interaction with the carbon target, aluminum baffle, and downstream matter. 
%With the nominal simulation, we need to generate a Gaussian beam with a mean of 0 cm and 250 million POTs in the horizontal direction if we are to obtain the same number of interactions as the nominal simulation. 
 %In contrast, a new Gaussian with 250 million POT has to be generated every time for 540 million interactions from the nominal simulation, which requires substantial computing power.

\section{Validation}\label{sec:val}
%A validation of the uniform beam simulation results has been performed against the nominal simulation results. 
%\textbf{\textcolor{red}{When comparing two distributions, add quantitative value from ks test: $Double_t ks = h1->KolmogorovTest(h2);$}}
Validation studies of the proposed beam simulation technique are an important requirement to prove the capability of reproducing the nominal simulation results. 
Therefore, two studies with the NuMI simulation have been carried out to validate the uniform beam technique against the nominal simulation:
\begin{itemize}
    \item Testing statistical reproducibility.
    \item Testing kinematics reproducibility.
\end{itemize}
%The results from the uniform beam simulation studies have been compared with the nominal beam simulation to validate our proposed simulation technique. 
In the validation studies, the uniform beam technique and multiple decays of hadrons were combined to increase the statistics of the muon events in the simulation.
The uniqueness of the uniform beam simulation is in that we can draw as many Gaussian beam profiles horizontally or vertically.
%We have created a uniform beam simulation sample with 1 billion protons-on-target (POT). 
\subsection{Testing Statistical Reproducibility}
We generated five Gaussian beams using the nominal simulation, each containing 250 million POT, and one uniform simulation containing a total of 1 billion POT to compare beam profiles. We drew five Gaussian weighted proton beam samples from the uniform simulation sample with beam centers at  -0.2\,cm, -0.1\,cm, 0\,cm, 0.1\,cm, and 0.2\,cm along the horizontal direction and compared them to the nominal Gaussian beams.
%The statistical reproducibility study is performed by using independent simulation samples from the uniform beam data and nominal beam simulation data. 
%Using the uniform beam data sample, we generated 5 Gaussian weighted proton beam samples with the beam center at -0.2 cm, -0.1 cm, 0 cm, 0.1 cm, and 0.2 cm along the horizontal direction. 
\begin{figure}%[htpb!]
\begin{center}%\setlength{\unitlength}{0.5cm}
\includegraphics[width=\linewidth]{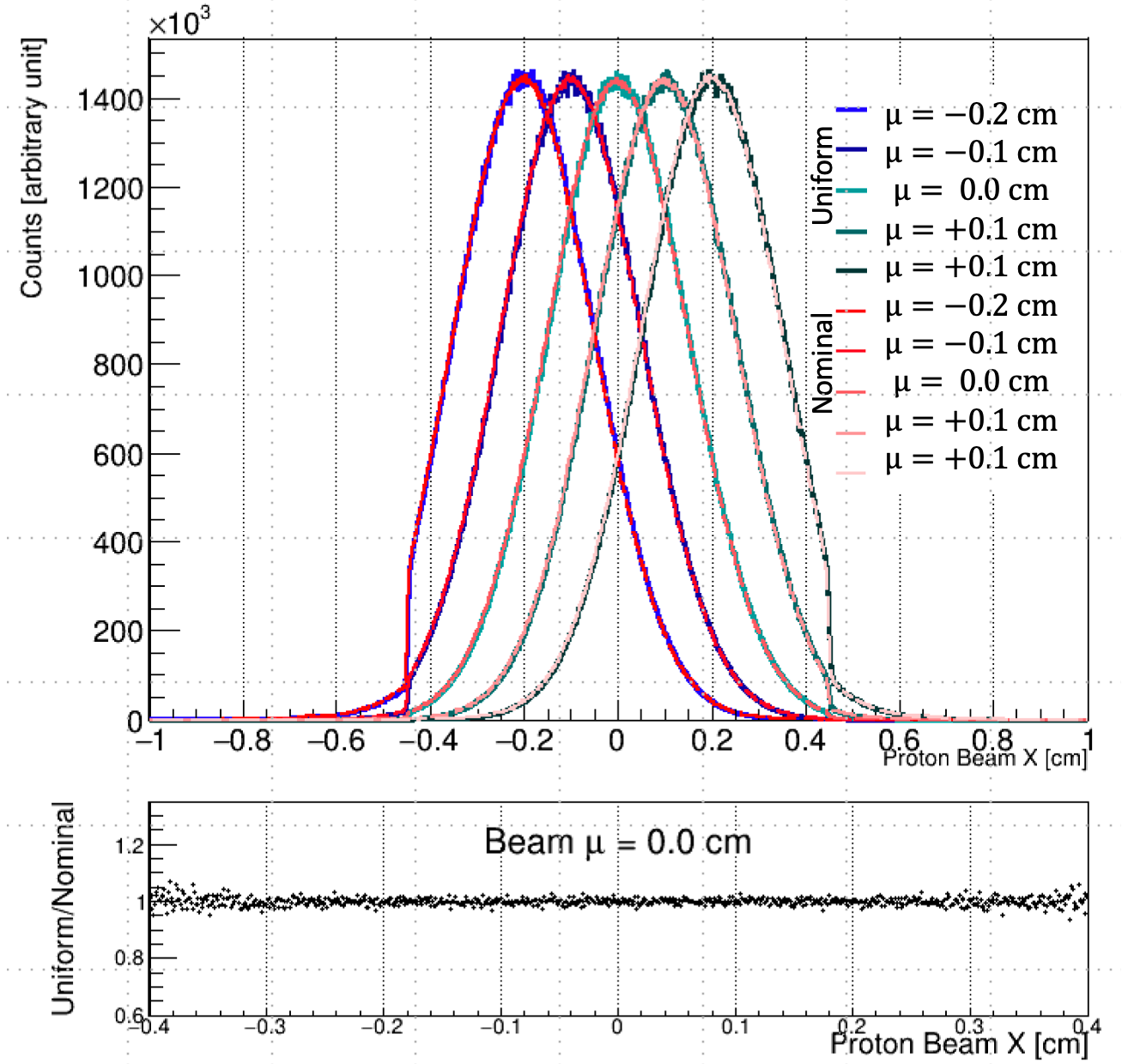}
\caption{Comparison of the proton beam profiles drawn from uniform and nominal simulations in the horizontal direction with the mean spanning from $-0.2$ to $+0.2$\,cm. %The red line represents a Gaussian slice from the uniform beam simulation, whereas the green line shows the same Gaussian slice from the nominal beam simulation. 
All plots shown are area normalized.}\label{fig:beamprof}
\end{center}
\end{figure} 
Figure~\ref{fig:beamprof} shows a comparison of the proton beam profiles from the uniform beam simulation (cyan) and Nominal beam simulation (red) at five different beam positions along the horizontal direction. The beam profiles have been plotted only for the incident protons that have corresponding muon candidates from the decay of hadrons which were created by the proton-matter interactions. 
The uniform and nominal simulations match up with all five beam profiles.
The bottom plot shows the ratio of two beam profiles for the mean of 0.0\,cm based on uniform and nominal simulations. A ratio of one confirms an excellent agreement.
%This plot describes the corresponding incident protons that provides muon candidates from the decay of hadrons those created from  proton-matter interactions. 

\begin{figure}%[htpb!]
\begin{center}%\setlength{\unitlength}{0.6cm}
\includegraphics[width=.85\linewidth]{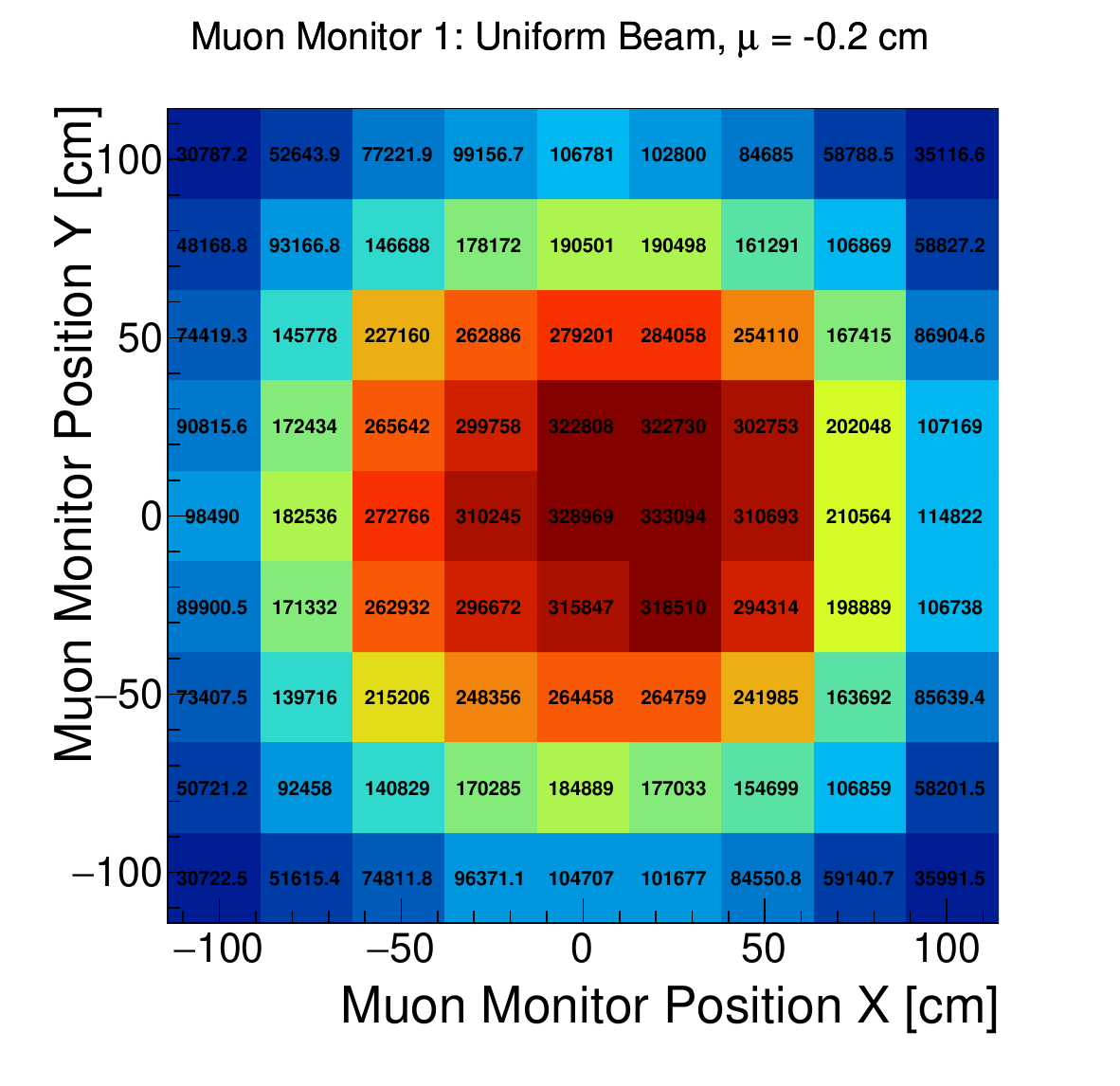}
\includegraphics[width=.85\linewidth]{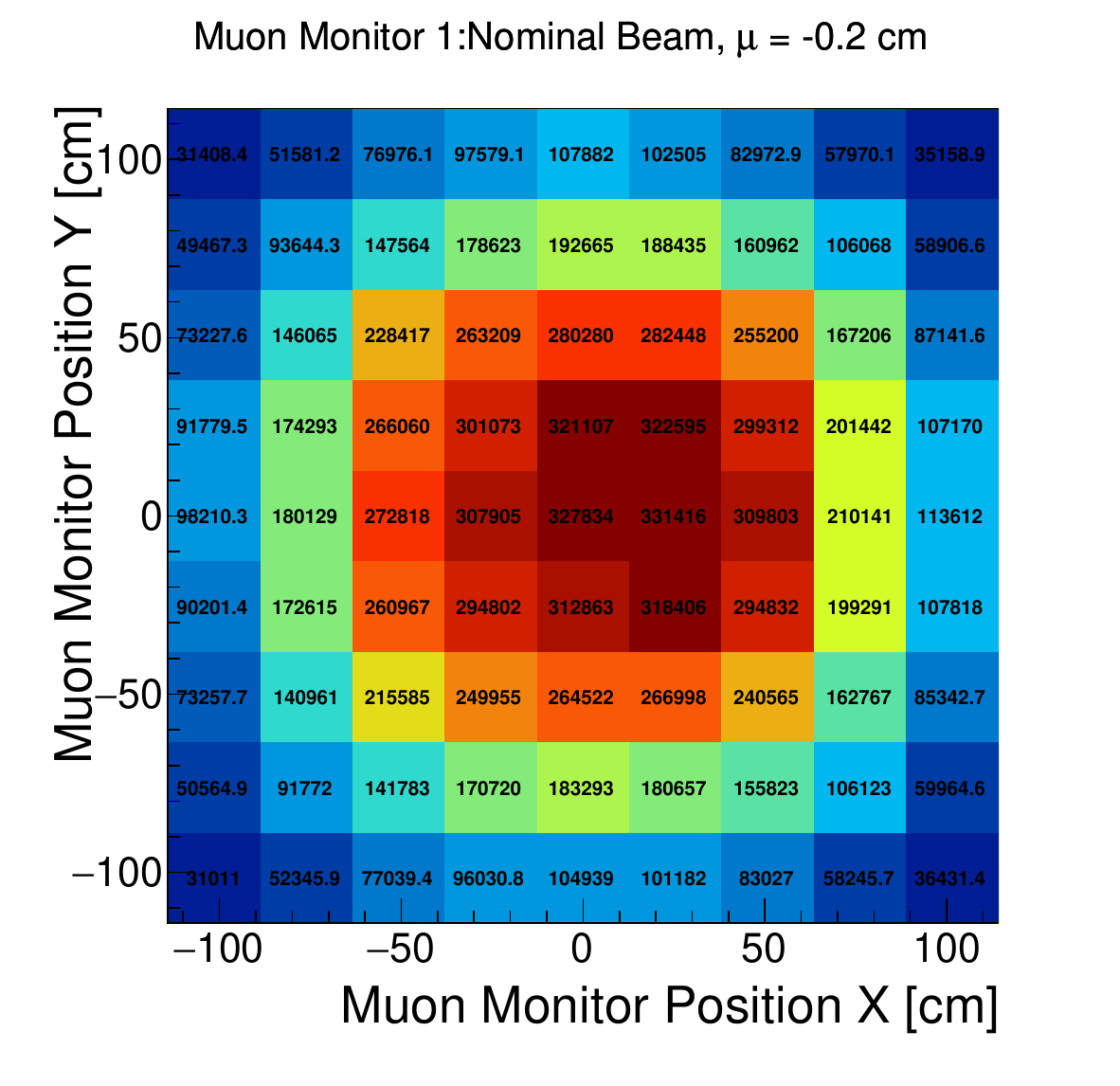}
\includegraphics[width=.85\linewidth]{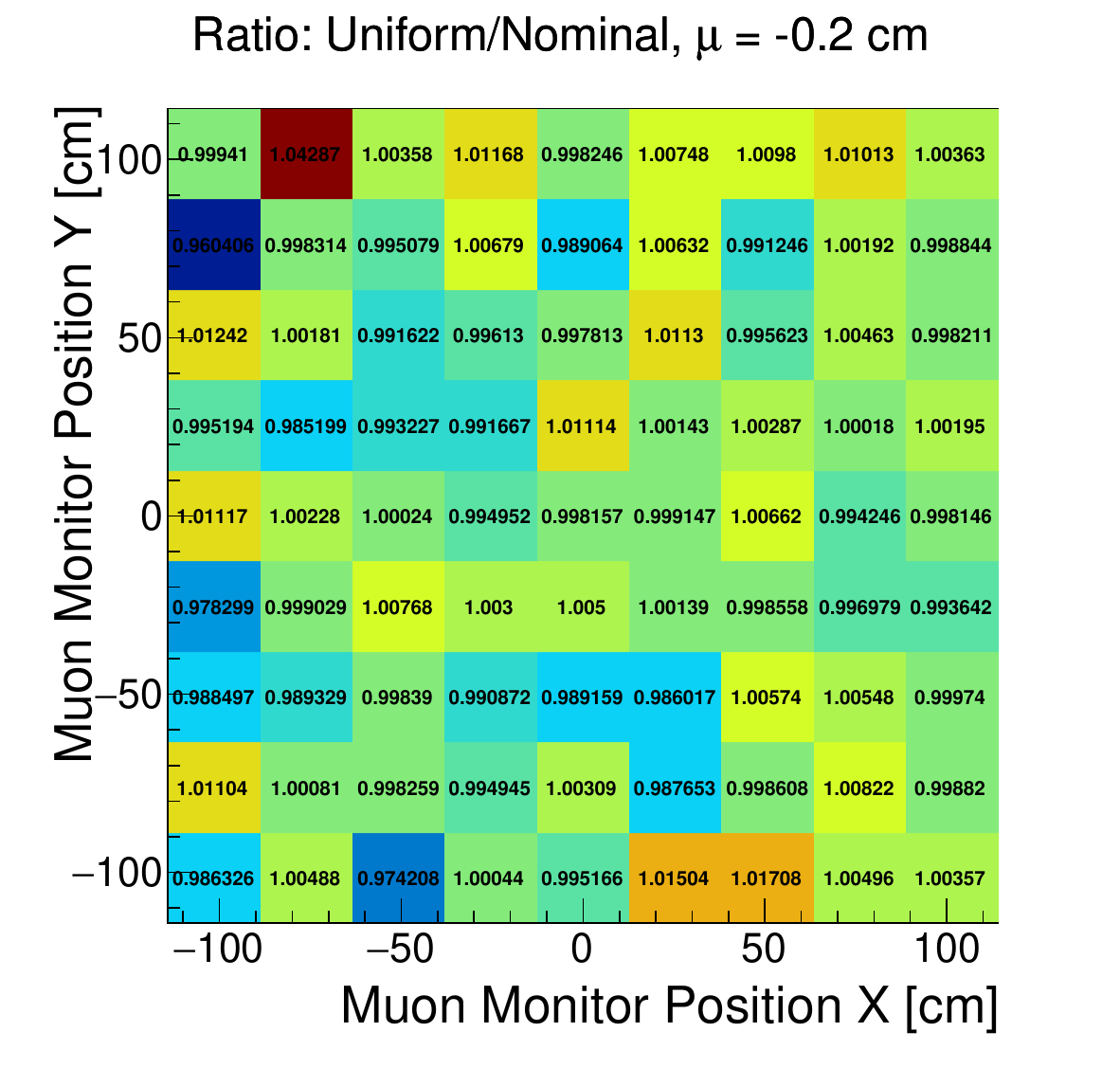}
\caption{NuMI muon monitor 1 responses from two different simulations: uniform (top) and nominal (middle). Ratio of the muon monitor 1 responses between the uniform and the nominal simulations (bottom).}\label{fig:mm1response}
\end{center}
\end{figure}

%Fig. ~\ref{fig:beamprof} shows a comparison of the proton beam profile in the horizontal direction between uniform beam simulation (cyan) and nominal beam simulation (red). It is evident that Gaussian profiles with means at -0.2 cm, -0.1 cm, 0 cm, 0.1 cm, and 0.2 cm have an excellent match between the two simulations.
%\textcolor{red}{In addition, the momenta were plotted in order to compare the two simulations.}
%\textcolor{red} {Add momenta plot from Yiding, comparing uniform with nominal.}\\
%Following that, we plotted the response to muon monitor 1 for a Gaussian beam generated from the uniform beam simulation with mean = 0 cm. The same response has been plotted for a Gaussian beam with mean = 0 cm with the nominal simulation. Muon monitor 1 behaves the same in uniform beam simulation and nominal simulation for the majority of its central pixels. There are some pixels on the edge of muon monitor 1 that show some difference between the two simulations, but the maximum difference is less than 3\%. 
Figure~\ref{fig:mm1response} illustrates an example of NuMI muon monitor 1 response with a given Gaussian beam generated from the uniform beam simulation. The same Gaussian beam was also generated using the nominal beam simulation. A ratio between those two simulations for each pixel has also been shown.
We plotted muon monitor responses only for incident protons decaying into muon candidates. 
Statistical fluctuations can explain a maximum difference of $\sim$4\% between two simulations in some of the edge pixels of a muon monitor. 
We demonstrated that the statistical fluctuations affect muon monitor pixel responses even within the same nominal simulation. We generated five different nominal simulation samples using five unique random seeds and then calculated the ratio of the muon monitor 1 responses between two Nominal simulation samples with two different seeds. It has been shown that even within the same nominal simulation, some edge pixels fluctuated based on the random seed by $\sim$ 4\% at most. 
\begin{figure}%[htpb!]
\begin{center}%\setlength{\unitlength}{0.6cm}
\includegraphics[width=.85\linewidth]{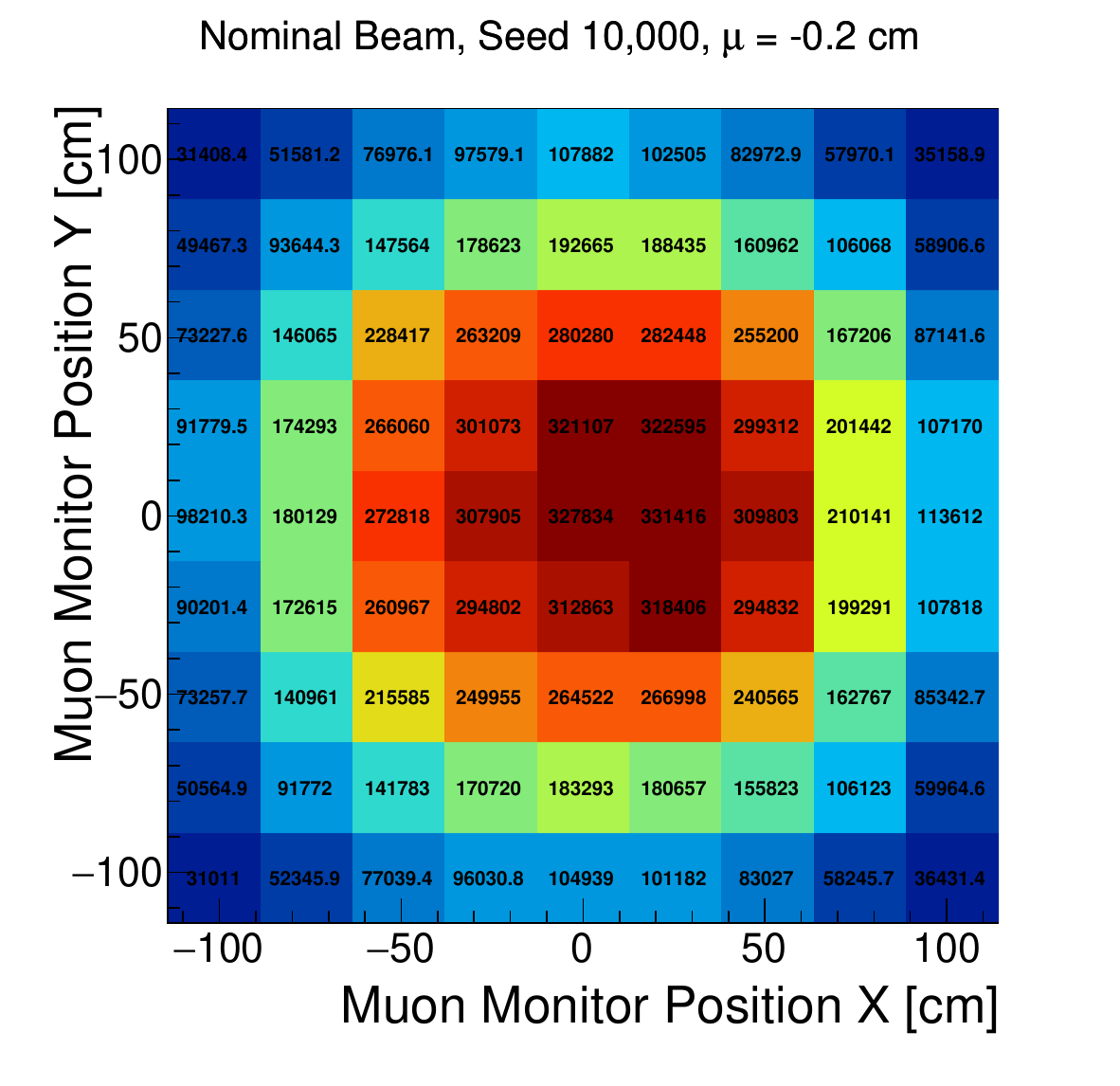}
\includegraphics[width=.85\linewidth]{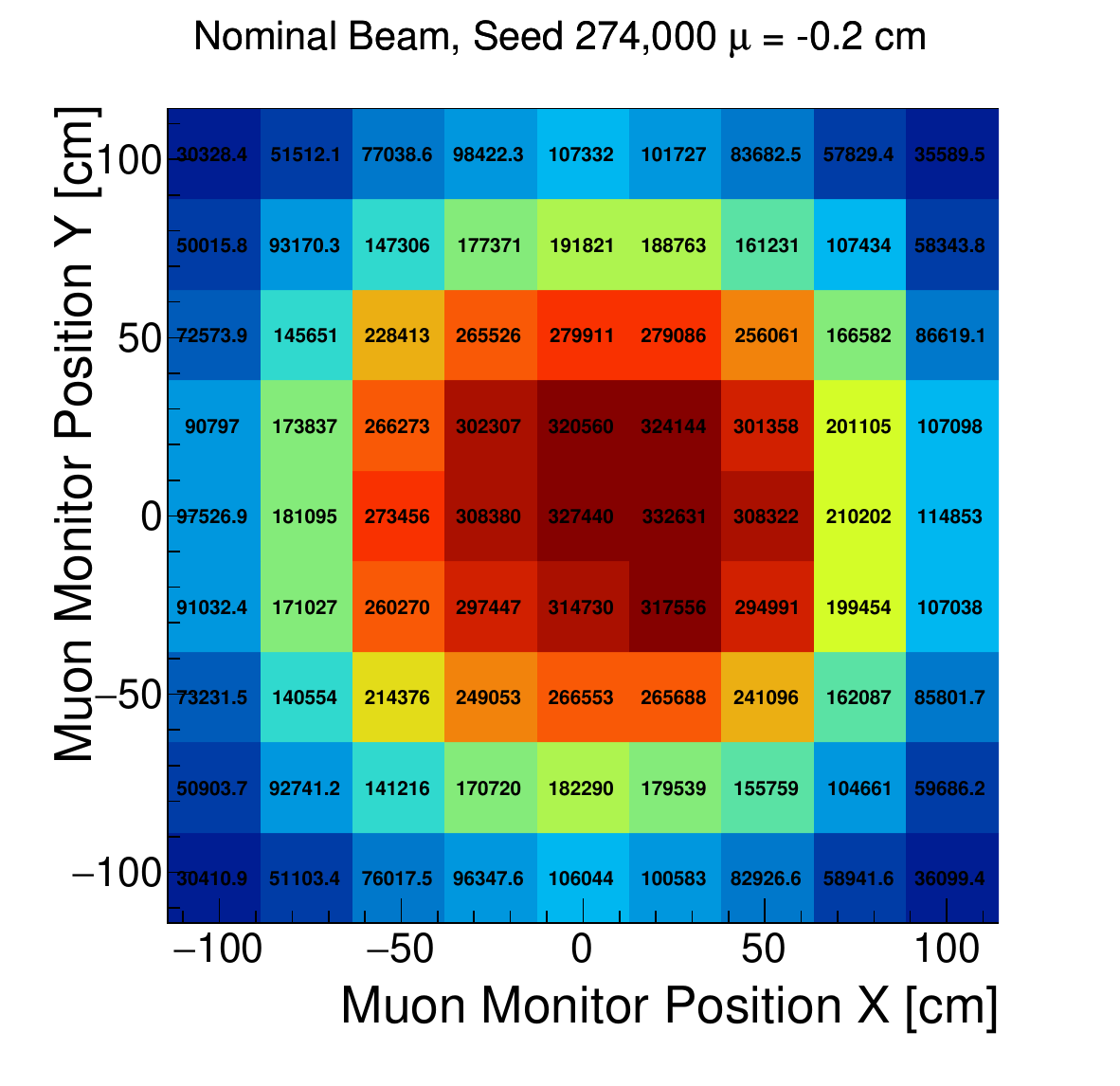}
\includegraphics[width=.85\linewidth]{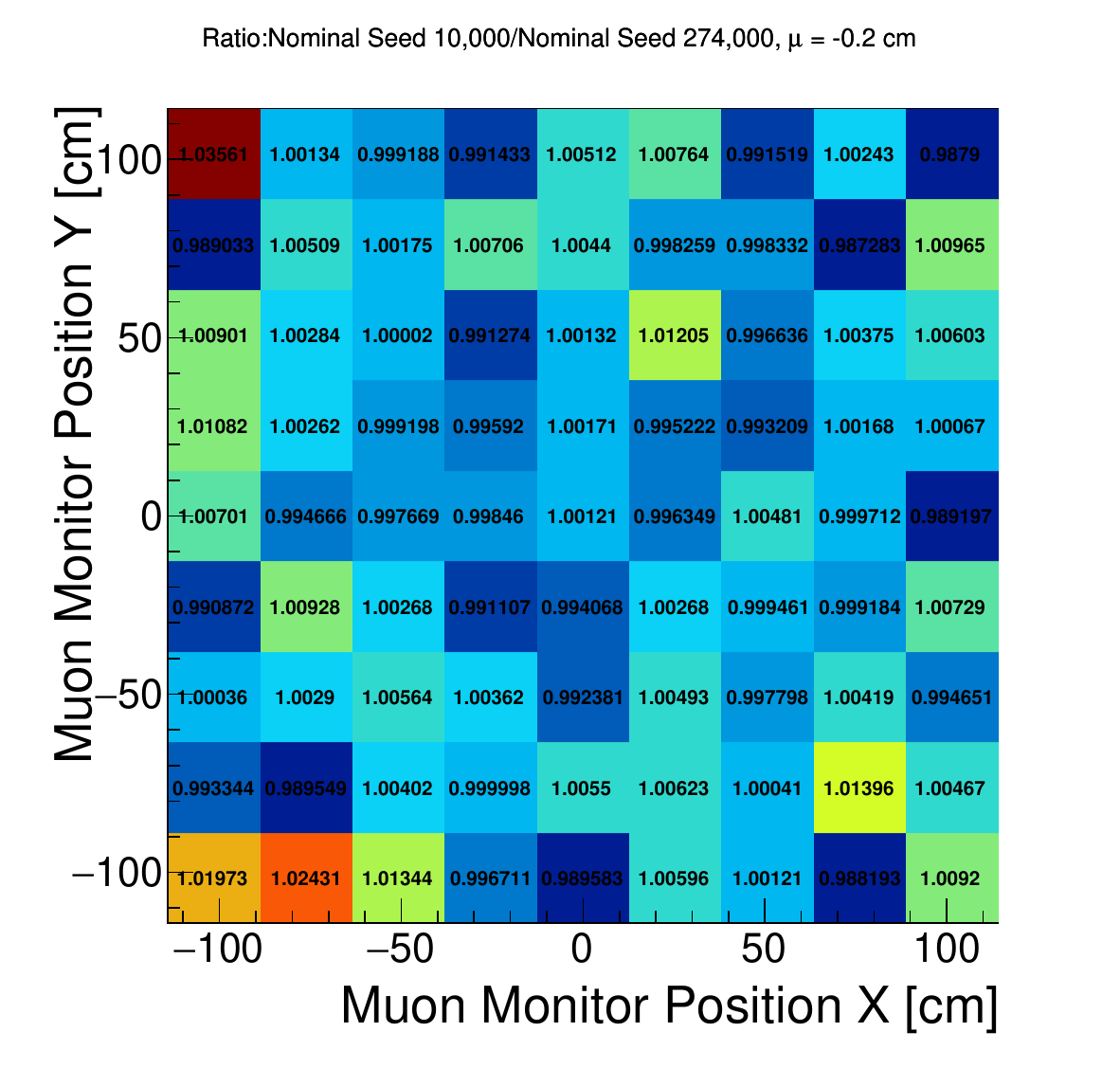}
\caption{NuMI muon monitor 1 responses from nominal simulations with two different seeds (top and middle, respectively). Ratio of the muon monitor 1 responses between nominal simulations with two different seeds (bottom).}\label{fig:mm1response_nominal}
\end{center}
\end{figure}
It is consistent with the difference in edge pixel response between uniform and nominal simulations on muon monitor 1. Hence, the $\sim$ 4\% difference is not an artifact of the uniform simulation technique, rather stems from statistical fluctuations. Increasing statistics will reduce this fluctuation.
Figure~\ref{fig:mm1response_nominal} top distribution shows muon monitor 1 pixel responses for two nominal simulation samples created with different random seeds, while the bottom distribution shows the ratio between the two simulations. There is a maximum difference of $\sim$ 4\% in one of the corner left edge pixels. 

\begin{figure}%[htpb!]
\begin{center}%\setlength{\unitlength}{0.6cm}
\includegraphics[width=\linewidth]{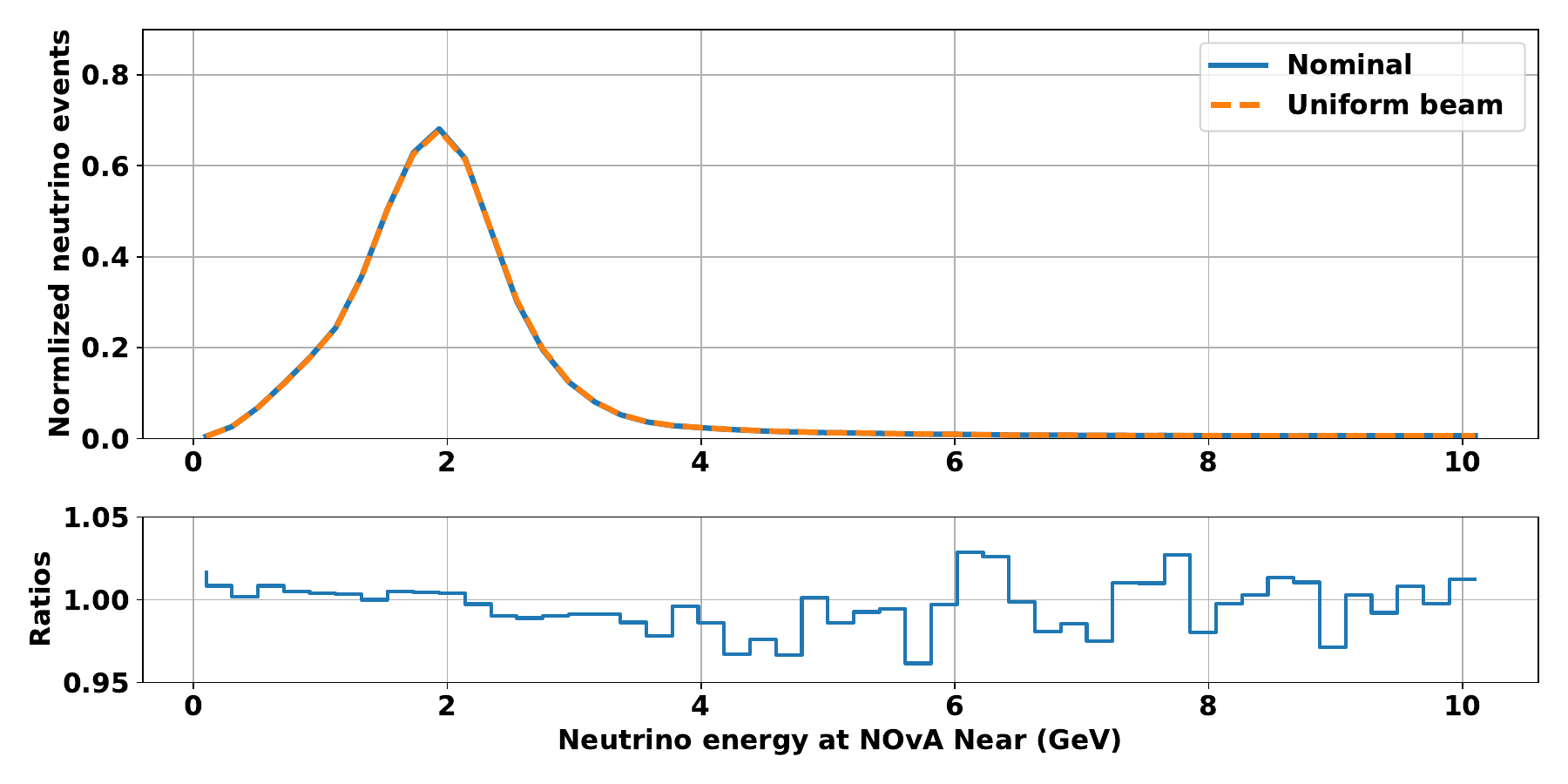}
\caption{Neutrino energy spectra at the NOvA Near Detector from nominal and uniform beam simulations (top). Ratio of the neutrino energy spectra at the Near Detector between nominal and uniform simulations (bottom).}\label{fig:nuenergy}
\end{center}
\end{figure}

%Table \ref{tab:1} shows the comparison between different muon flux centroids on muon monitor 1 calculated with nominal simulation samples each with a different random seeds. The maximum \% difference obtained between the muon flux centroids using two different seeds is X\%. 
%\begin{table}[]
%\begin{tabular}{ |p{2cm}||p{2cm}|p{2cm}| }
 %\hline
 %\multicolumn{3}{|c|}{Nominal Simulation} \\
 %\hline
 %Random seeds & Muon flux centroid &\% difference\\
 %\hline
%Seed 1   & XX    &XX\\
%Seed 2   & XX    &XX\\
%Seed 3   & XX    &XX\\
%Seed 4   & XX    &XX\\
%Seed 5   & XX    &XX\\
 %\hline
%\end{tabular}
 %\caption{XXX}
  %  \label{tab:1}
 %\end{table}
Finally, Fig.~\ref{fig:nuenergy} illustrates the NOvA Near Detector neutrino energy spectrum plotted from nominal and uniform simulations. As can be seen from the bottom plot, in the region of interest, the ratio of the two spectra stays close to 1 for the energy range of interest. 

\subsection{Testing Kinematics Reproducibility}
 \begin{figure}[htpb!]
\begin{center}%\setlength{\unitlength}{0.6cm}
\includegraphics[width=.9\linewidth]{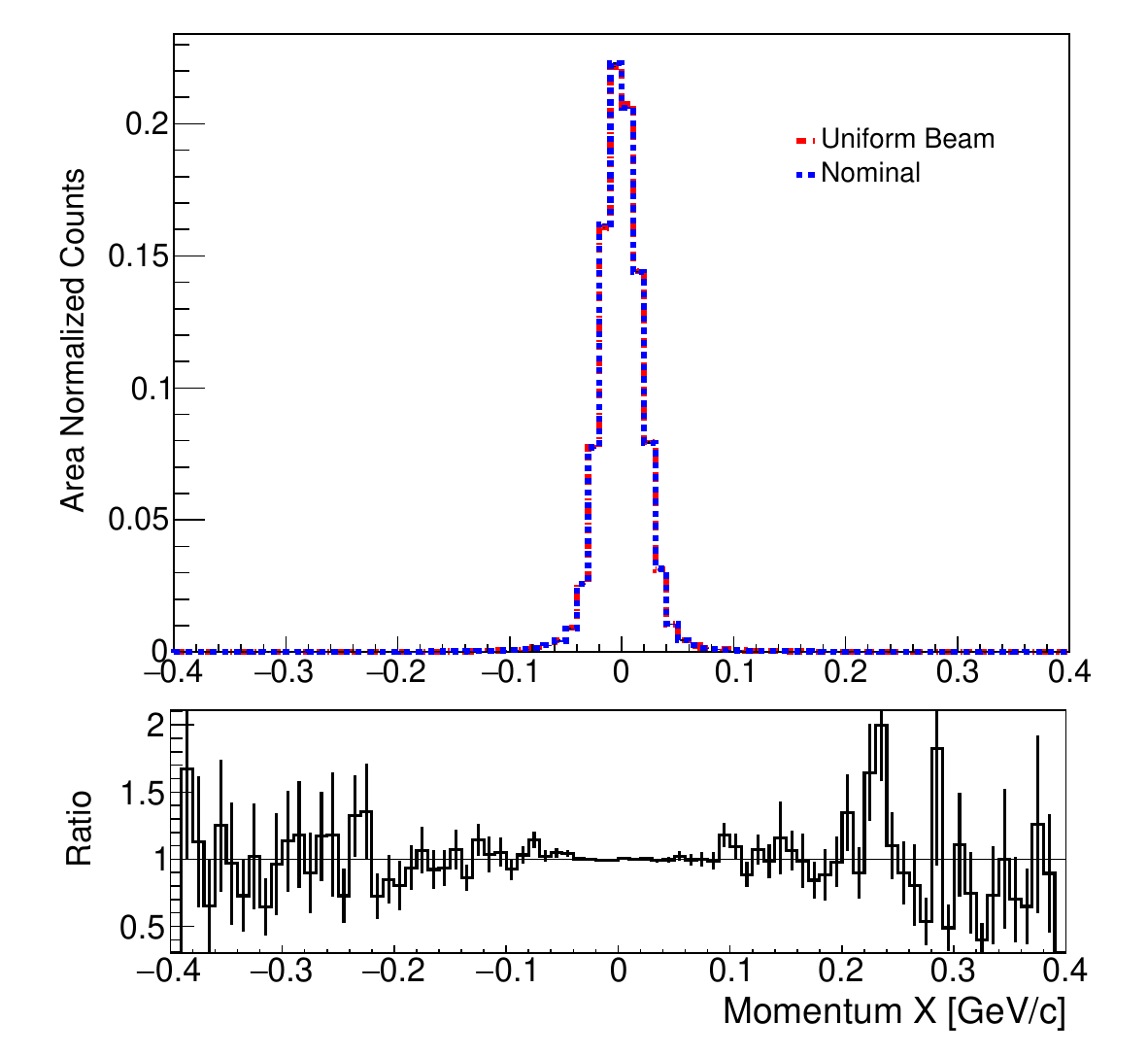}
\includegraphics[width=.9\linewidth]{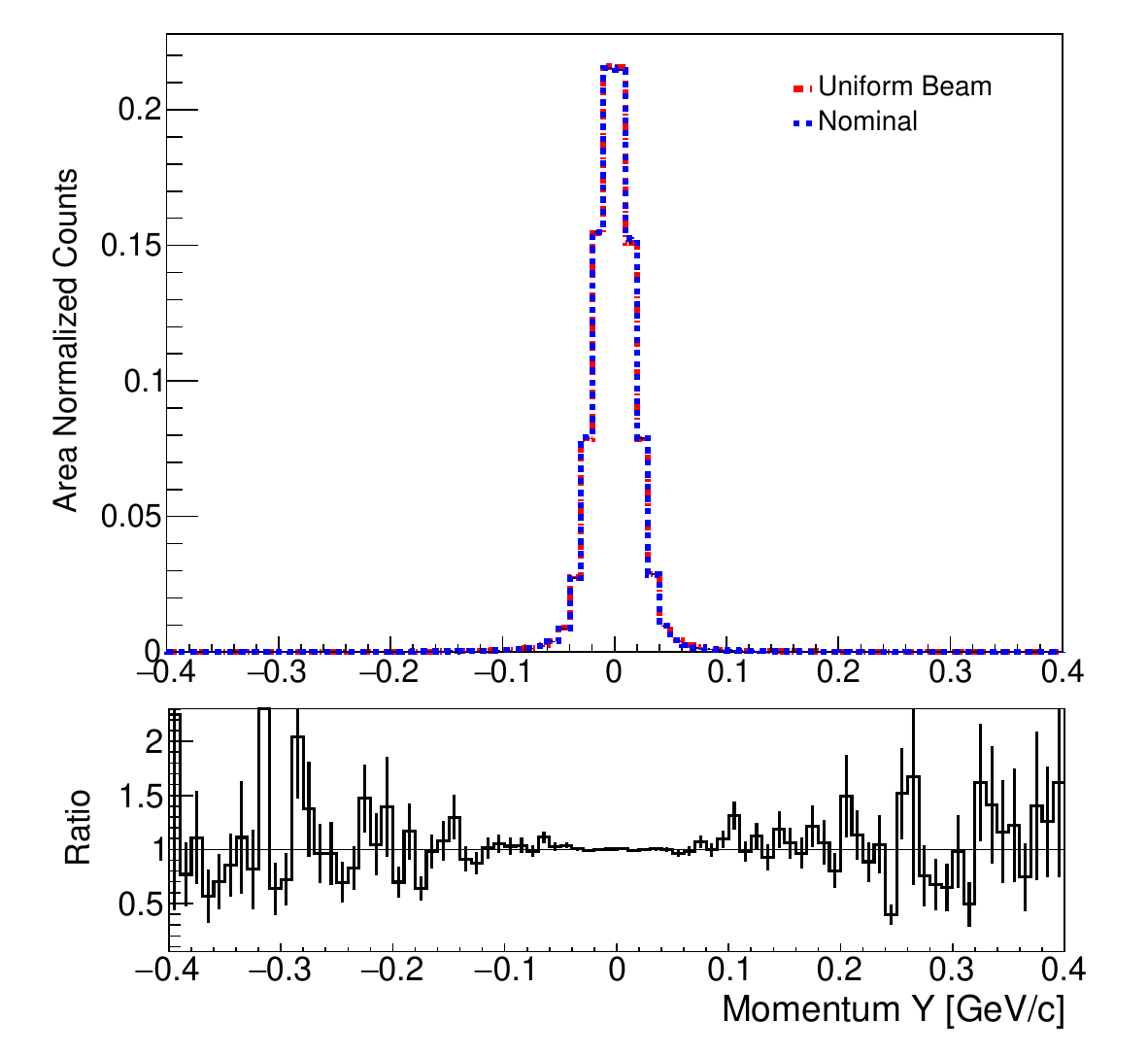}
\includegraphics[width=.9\linewidth]{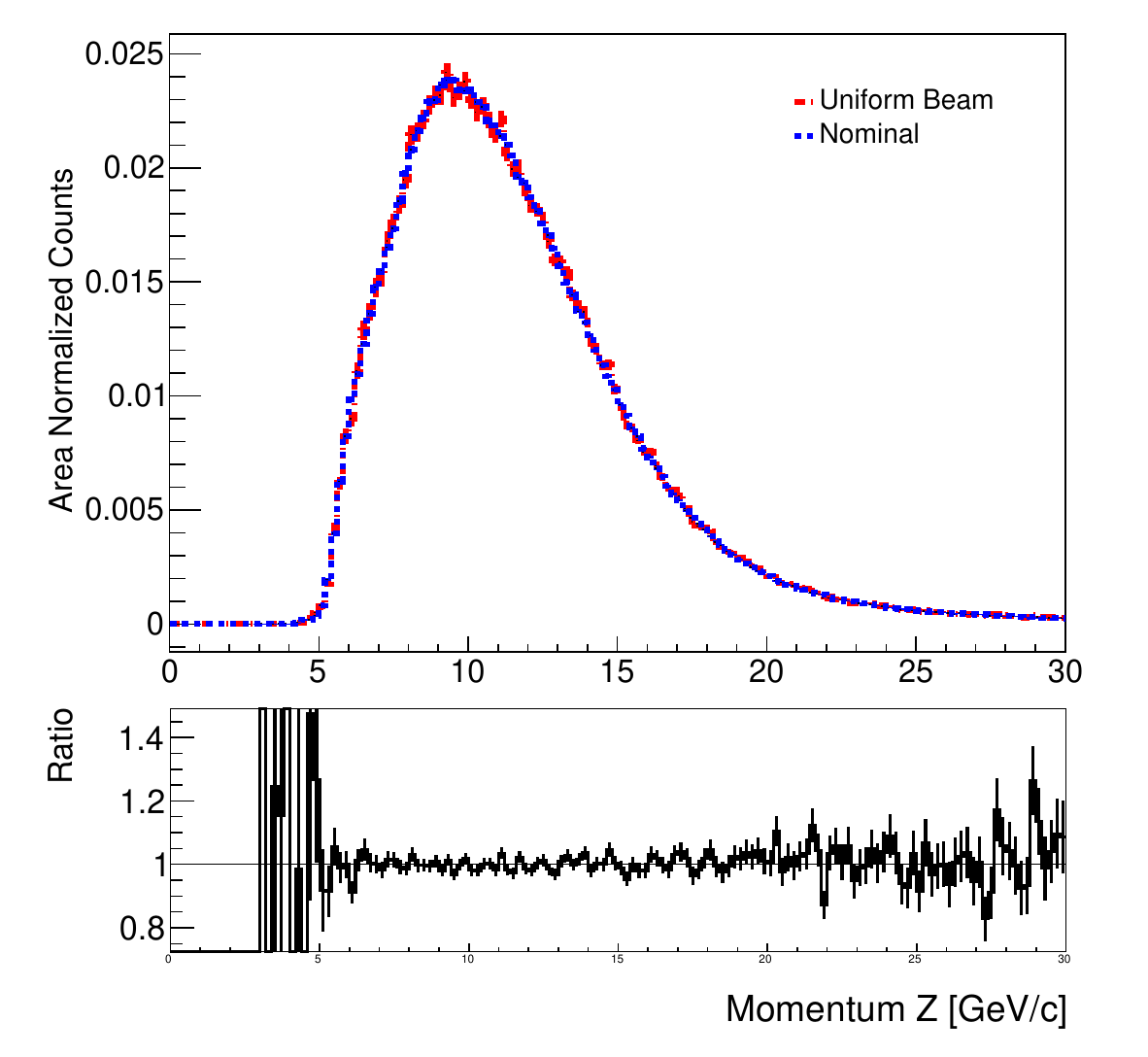}
\caption{A comparison of the muon momenta along the X (top),Y (middle) and Z (bottom) axes produced at the hadron decay with uniform (red) and Nominal (blue) simulations for a Gaussian beam of $\mu$ = 0.0\,cm.}\label{fig:muon_mom}
\end{center}
\end{figure}
We have tested the reproducibility of the kinematics of the secondary particles after applying the uniform beam technique. 
Pion decay kinematics and decay pipe size determine the size of the muon beam at the Muon Monitors.
In this study, we have used 0.15\,cm width Gaussian beam samples, with mean set at $\mu$ = 0.0\,cm from uniform and nominal simulations to compare the muon momentum distributions at the hadron decay along $x$, $y$, and $z$ directions at the muon monitor 1. %These comparisons show that muon production momenta are consistent.
 %There is very good agreement between the momenta of the muon production at the hadron decay from the Uniform and Nominal simulations.
%Comparison of the momenta of the production of the muons at the hadron decay from the Uniform and the Nominal simulation show an excellent agreement. 
%Comparing the momenta of the production of muons at the hadron decay from Nominal and the Uniform beam simulations provides a solid validation for the reproducibilty of the nominal simulation information after applying the proposed technique on the uniformly distributed sample. 
Figure~\ref{fig:muon_mom} shows the distributions of the momenta of the muons produced at the hadron decay. In the red distributions, the momenta are plotted using the uniform beam simulation, whereas in the blue distributions, they are plotted using the nominal simulation. Muon production momenta are shown to be consistent between uniform and nominal simulations, which is validated by the ratio $~$ 1 between the momenta distributions from these simulations.
\section{Computing Resources}\label{sec:resource}
%Running uniform beam simulation technique to generate large simulation samples reduces the usage of computing resources. 
As a result of using uniform beam simulations to generate large simulation samples, the amount of computing resources needed to run the simulations can be reduced considerably.
Here we show a comparison between the amount of computing resources used to generate nominal Gaussian and uniform beam simulation samples, which is shown in Table.~\ref{tab:cpu_time}.
%The usage of the computing resources to generate nominal Gaussian and uniform beam simulation samples can be compared as shown in Tab.~\ref{tab:cpu_time}.
%Based on the table, it can be seen that running the nominal simulation to generate one random sample takes about an hour, whereas running the uniform simulation to create four times as many statistics takes almost the same amount of time. 
%In the case of the uniform simulation, we create the Gaussian beam post-processing. 
%In order to generate 1000 random samples using nominal simulation, it would take approximately 1000 hours of CPU time. 
The table shows that running a nominal simulation to generate one random sample takes about an hour, whereas running a uniform simulation to generate four times as many statistics takes almost the same time. The fractions of protons interacting with matter are smaller in uniform beam simulation than in nominal simulation. For uniform simulations, we create Gaussian beam post-processing. Using nominal simulation, however, it would take approximately 1000 hours of CPU time to generate 1000 random samples.
Nevertheless, if the proposed weighting technique is applied to a high POT single uniform beam simulation sample, the CPU time required to generate 1000 random Gaussian samples can be significantly reduced. 
%To generate 1000 random samples running nominal simulation takes 1000 CPU hours approximately. However, the CPU time can be significantly reduced to generate 1000 random Gaussian samples by using the proposed weighting technique on a high POT single uniform beam simulation sample.  
\begin{table}[h!]
\centering
\begin{tabular}{ c| >{\centering\arraybackslash}m{1.5cm} | >{\centering\arraybackslash}m{2.0cm} | >{\centering\arraybackslash}m{1.8cm} }
\hline
 & POT & $<$CPU Time$>$ (min) & $<$Memory$>$ (MiB) \\
\hline
  Nominal & 250M & 59 & 51.2 \\ 
Uniform & 1000 M & 67 & 52.7 \\ 
\hline
\end{tabular}
\caption{A comparison of the average CPU time and memory usage required to generate one sample of uniform beam simulation versus one sample of nominal beam simulation.\label{tab:cpu_time}}
\end{table}
%\begin{table}[!h]
%\begin{center}
%\begin{tabular}{ c|c|c|c } 
%\hline
 %& POT & $<$CPU Time$>$ (min) & $<$Memory$>$ (MiB) \\
%\hline
 % Nominal & 250M & 59  & 51.2 \\ 
%Uniform & 1000 M & 67 & 52.7 \\ 
%\hline
%\end{tabular}
%\caption{Comparison of the average CPU time and memory usage to generate single samples running uniform beam simulation vs nominal simulation.\label{tab:cpu_time}}
%\end{center}
%\end{table}

\section{Applications of the Technique }\label{sec:app}
Simulation samples were generated for selected proton beam position configurations with different horn current settings to test the uniform beam simulation technique. This section presents the simulation technique for scanning targets by generating uniform beams along horizontal and vertical positions at the target for 200 kA and 180 kA horn currents, respectively. Simulated samples are generated using the combined g4numi and muon monitor packages. 

\subsection{Beam Based Alignments of Beamline Components}
In order to ensure that the target-baffle system is aligned, the proton beam is scanned across the target-baffle system horizontally and vertically at the beginning of every run period and the responses on the muon monitors are observed. Target scans demonstrate how muon monitors respond to variations in beam position horizontally and vertically. Muon monitors show linear responses to the beam centroid.  
To align the horn, the target is removed and the beam is scanned along the horn cross-hair to observe the responses on the hadron monitor. A similar scan is presented here using our proposed technique to generate 100 Gaussian beam profiles with $\sigma = 0.15$ cm along the horizontal axis from -0.8 cm to +0.8 cm. It should be noted that in this simulation the cross-hair feature of the horn is not included. 

%To check the alignment of the target-baffle system, horizontal and vertical target scans are generally performed with the primary proton beam at the beginning of a run period. To align the horn, the target is removed and the beam is scanned along the horn cross-hair to observe the responses on the hadron monitor. %A similar scan is presented here using our proposed technique to generate 100 Gaussian beam profiles with $\sigma = 0.15$ cm along the horizontal axis from -0.8 cm to +0.8 cm. 
\begin{figure}
\begin{center}%\setlength{\unitlength}{1.0cm}
  %\begin{tabular}{@{}cc@{}}
\includegraphics[width=\linewidth]{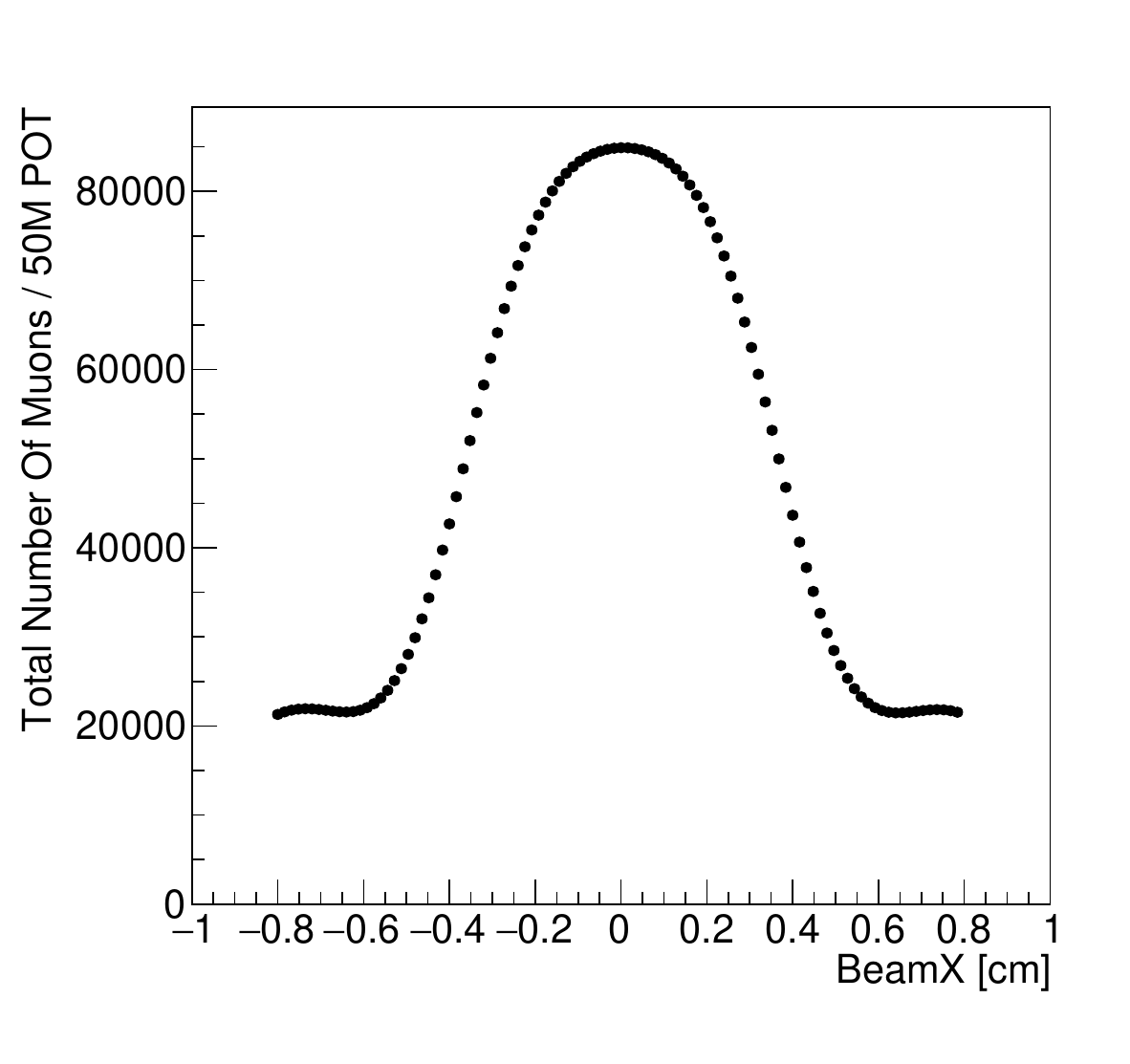} 
\includegraphics[width=\linewidth]{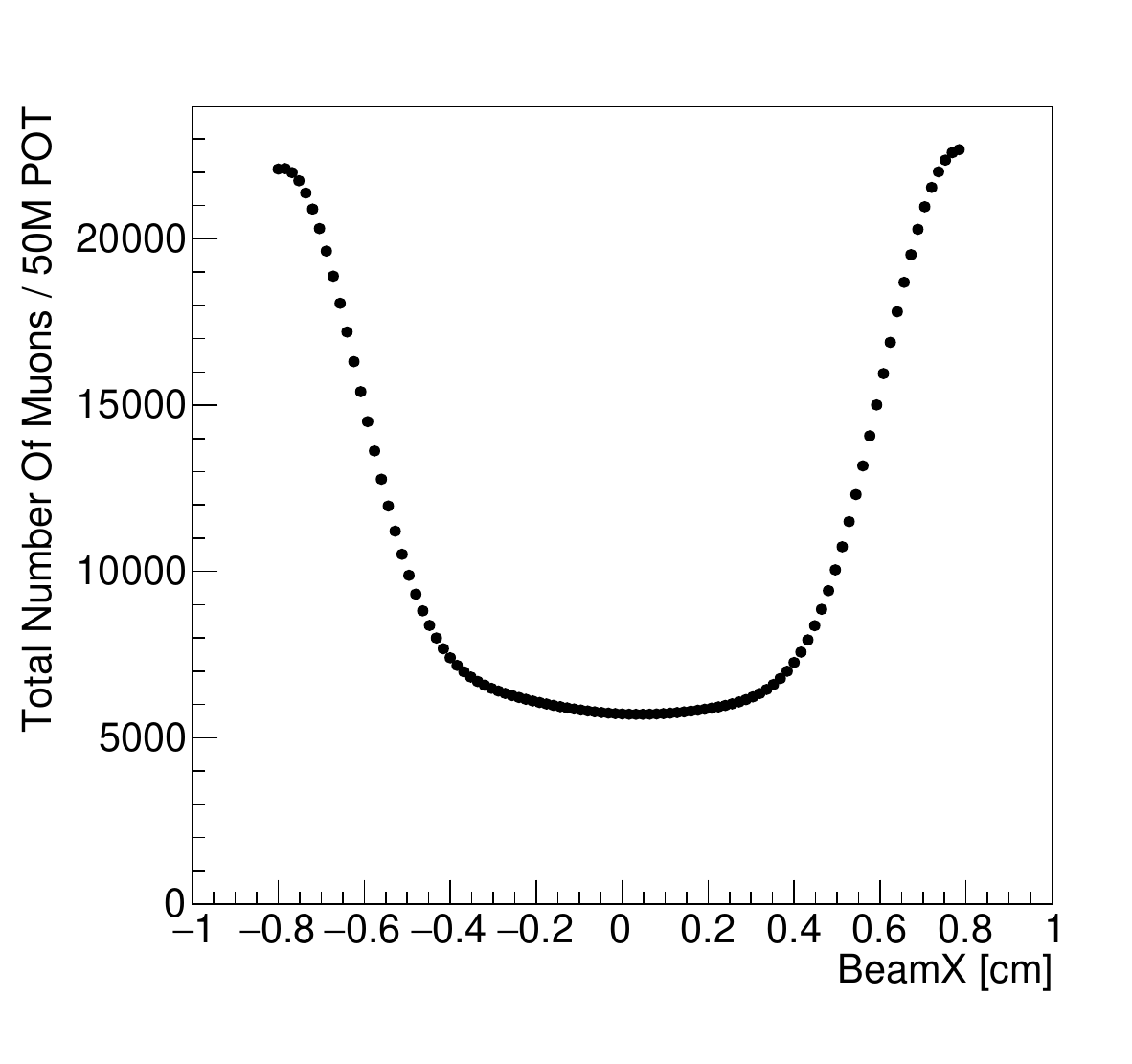} 
    %\end{tabular}
\caption{Total number of muon on muon monitor 1 from the beam scan with including the target (top) and removing the target (bottom).}\label{fig:tgt_scan}
\end{center}
\end{figure}
The demonstration has been performed with and without the target in the beamline. The horn current has been set to +200kA for this study. Muon monitor 1 shows a variation in the total number of muons as a function of beam position on the target. 
Fig.~\ref{fig:tgt_scan} shows the total number of muons detected at muon monitor 1 during the beam scan with (left) and without (left) the target. By simulating the interactions of the beam with the baffle, target, and other beamline materials, we are able to observe muons at muon monitor 1. A similar study of the target scan with the NuMI beam has been described in chapter 10: Beam Monitoring Measurements of the Accelerator Systems and Instrumentation for the NuMI Neutrino Beam of reference 10.\cite{Zwaska:2005be}.
  %--------------------------------
\subsection{Muon Flux Correlation Studies}
A study on the correlation between the muon flux at the muon monitors and the variations in the proton beam can be carried out by measuring the muon flux centroid against the variations in the proton beam along horizontal and vertical directions on the target, respectively. This study can be extended by varying the horn current configurations to study the horn focusing effects on the muon flux. 

\begin{figure}%[htp!]
  \begin{center}%\setlength{\unitlength}{1.0cm}
  \includegraphics[width=\linewidth]{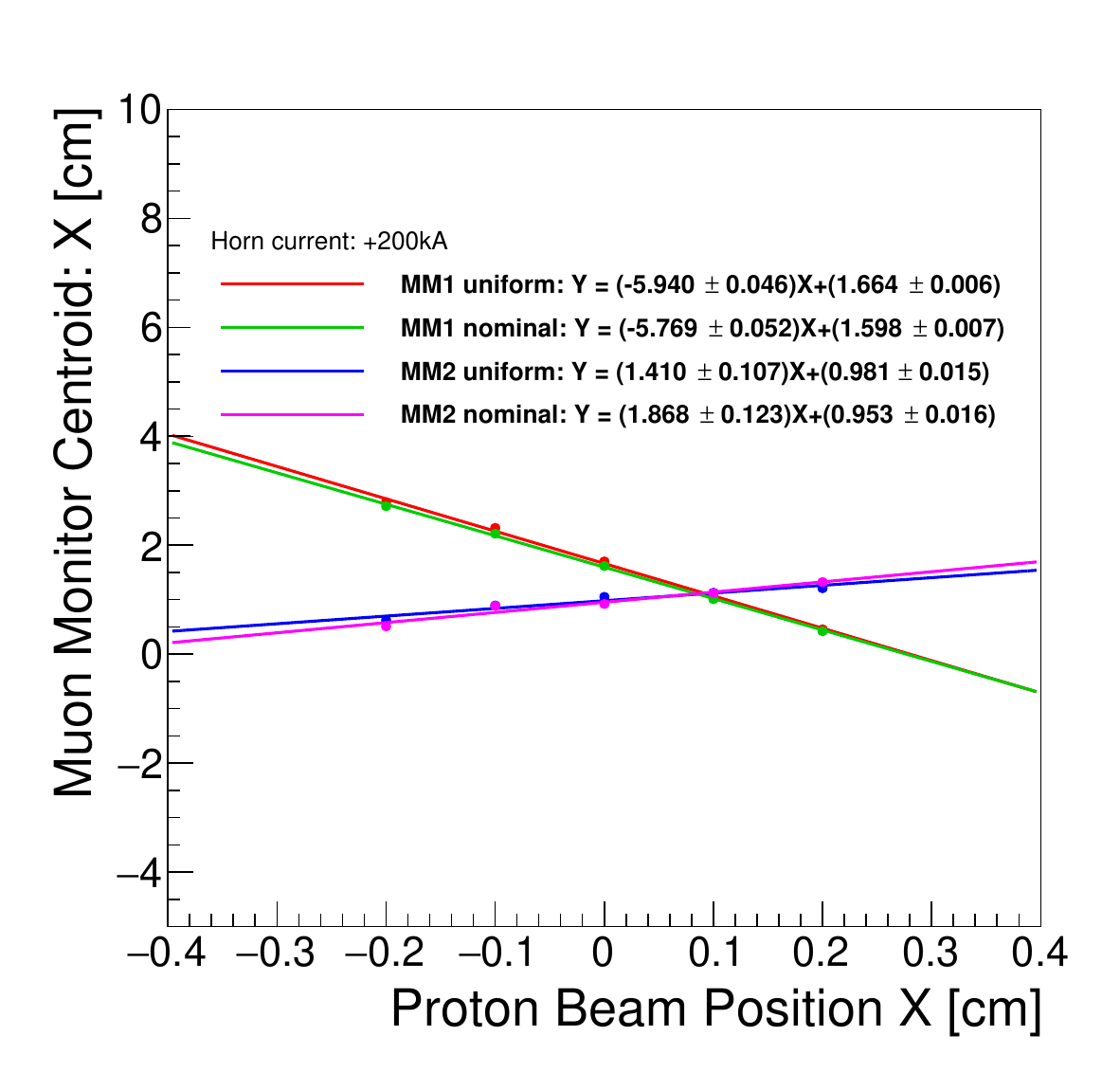}
  \includegraphics[width=\linewidth]{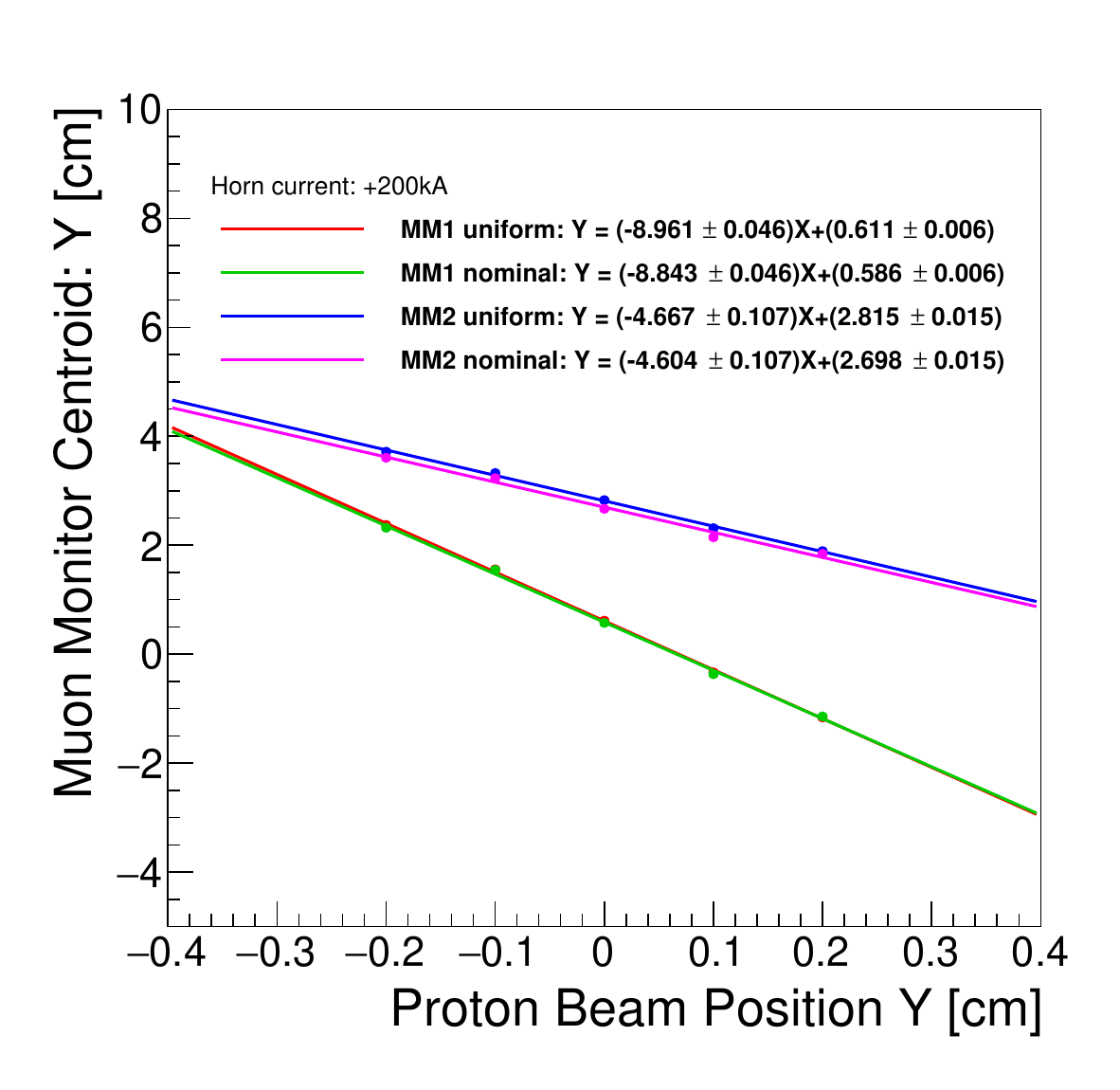}
   % \end{tabular}
    \caption{The muon flux centroid estimated with uniform beam simulation and nominal simulation respectively, on muon monitor 1 (uniform: red, nominal: green) and muon monitor 2 (uniform: blue, nominal: purple) in a horizontal beam scan (top) and vertical beam scan (bottom) for a +200 kA horn current as a function of the proton beam position at the target.}\label{fig:mmx_sim}
    \end{center}
  \end{figure}

\begin{figure}%[htp!]
  \begin{center}%\setlength{\unitlength}{1.0cm}
  \includegraphics[width=\linewidth]{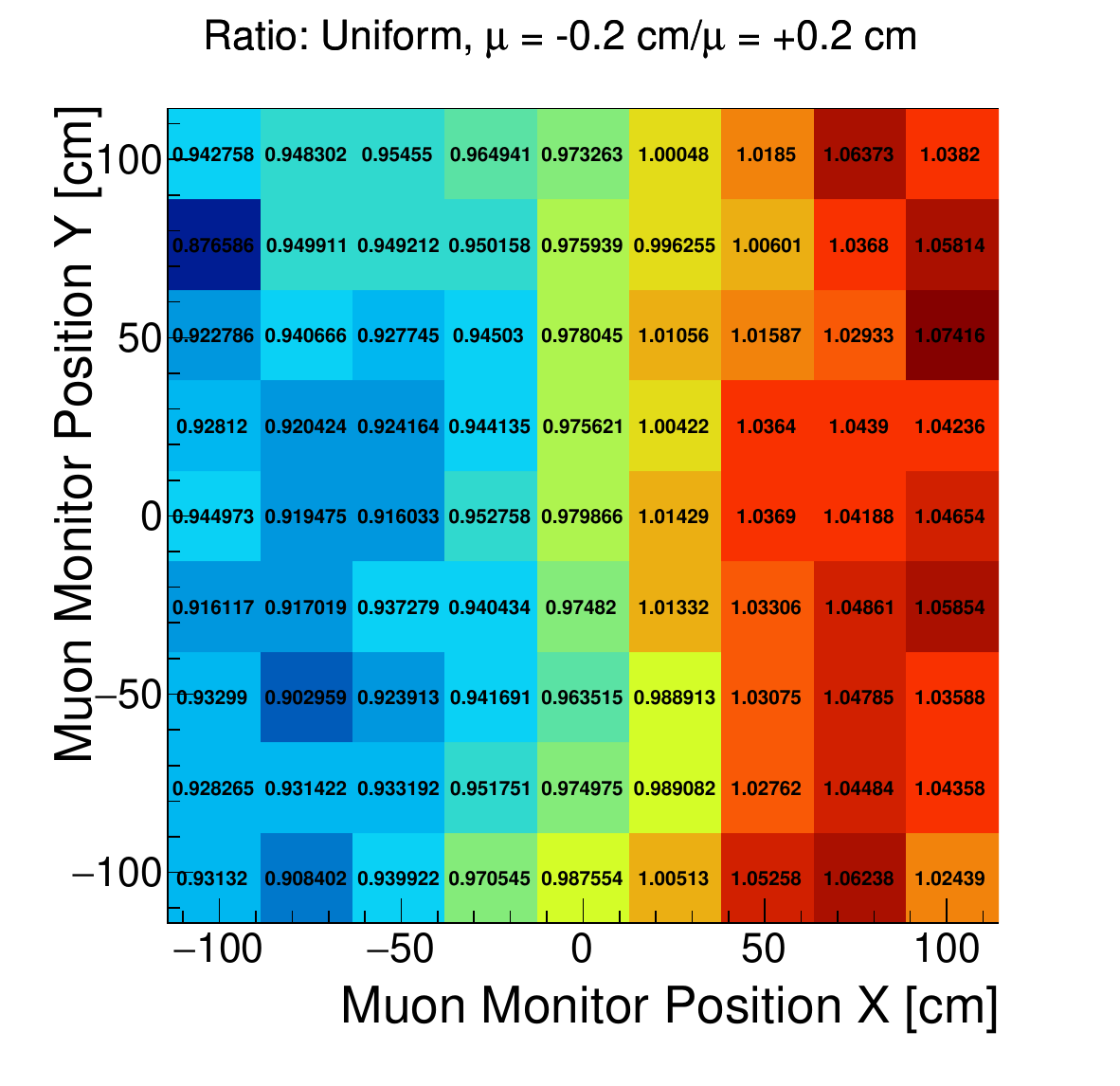} \includegraphics[width=\linewidth]{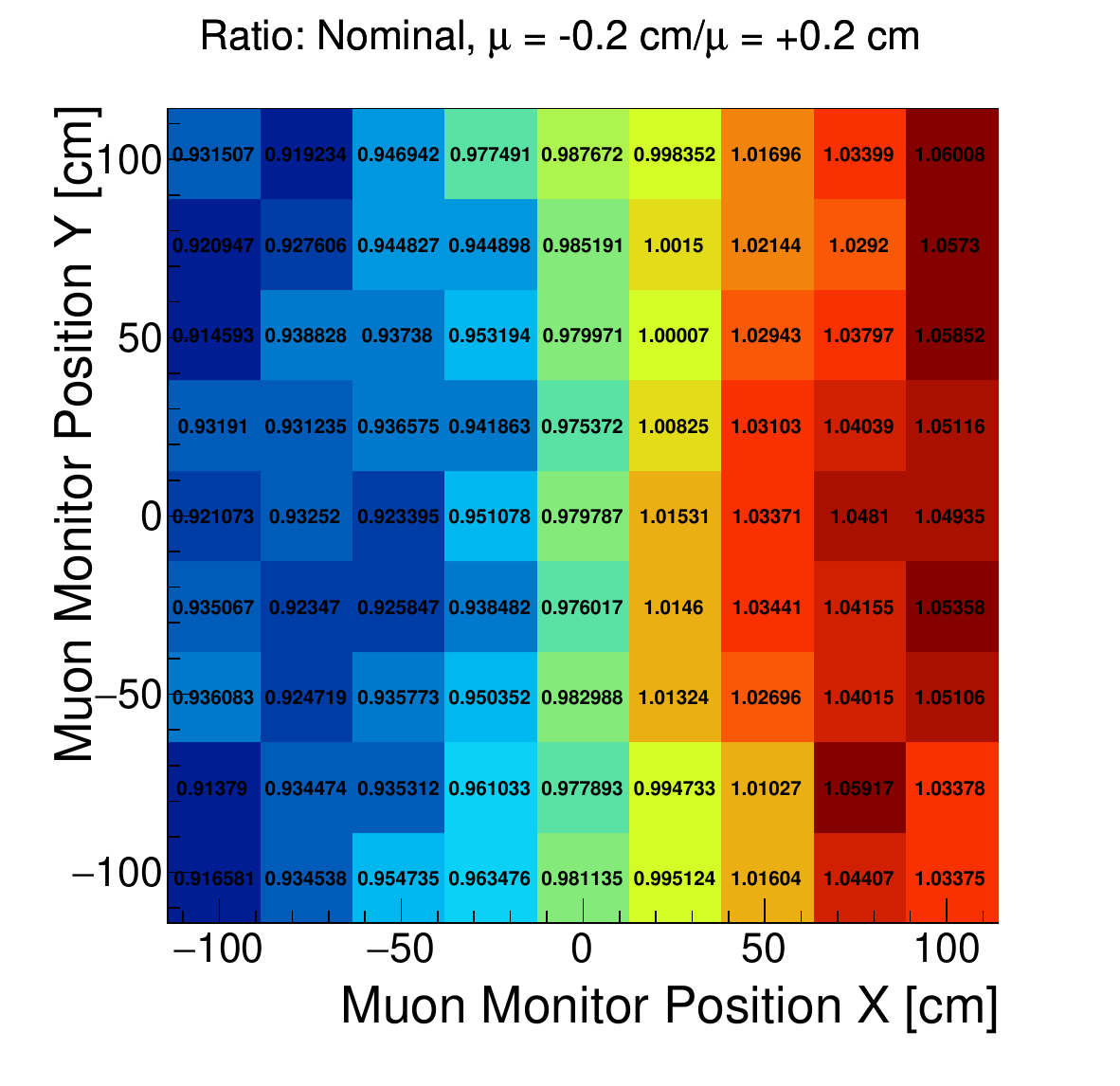}
   % \end{tabular}
    \caption{Ratio of muon flux in each pixel on muon monitor 1 for proton beam $\mu$ = -0.2 cm and $\mu$ = +0.2 cm respectively in the horizontal direction for a +200 kA horn current. Uniform beam simulation (top). Nominal beam simulation (bottom).}\label{fig:mmx_leftright}
    \end{center}
  \end{figure}

In order to demonstrate this correlation with simulation, we have generated Monte Carlo samples each with 1 billion protons on target (POT) for +200 kA horn current setting. 
We generate horizontal beam scan samples by throwing protons uniformly along the horizontal axis from -1.0 cm to +1.0 cm while keeping the vertical beam as a random Gaussian with a 0.15 cm beam width and zero centroid. In order to generate a vertical beam scan sample, the axis selections for the uniform beam and Gaussian beam are changed. 
%We select Gaussian samples along the horizontal or vertical positions on the target from -0.2 cm to 0.2 cm as shown in Fig.~\ref{fig:all_gauss}. 
Each iteration uses a centroid separation of 0.02 cm and a Gaussian width of 0.02 cm. The Gaussian width of the proton beam has been set to 0.15 cm for the current scan. 
The study results show a linear correlation to the beam position variations. This linear correlation is unique to the muon monitor locations and horn focusing configurations. 
%To verify the uniform beam simulation technique, we compare the results of the uniform simulation beam scan with the results of the nominal simulation beam scan. 
 %\begin{figure}[htpb!]
  %\begin{center}\setlength{\unitlength}{1.0cm}
   %\includegraphics[width=.4\textwidth]{beamx_gauss}
%\caption{Example of the weighted proton beam position distributions for each Gaussian beam selections for the beam scan study.}\label{fig:all_gauss}
 %   \end{center}
  %\end{figure}
Fig.~\ref{fig:mmx_sim} shows the responses of the muon flux centroid along the horizontal (top) and the vertical (bottom) axes at the muon monitor 1 (red) and 2 (blue) as a function of the proton beam positions for +200 kA horn current plotted with the uniform beam simulation. 
The equivalent plots with the nominal simulation are shown at muon monitor 1 (green) and 2 (purple).
The muon flux centroid on each muon monitor is calculated by taking the average of the projected distribution on the horizontal axis from 81 muon monitor pixels of ionization chamber signals. The distributions have been fitted to a linear function. The fitted slopes from the uniform beam simulation are compared with the slopes calculated using nominal simulation.
Another example of the muon monitor response to the primary beam position variation can be demonstrated as the ratio of the muon flux at muon monitor 1 for proton beam at $\mu$ = -0.2 cm to the beam at $\mu$ = +0.2 cm in the horizontal direction as shown in fig.~\ref{fig:mmx_leftright}.
%This ratio comparison demonstrate the muon flux behaviour at the muon monitor 1 due to the horn focusing mechanism. 
By observing this ratio we can see the muon flux behavior at the muon monitor 1 due to the horn focusing mechanism at work \cite{yonehara2023exploring}. The same flux behavior is observed in both uniform and nominal simulations.
It can be seen that the muon flux moves from right to left on muon monitor 1 as the primary beam moves from left to right.
This is due to the fact that hadrons are over-focused at muon monitor 1 hence produces the negative slope as shown in fig.~\ref{fig:mmx_sim} (top).
%This is due to the fact that the chromaticity of the hadron momentum experiencing the over focusing through the horn magnetic field  negative slope representation on muon monitor 1, as shown in fig.~\ref{fig:mmx_sim} (left).
%This indicates that the muon flux moves from right to left on muon monitors when the primary beam moves from left to right by following the negative slope representation as described on muon monitor 1 in fig.~\ref{fig:mmx_sim} (left).
%\textcolor{red}{Fig. ~\ref{fig:mmx_sim} illustrates how muon monitors respond to beam position variations horizontally and vertically during a target scan. Different slopes demonstrate how the focusing effect impacts muons with different energies.For example, the muon flux ratio at muon monitor 1 for proton beams of $/mu$ = -0.2 cm and $/mu$ = +0.2 cm in the horizontal direction indicates a correlation between the pixel response of the muon monitor and proton beam position as it varies from left to right as shown in fig.~\ref{fig:mmx_leftright}.The same correlation is observed in both uniform simulation as well as nominal simulation.
%}

\begin{figure}
  \begin{center}\setlength{\unitlength}{1.0cm}
  %\begin{tabular}{@{}cc@{}}
   \includegraphics[width=\linewidth]{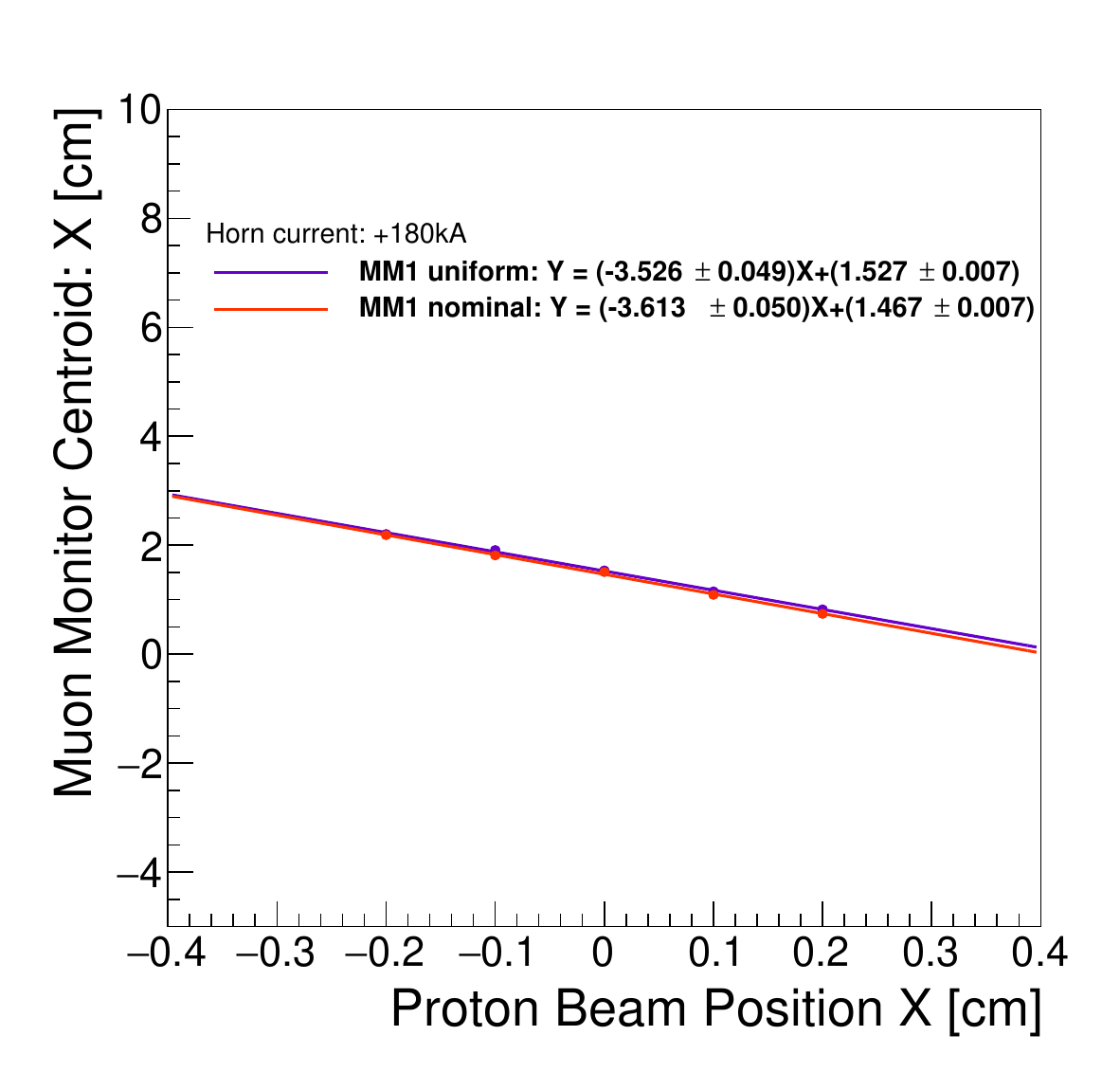} 
    \includegraphics[width=\linewidth]{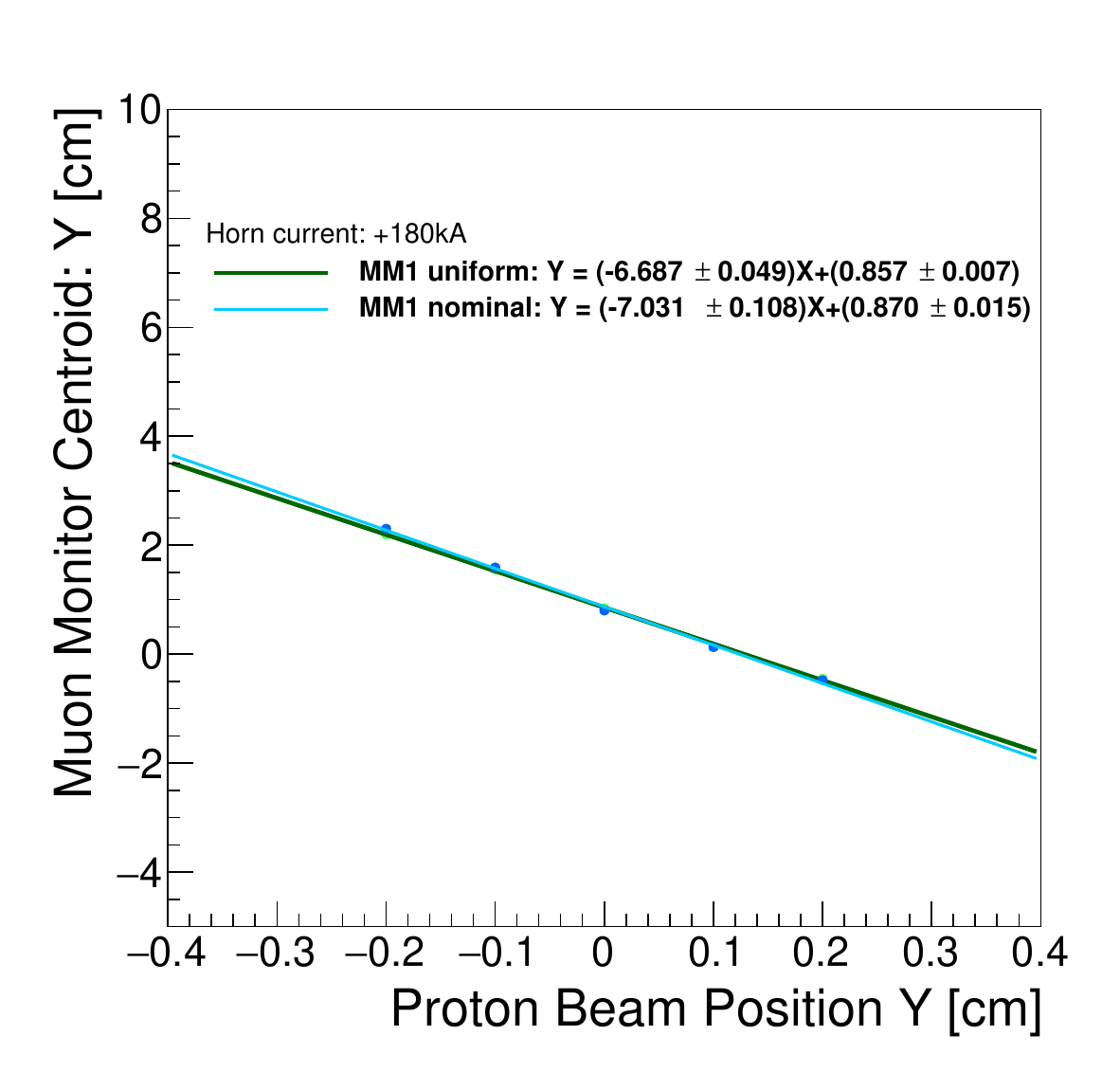}
    %\end{tabular}
\caption{The uniform simulation estimates of the muon flux centroid along the horizontal (top) and vertical (bottom) projections on muon monitor 1 for 180 kA horn current setting as a function of the proton beam position at the target.}\label{fig:mmy_sim}
    \end{center}
  \end{figure}

In order to repeat the beam scan studies for a different horn current setting of +180 kA, we repeat the same analysis procedure described earlier. Fig.~\ref{fig:mmy_sim} shows the muon flux centroid along the horizontal (left) and vertical (right) axes at muon monitor 1 as a function of the proton beam positions for a horn current of +180 kA, using both the uniform beam and nominal simulations. 
   %  \begin{figure}[htpb!]
  %\begin{center}\setlength{\unitlength}{1.0cm}
   %\includegraphics[width=.4\textwidth]{Ytrgt_FitMM}
    %\includegraphics[width=.4\textwidth]{Ytrgt_FitMM_180kA}
%\caption{The measured muon flux centroid along the vertical projection on Muon monitor 1 (blue) and 2 (red) for 200 kA (top) and 180 kA (bottom) horn current settings as a function of the proton beam position at the target.}\label{fig:data_y}
 %   \end{center}
  %\end{figure}
 \\
Two different horn current settings produced consistent slopes in both horizontal and vertical scans of both simulations.

\begin{figure}
\begin{center}%\setlength{\unitlength}{1.0cm}
%\begin{tabular}{@{}cc@{}}
\includegraphics[width=\linewidth]{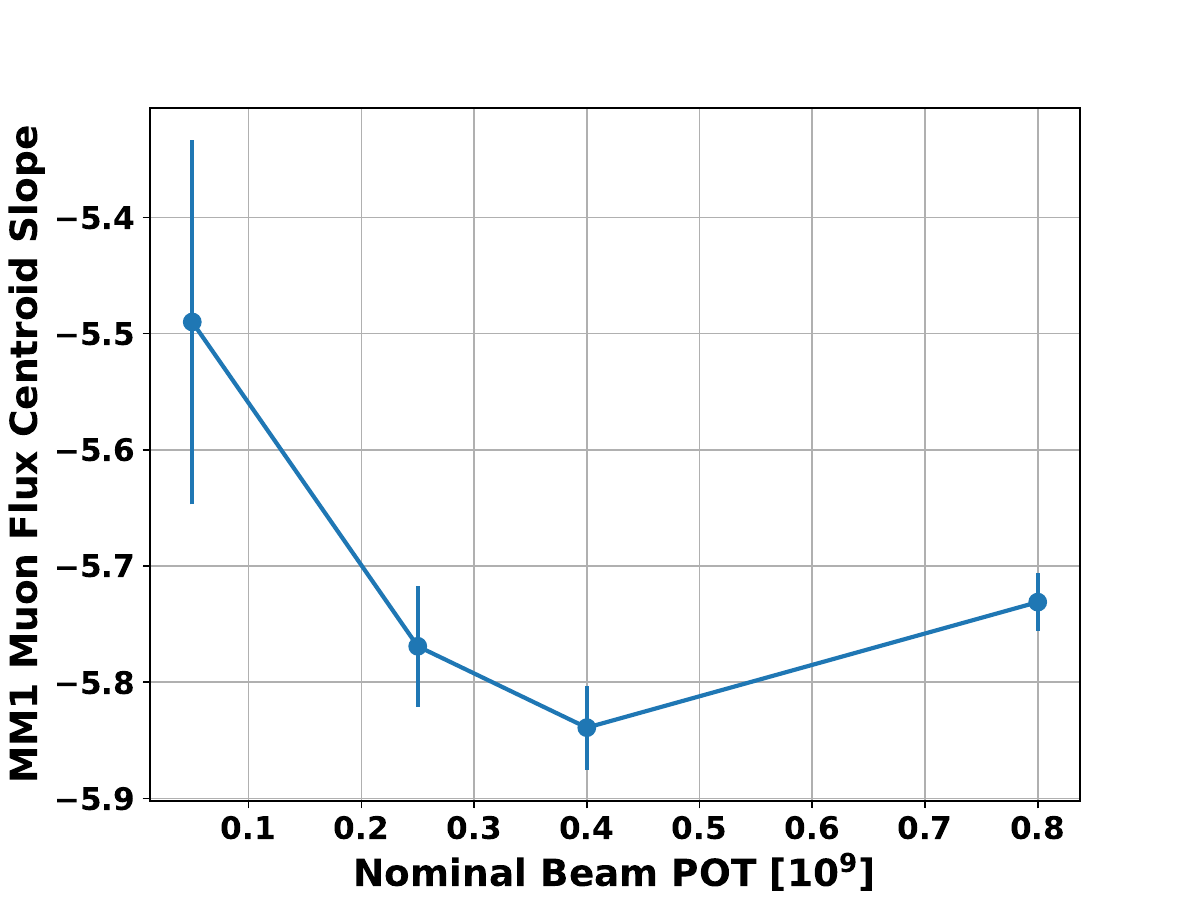} 
\includegraphics[width=\linewidth]{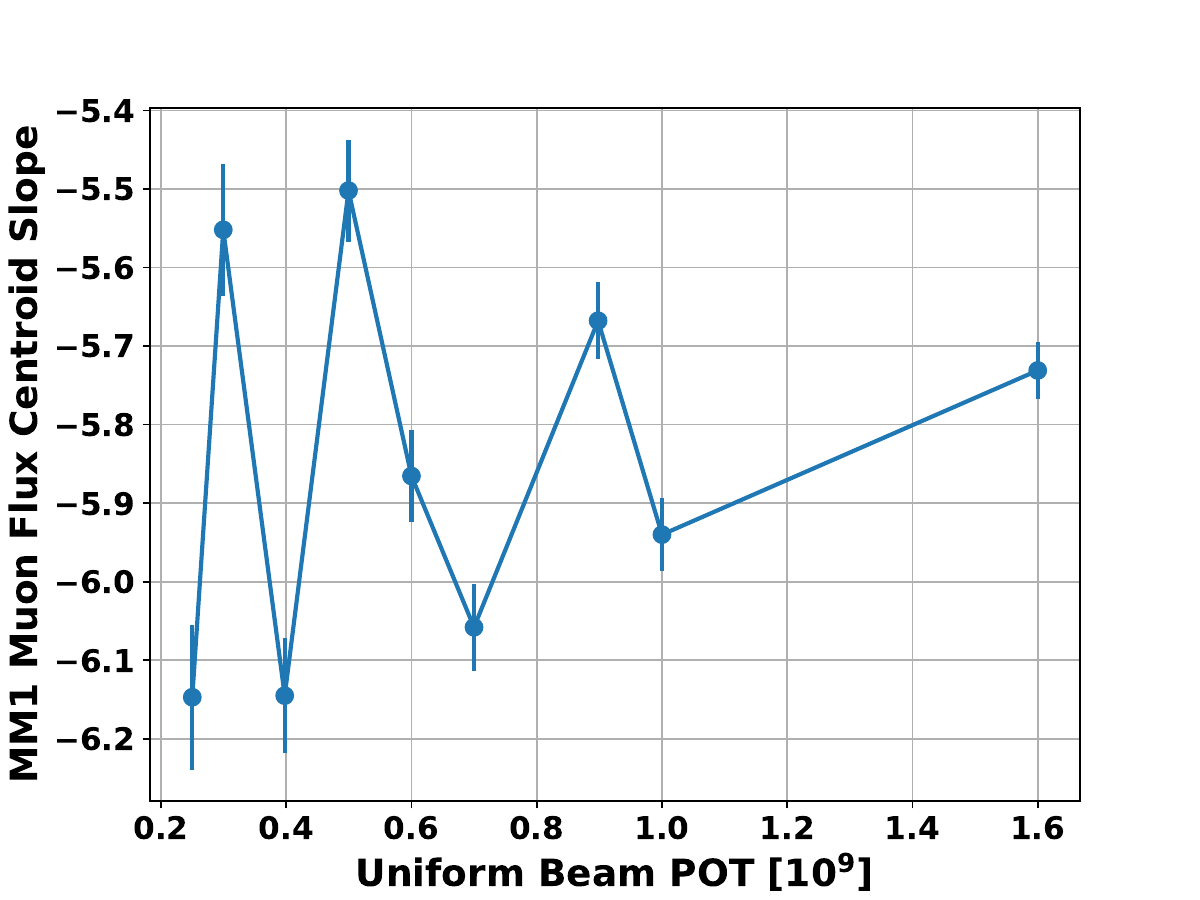} 
    %\end{tabular}
\caption{ The measured slope of the muon flux centeroid against the proton beam position as a function of the protons on the target (POT). The slope of muon flux centeroid from the nominal simulation (top) and the uniform beam simulation (bottom).}\label{fig:uni_nomi_pot}
\end{center}
\end{figure}

In addition, we have studied the statistical effects on the measured slope of the muon flux centeroid against the proton beam position. The simulation samples used in this study are independent and have different POTs. Fig.~\ref{fig:uni_nomi_pot} shows slopes as a function of POT for nominal simulation and uniform beam simulation. A statistical fluctuation is observed as a function of POT variation. 
Here, the POT of any Gaussian beam that we used to estimate the slope from the uniform beam simulation data is equivalent to $1/4$ of the total uniform beam simulation POT as described in Sec.~\ref{sec:simdata}.
  %-----------------------------------
\subsection{Beam Spot Size Scans}
In order to establish a correlation between neutrino and muon beam profiles, it is essential to understand how beam spot size affects muon monitor observations. 
\begin{figure}%[htp!]
  \begin{center}%\setlength{\unitlength}{1.0cm}
  \begin{tabular}{@{}cc@{}}
   \includegraphics[width=.23\textwidth]{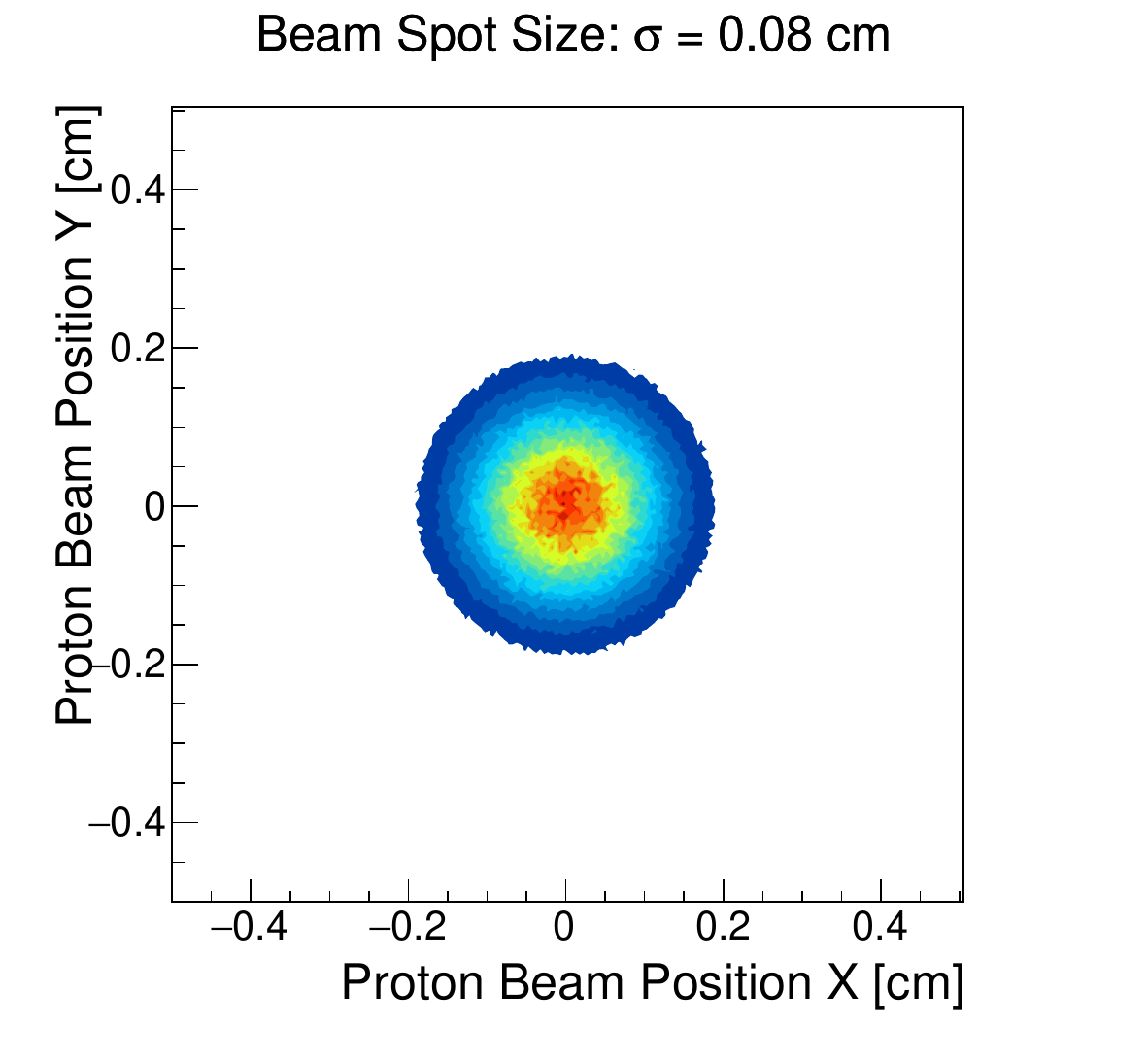} &
    \includegraphics[width=.23\textwidth]{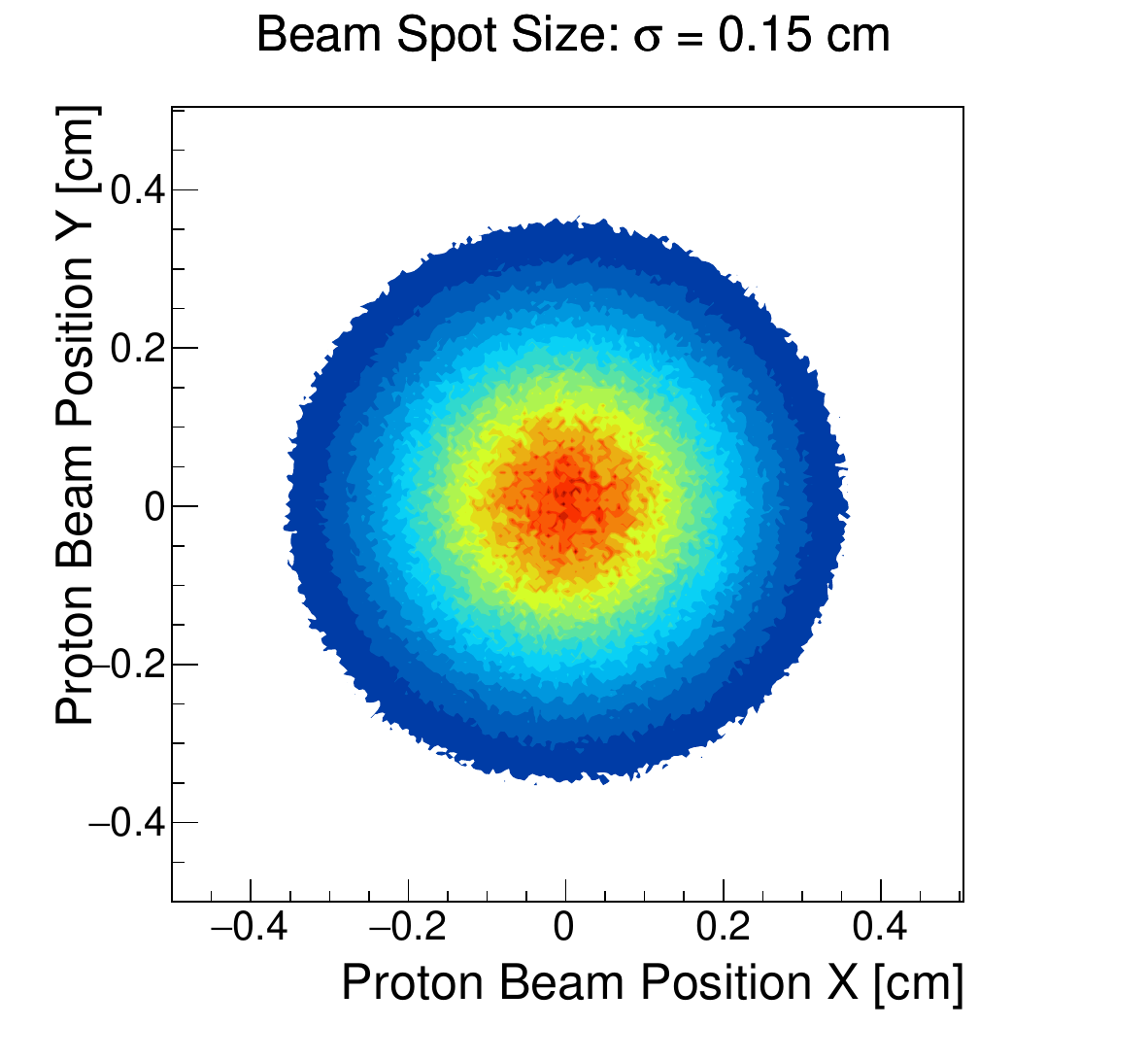} \\
    \includegraphics[width=.23\textwidth]{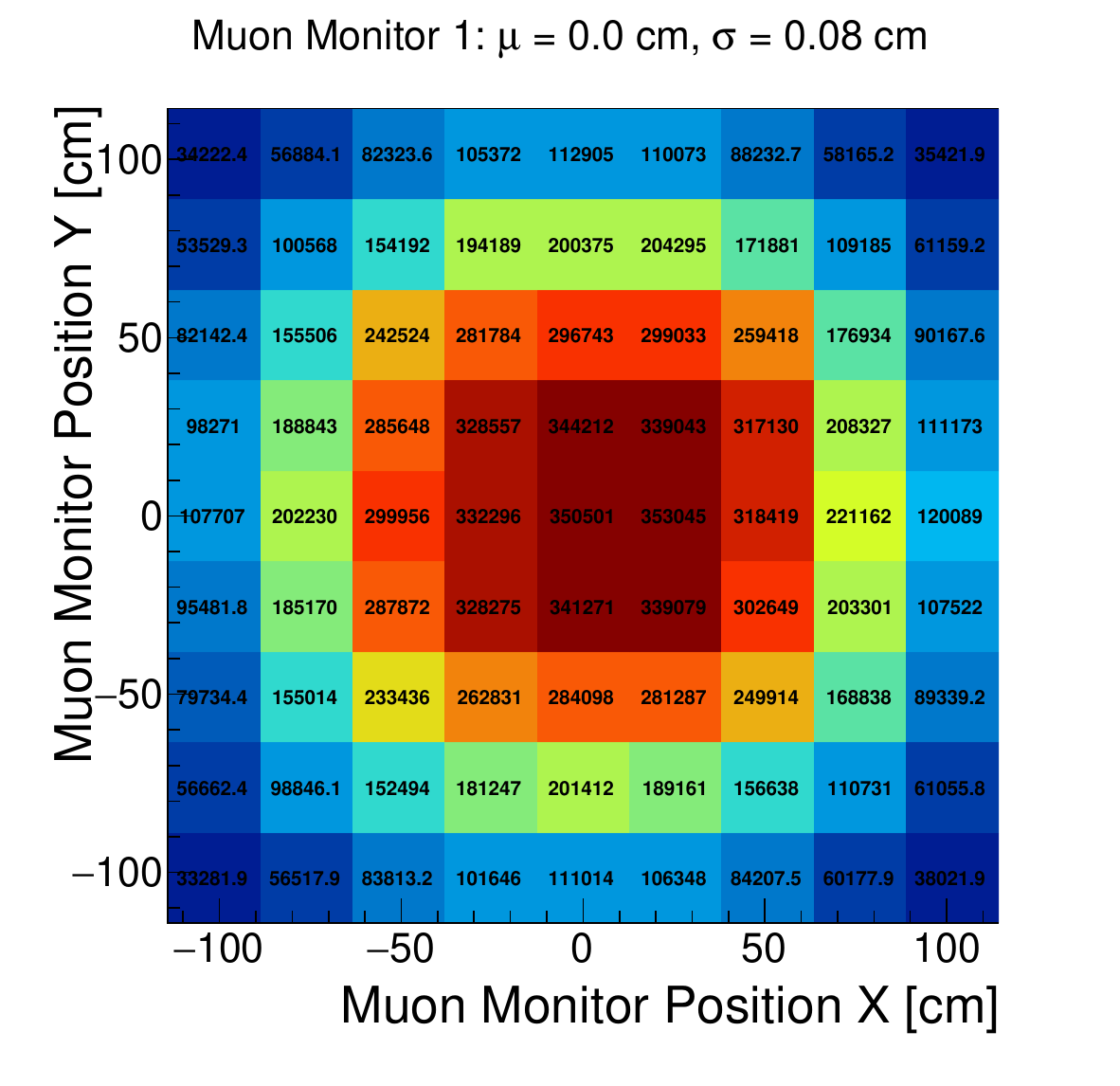} &
    \includegraphics[width=.23\textwidth]{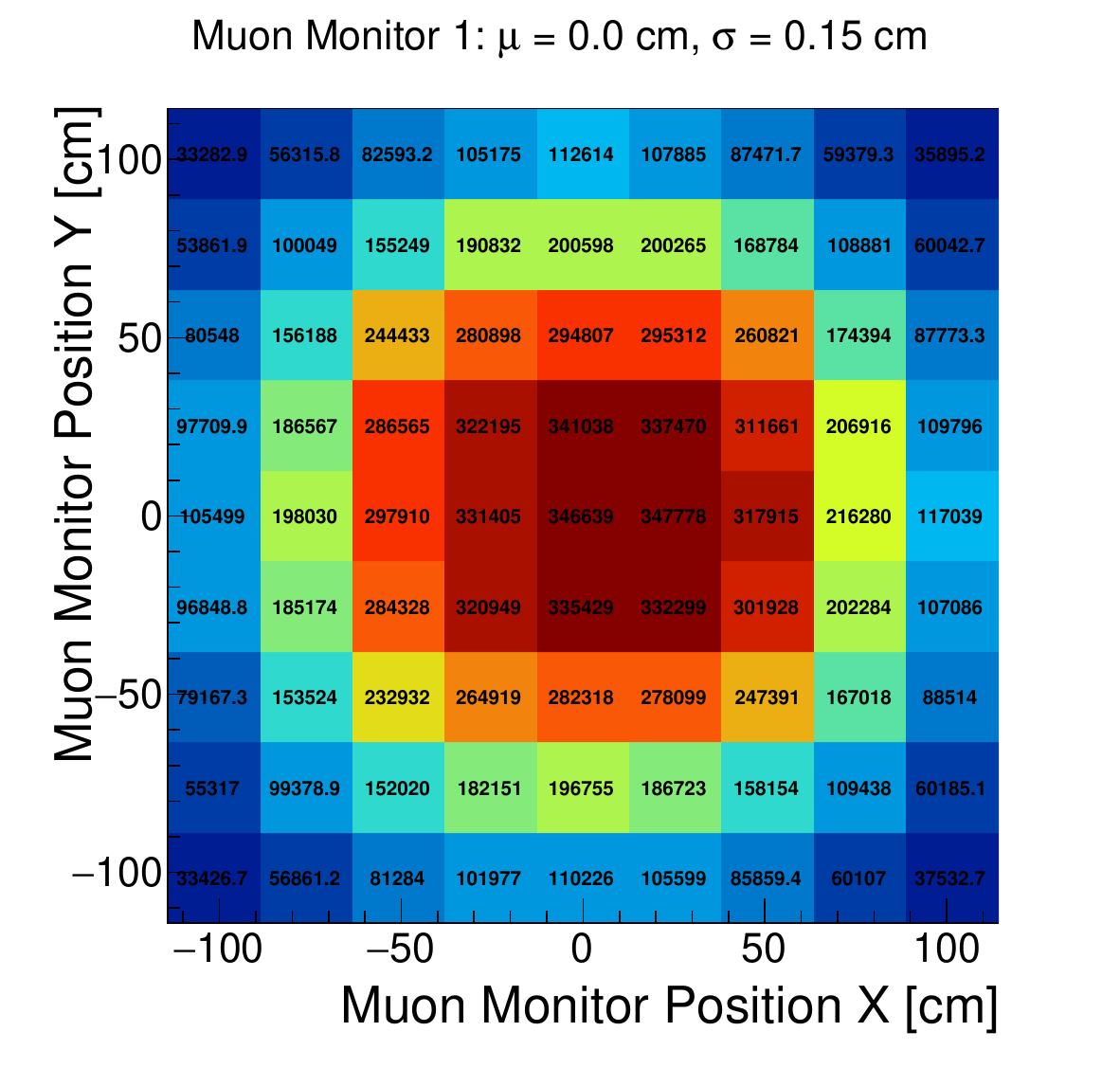} 
    \end{tabular}
\caption{The beam spot size selections: $\sigma = 0.08$ cm (top left) and $\sigma = 0.15$ cm (top right) and corresponding muon monitor responses for $\sigma = 0.08$ cm (bottom right) and $\sigma = 0.15$ cm (bottom left). }\label{fig:spot}
 \end{center}
\end{figure}
With the uniform beam simulation technique, we have produced Gaussian beams by gradually changing the Gaussian width $\sigma$. We have generated a simulation sample with 1 billion POT of uniformly distributed proton throws along horizontal and vertical directions ranging from -1.0 cm to +1.0 cm. 
\begin{figure}%[htpb!]
  \begin{center}\setlength{\unitlength}{1.0cm}
  %\begin{tabular}{@{}cc@{}}
   %\includegraphics[width=.28\textwidth]%{AllNewPlots/uniform_beamspot.pdf} &
    \includegraphics[width=\linewidth]{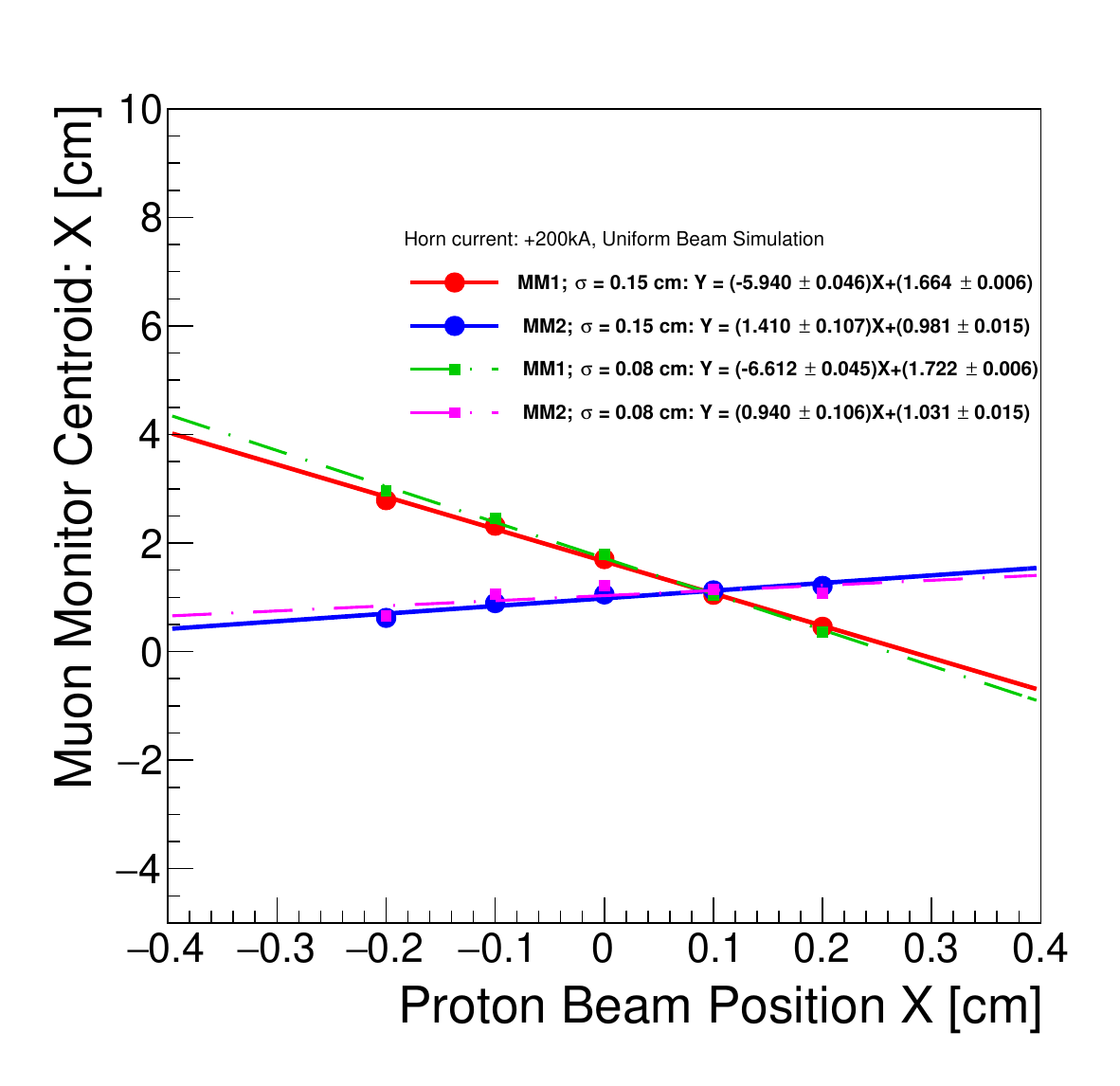} 
    %\end{tabular}
\caption{The muon flux centroid estimated along the horizontal projection on Muon monitor 1 and 2 for 200 kA for two different beam spot sizes: 0.15 cm and 0.08 cm.}\label{fig:spotslope}
    \end{center}
  \end{figure}
We have studied the muon flux distributions at monitor 1 for beam spot size change of 0.08 cm, and 0.15 cm.
Fig.~\ref{fig:spot} shows the beam spot sizes of 0.08 cm (top left) and 0.15 cm (top right) and the corresponding muon monitor responses in the contour plots (bottom left and right) respectively.
Fig.~\ref{fig:spotslope} shows the muon flux centroid estimation for different proton beam positions with changing beam spot sizes. According to these observations, muon flux centroids and standard deviations do not change as beam spot size changes. Simulation studies indicate that changing the beam spot size will not affect muon monitor measurements. 
 %-----------------------------------
%\subsection{Application of machine Learning: predict neutrino flux}
%The hard-to-measure beam parameters can be predicted using machine learning (ML) algorithms based on simulations (Horn tilting angle, target offset). It is possible to predict neutrino flux using the ML algorithm with uniform beam simulations.
%Simulation-based machine learning requires a lot of high statistical data samples. Multiple decays and uniform beam simulations allow us to collect enough MC data points. 
%Based on muon monitor simulations, ML can predict situations that cannot be measured in reality. For example, horn currents at low levels and a proton beam striking a baffle etc. Based on simulation, machine learning allows us to predict beam parameters that are difficult to measure (Horn tilting angle, target offset).
%It has also been possible to predict neutrino flux at the NOvA Near detector using simulated muon events from 81 pixels from each muon monitor.  A neural network was used to make prediction of the neutrino flux. 
%Fig.~\ref{fig:ML} shows the predicted and true neutrino energy spectra at the N0vA Near detector where the prediction matches the training sample well. 
%\begin{figure}[htpb!]
 % \begin{center}\setlength{\unitlength}{1.0cm}
 %  \includegraphics[width=.4\textwidth]{AllNewPlots/neutrinoflux_ML.png}
 % \caption{XXX.}\label{fig:ML}
 %   \end{center}
 % \end{figure}

\section{Application in LBNF simulation}\label{sec:dunesim}
Neutrino beam instrumentation (NBI) for LBNF provides measurements of the secondary beam for commissioning, alignment, monitoring, and hardware protection. This instrumentation complements the primary beam instrumentation as well as the neutrino detectors. It includes a hadron alignment detector system (HADeS) for measuring the remaining secondary particles in the Decay Pipe and a muon monitor detector system (MuMS) for monitoring muons downstream of the Hadron Absorber.
HaDeS is based on the NuMI hadron monitor \cite{Zwaska:2006px}, which will be used to align the beamline components by analyzing the residual beam after scanning the primary beam across the target, baffle, and horn. The NuMI hadron monitor consists of a linear array of parallel-plate ionization chambers of the size of 1 m$^2$, $7\times7$ array. HaDeS will implement LBNF-specific modifications in terms of higher channel counts to account for smaller pixel sizes compared to NuMI. 
During beam commissioning, beam alignment, and diagnosing failures, the HaDeS will be lowered into the beam for low-intensity alignment scans. A number of purposes are accomplished by the MuMS, including monitoring of long-term target degradation under beam irradiation by comparing muon fluxes at different energies. In addition to monitoring the condition of a target in detail, this monitor bypasses the need to wait for neutrinos or other tertiary beam monitors. Similar detector technology will be used for MuMS as for HaDeS, following NuMI's lead.\par
\begin{figure}[htp!]
  \begin{center}%\setlength{\unitlength}{1.0cm}
  %\begin{tabular}{@{}cc@{}}
   \includegraphics[width=0.85\linewidth]
    {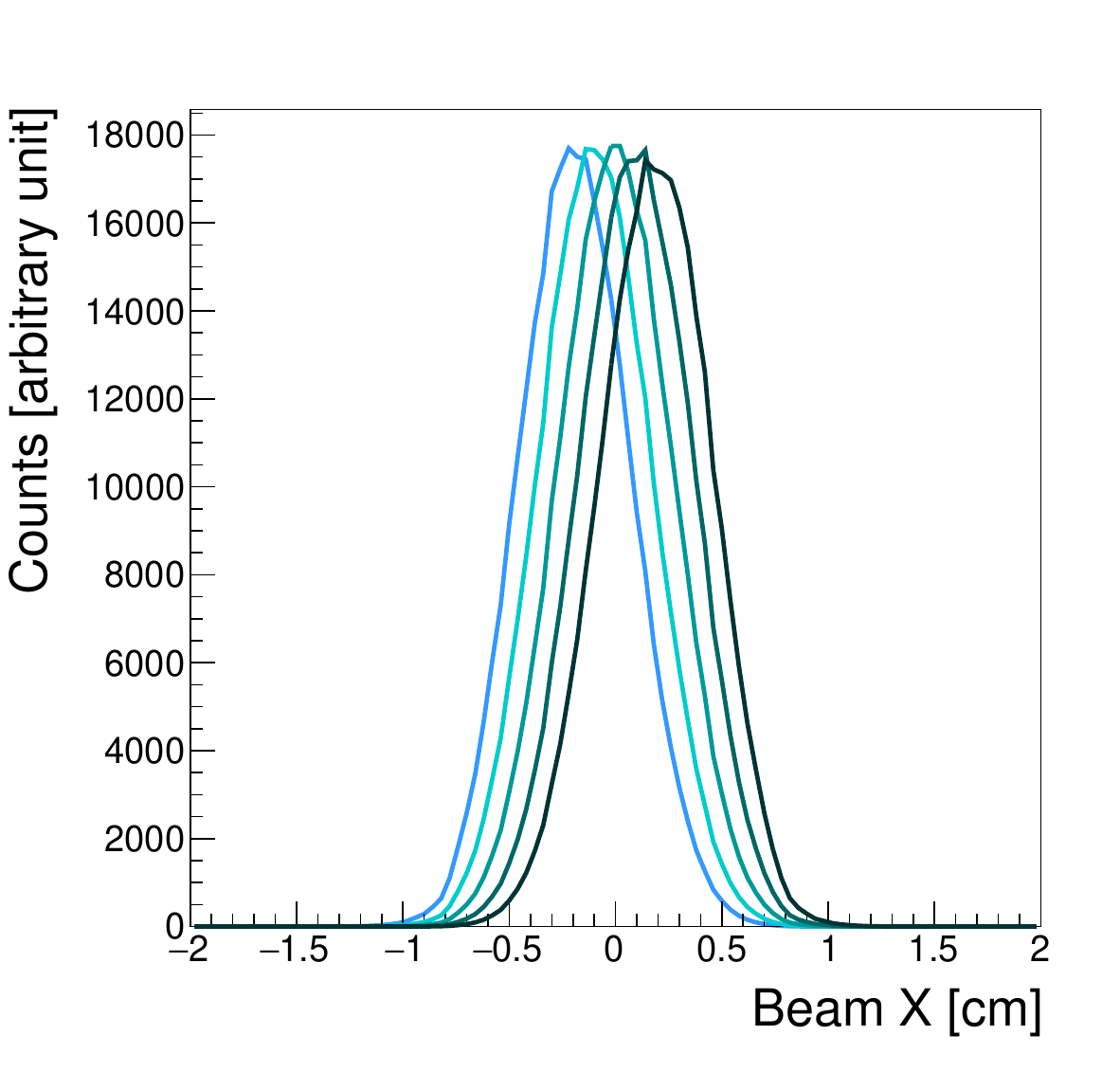}
    \includegraphics[width=0.85\linewidth]{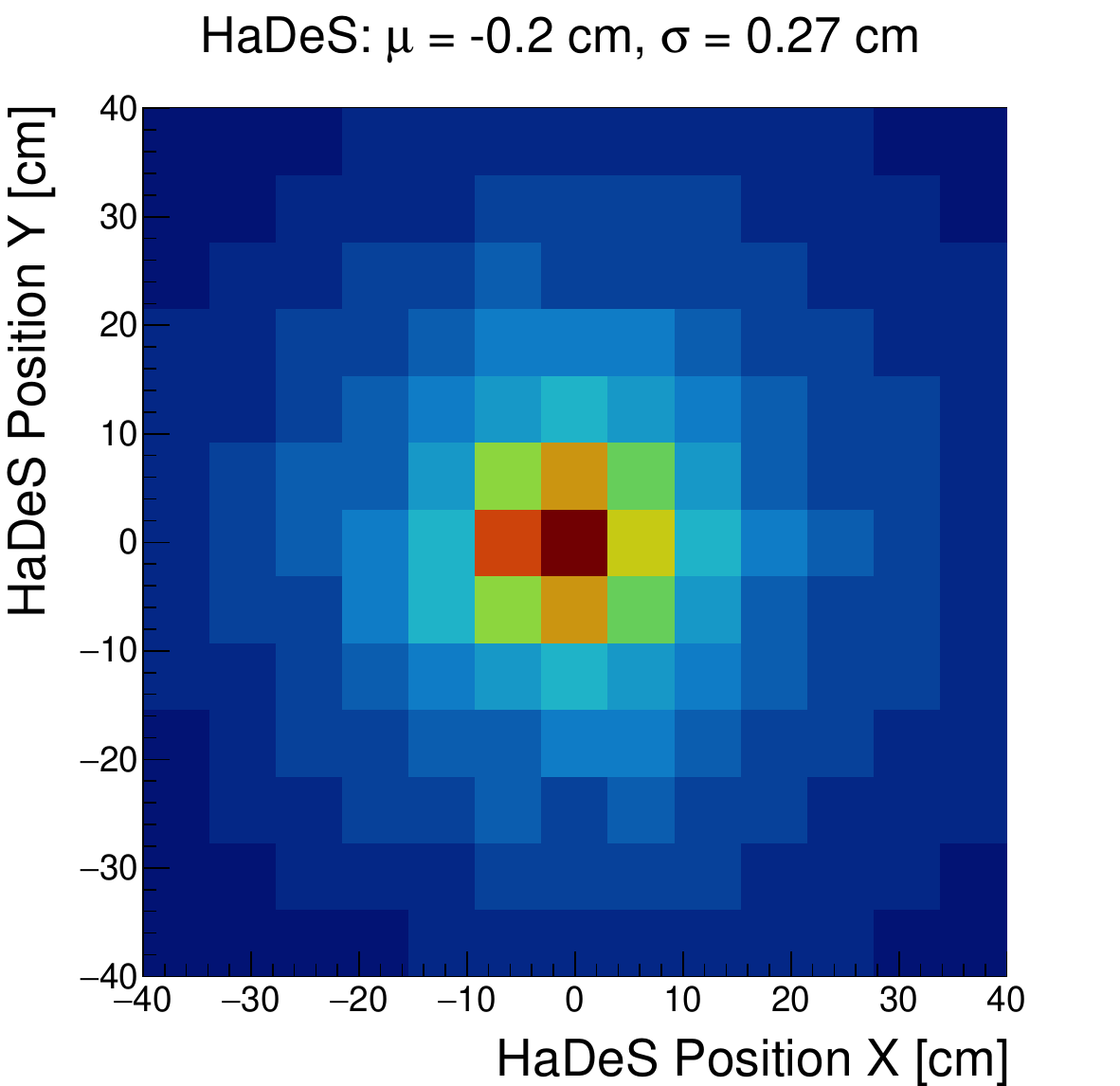}
    \includegraphics[width=0.85\linewidth]{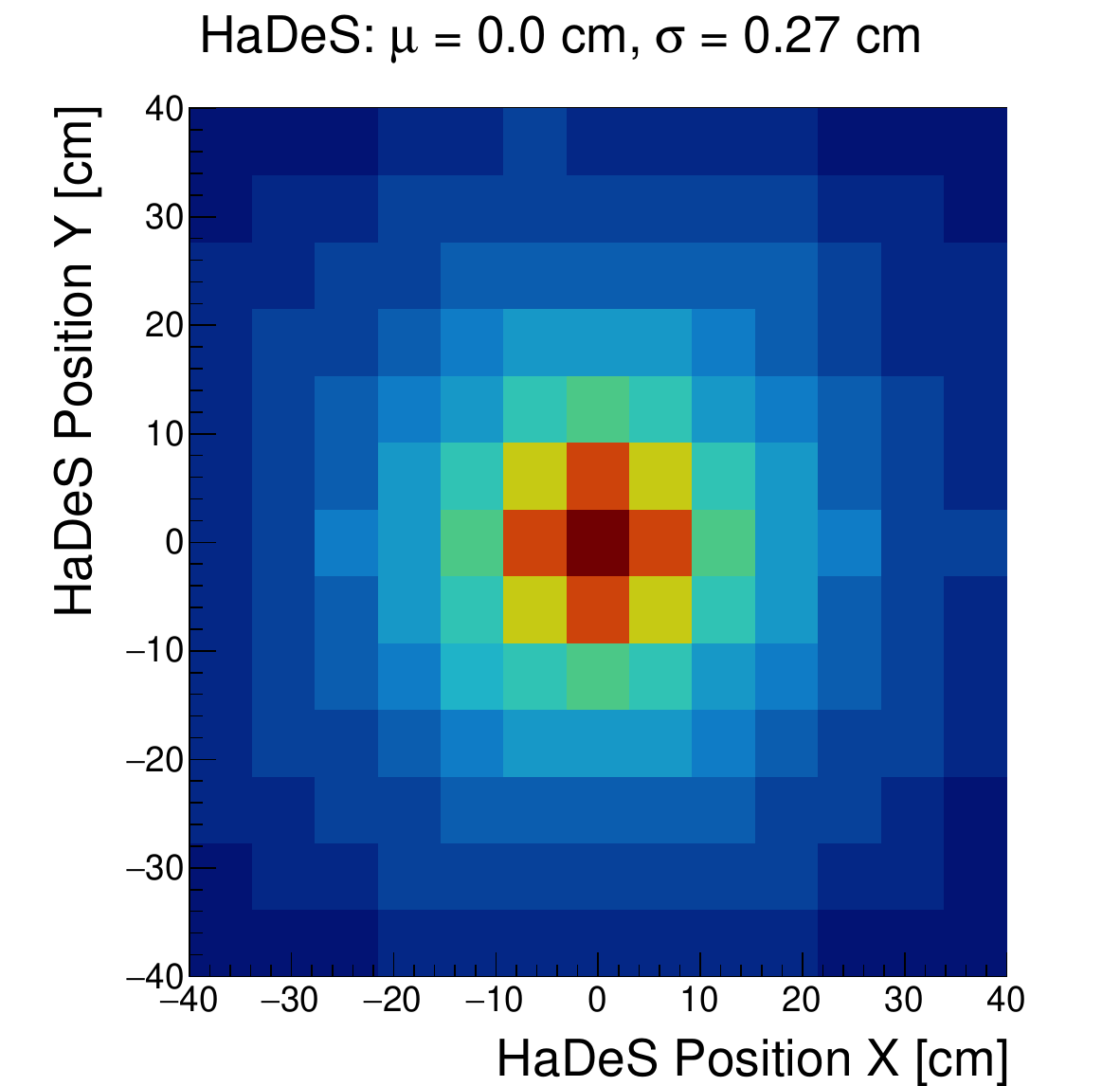}
    %\end{tabular}
\caption{The horizontal proton beam positions (top) for recorded beam interactions that have a neutrino candidate at the downstream neutrino detectors. Beam $\sigma = 0.27$ cm. Distribution of charges particles at the LBNF HaDeS location (middle) for a Gaussian proton beam with $\mu$ = -0.2 cm. Equivalent distribution (bottom) for a Gaussian proton beam with $\mu$ = 0.0 cm.}\label{fig:dunegauss}
    \end{center}
  \end{figure}

The LBNF beam simulation (g4lbnf) is a dedicated Geant4 beamline simulation based on the optimized design of the beamline. 
In the nominal g4lbnf simulation, a Gaussian beam is used with beam $\sigma$ set at 0.27 cm. We have implemented the uniform beam simulation technique in g4lbnf and drawn five Gaussian distributions for the proton beam in the horizontal direction from the uniform beam by changing the proton beam mean $\mu$ from -0.2 cm to +0.2 cm while keeping the Gaussian width $\sigma$ fixed at 0.27 cm as shown in Fig.~\ref{fig:dunegauss} top. 
The plots shown below as a demonstration of the uniform method have been generated using a simulation sample that contains 1000 jobs, each containing 10000 POT. This is just an example. It should be noted that this is only an example. POT can be simulated in different amounts, depending on the need of the user. 
%However, it is important to remember which POT/file has been requested, since it could be required later to normalize the plots.
The bottom two plots here show the distributions of the charged particles in a virtual detector at the location of the HaDeS for the Gaussian beams with $\mu$ = -0.2 cm, and, 0.0 cm respectively. Only those events for which a beam interaction has been recorded are included in the Gaussian distribution.
During beam-based alignment, low intensity beam will be scanned across the face of the HaDeS to determine beam direction. In order to determine if the pixel map of the hadron monitor is accurate, we can compare the HaDeS distributions from simulation with data.
\begin{figure}[htp!]
  \begin{center}\setlength{\unitlength}{1.0cm}
  \begin{tabular}{@{}cc@{}}
    \includegraphics[width=.45\textwidth]{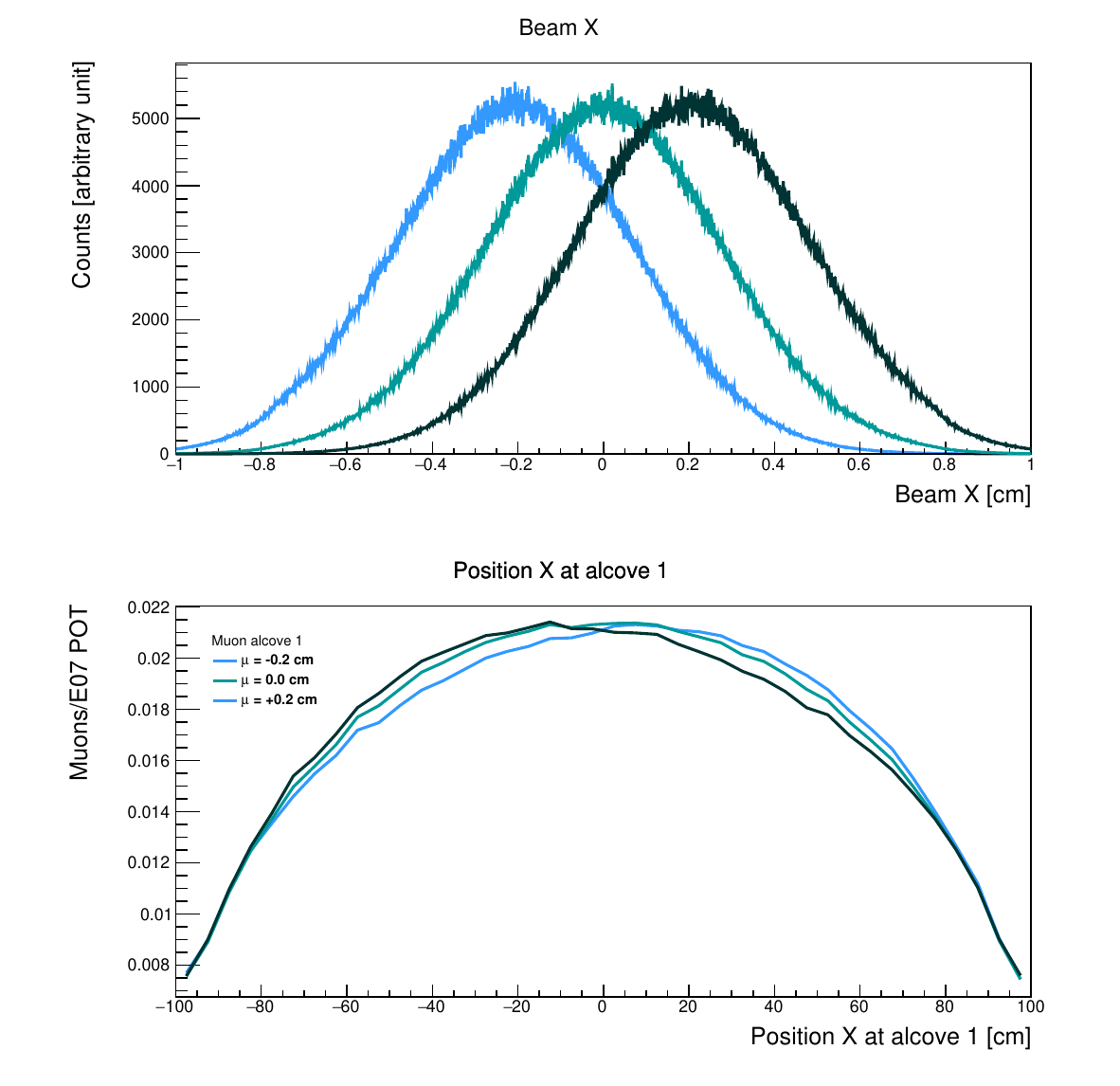} 
    \end{tabular}
\caption{Top: the Gaussian slices of the LBNF proton beam with mean at -0.2 cm, 0.0 cm and +0.2 cm from left to right. Bottom: muon beam profiles at the location of alcove 1 for the corresponding proton beams.}\label{fig:dunealcove}
    \end{center}
  \end{figure}
In Fig.~\ref{fig:dunealcove}, the top plot illustrates how Gaussian distributions are drawn horizontally after applying weights, only for the events for which beam interactions are recorded. On a virtual detector placed downstream of the absorber at the location of MuMS in muon alocve 1, the bottom plot shows the muon beam profiles after applying the same corresponding weights for each of the Gaussian beams.  
  
\begin{figure}[htp!]
  \begin{center}%\setlength{\unitlength}{1.0cm}
  %\begin{tabular}{@{}cc@{}}
    \includegraphics[width=\linewidth]{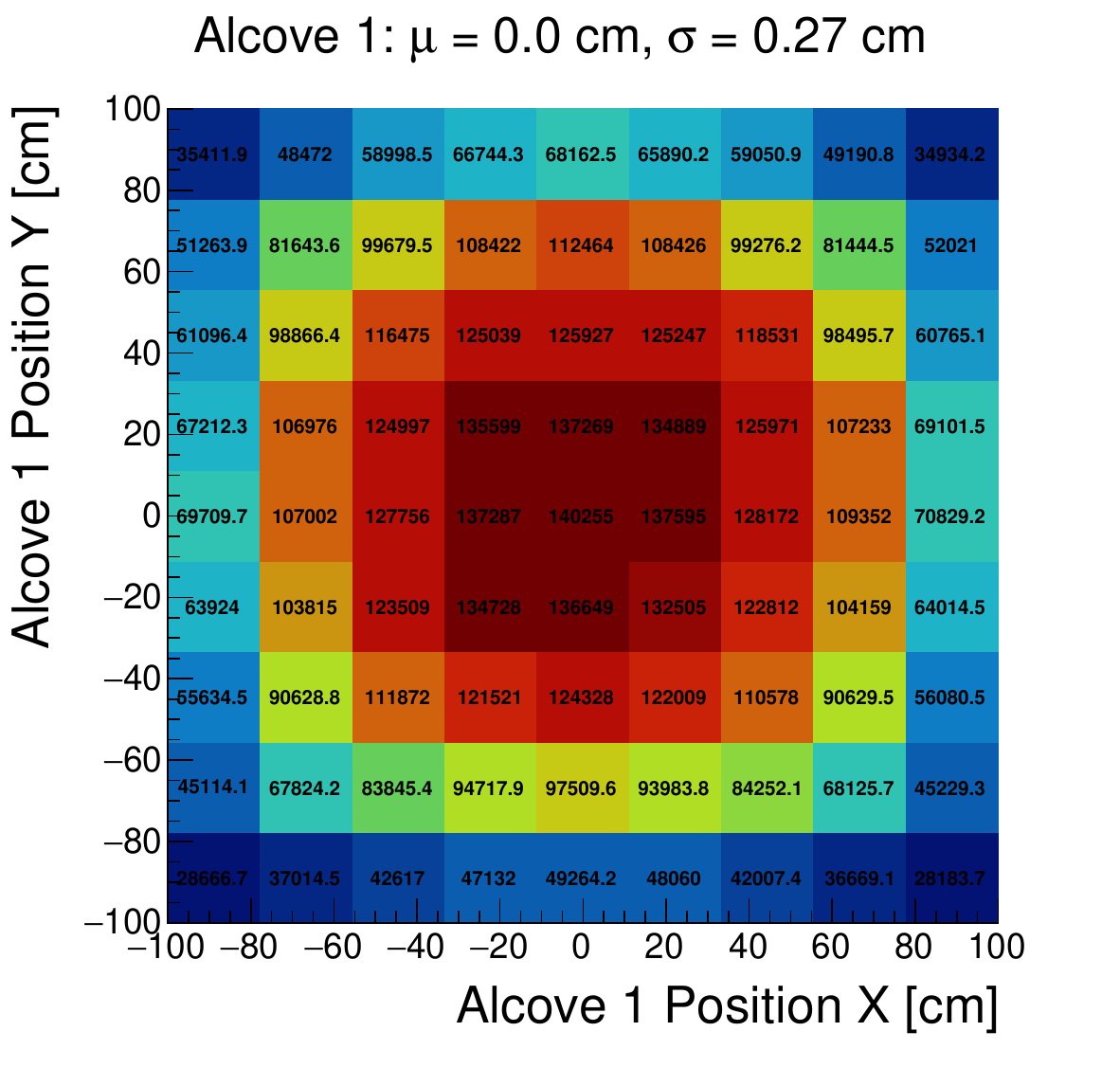}
    \includegraphics[width=\linewidth]{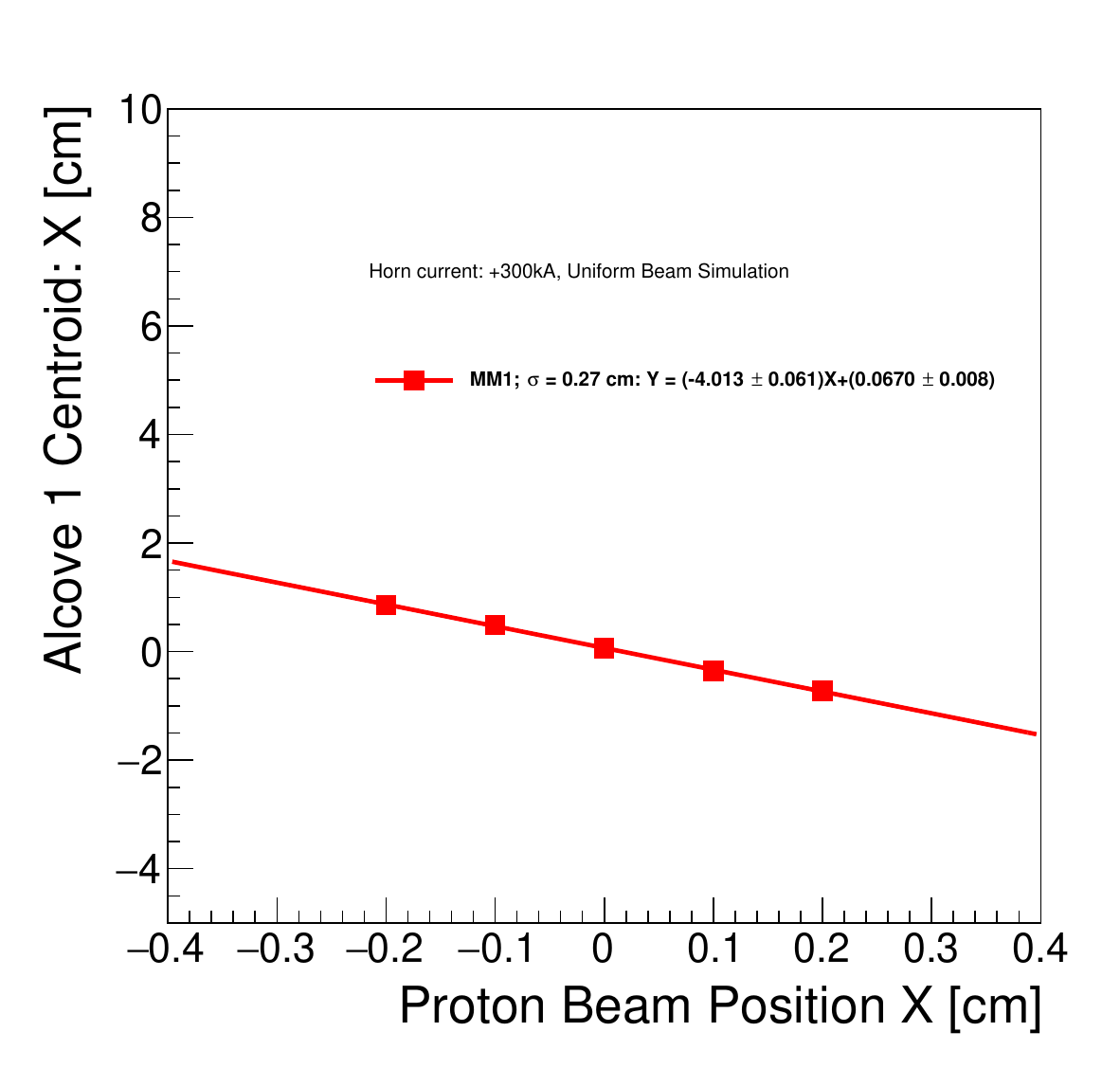}
    %\end{tabular}
\caption{The observed two dimensional muon distribution on a tracking plane at the LBNF alcove 1 location (top) for a given proton beam with $\mu = 0.0$ cm and $\sigma = 0.27$ cm. Muon flux centroid estimated at the alcove 1 location as a function of the horizontal proton beam position (bottom) at the target. }\label{fig:dunemm1}
    \end{center}
  \end{figure}

Each MuMS system detects muons that are focused differently depending on the energy of the parent pion. The two-dimensional event distribution on the alcove 1 tracking plane from a Gaussian proton beam slice with mean $\mu$ = 0.0 cm is shown in Fig.~\ref{fig:dunemm1}, left. The right plot shows the muon flux centroid estimation in a horizontal proton beam scan in the alcove 1 location. The horn current is 300 kA. The ability to perform beam scan studies using the uniform beam simulation technique will be valuable to understand impacts of the changing LBNF beam parameters on the beam response measurements. 
\section{Application of machine Learning: predict neutrino flux}\label{sec:machinelearning}
The application of Machine learning (ML) algorithms is becoming increasingly critical in particle accelerators and beamlines. Due to their efficiency and flexibility, artificial neural networks are ideally suited for complex simulations. A multi-layered artificial neural network (ANN) consists of an input layer, an output layer, and multiple hidden layers, each layer containing a number of nodes that are connected by randomly assigned weights. Every node in the layers produces an output according to its ``activation function''. The neural networks are then trained through repeated iterations to reduce estimated errors in the output by adjusting the weights of the connections between the nodes. Besides modeling nonlinear behavior, neural networks can also adapt to changes in the system over time. 

In addition to being useful for Fermilab's current neutrino beamlines, the development of machine learning algorithms for beam quality monitoring, anomaly detection, neutrino beam systematic studies, and neutrino beam quality monitoring will also be useful for future beamlines such as LBNF. 
In order to understand some of the rare ``anomaly" scenarios in the NuMI beamline, we need to generate simulated data. For example, g4numi simulation is used to generate simulated data for ML applications in the NuMI beamline. We can predict some ``hard-to-measure" beam parameters ML algorithms based on simulations such as Horn tilting angle, target offset. The muon monitors are sensitive to changes in the primary beam and variations in horn current. Machine learning applications can be built to monitor the quality of the NuMI neutrino beam based on the unique responses of muon monitors. 
%The fig.~\ref{fig:MLexample} below illustrates 
%The application of Machine Learning (ML) algorithms in predicting various beam parameters and neutrino fluxes is based on muon monitor pixel information. The muon event pattern from the muon monitor simulation serves as input for linear regression and neural network models, producing predictions for beam parameters or neutrino flux.
Utilizing Machine Learning (ML) algorithms to predict beam parameters and neutrino flux relies on muon monitor pixel information. The muon event pattern obtained from the muon monitor simulation is employed as input for a linear regression or neural network models, generating predictions for beam parameters and neutrino flux. In the following example, the neural network is trained using 241 muon monitor pixels as inputs.

The refinement of the neural network involves a crucial step known as hyper-parameter tuning, where the search for the optimal model architecture takes place. This optimization process ensures the effectiveness of the model in accurately predicting the desired outcomes.

%how different ML algorithms can be used to predict different beam parameters and neutrino fluxes based on muon monitor pixel information. The muon event pattern from the muon monitor simulation is used as input to a linear regression/ neural network and the outputs are the beam parameter predictions or neutrino flux prediction. 

%\begin{figure}%[htpb!]
  %\begin{center}%\setlength{\unitlength}{1.0cm}
  % \includegraphics[width=\linewidth]{ML_model.png} 
  %\caption{An example of application of Machine Learning Algorithms with Muon Monitor pixel inputs.}\label{fig:MLexample}
   % \end{center}
  %\end{figure}

%A uniform beam simulation sample was used for the training data samples, while three independent uniform beam simulation samples were used as the test data samples. 
%There are 6000 points in the training data sample, each containing beam parameters, muon events at pixels, and neutrino flux at the near detector.

Independent uniform beam samples were generated for ML model training, validation and testing. The dataset that has been used for training and validation purposes, consists of 6000 Gaussian beam profiles, each encompassing beam parameters, muon events at pixels, and the neutrino flux at the near detector. The uniform beam simulation and multiple decay simulation have been combined here to increase simulation efficiency. 
%Muon monitor observations depend on a number of parameters. %Taking pixel information from the muon monitor and disentangling it from other correlations is important. %Simulated data should be divided into training and test sets. %To understand correlations between muon monitor observations and beamline incidents, we need MC samples with large statistics, $\sim$300 million POTs in each sample. %In the past, the bottleneck has been obtaining large simulated samples running the nominal g4numi simulation, due to the lengthy sample preparation process. %Creating a nominal g4numi simulation sample with more than $\sim$ 300 million events takes $\sim$ 1 day. 

The challenge lies in disentangling muon monitor observations from various parameters and understanding correlations between these observations and beamline incidents. Large Monte Carlo (MC) samples with substantial statistics (approximately 300 million POTs in each sample) are crucial for these kind of studies. Historically, the bottleneck has been the time-consuming process of obtaining large simulated samples using the nominal g4numi simulation, which takes about a day for more than 300 million events.
%Since we need to scan over the beam parameter space, we need many large samples. 
Each time we change a beam parameter, we need to run the simulation again with large statistics for each change. Solutions such as GPU processing have been considered, but it becomes non-trivial as Geant4 does not support GPU processing. CPU processing on the DOE supercomputer National Energy Research Scientific Computing Center (NERSC) with 16000 cores would be an easier option. %However, by using uniform beam simulation, we can create large simulation samples more quickly and efficiently.
However, the use of uniform beam simulation proves to be an efficient strategy for creating large simulation samples quickly.

%To demonstrate the power of the Uniform beam simulation technique, we have generated a 6000 Monte Carlo data samples with Uniform simulation for ML applications. 
%Monte Carlo data samples with uniform beam simulation have been generated for ML applications to demonstrate the power of uniform beam simulation.
%The uniform beam simulation and multiple decay simulation have been combined to increase simulation efficiency. 
\begin{figure}[htpb!]
  \begin{center}%\setlength{\unitlength}{1.0cm}
   \includegraphics[width=\linewidth]{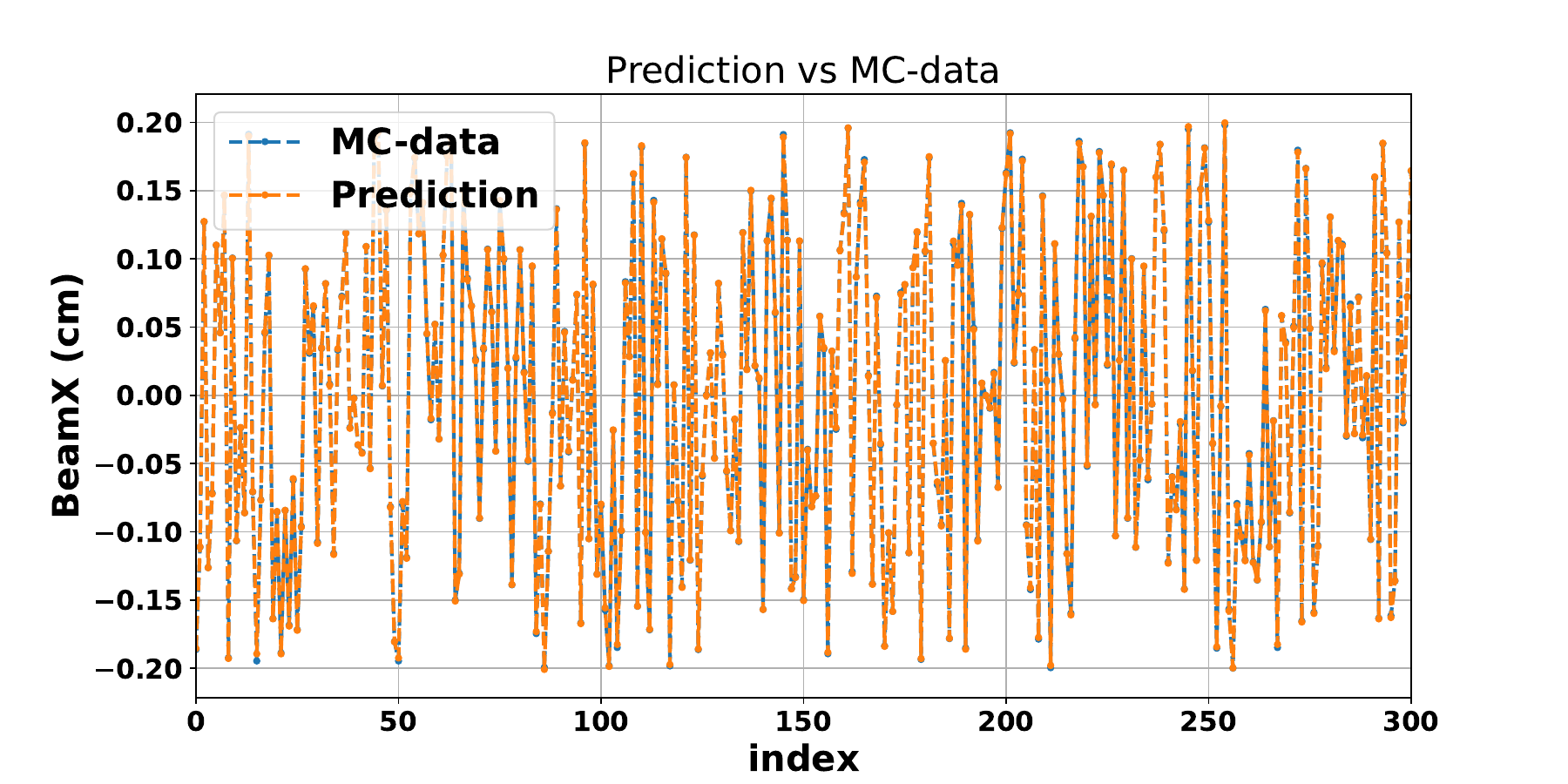}
   \includegraphics[width=\linewidth]{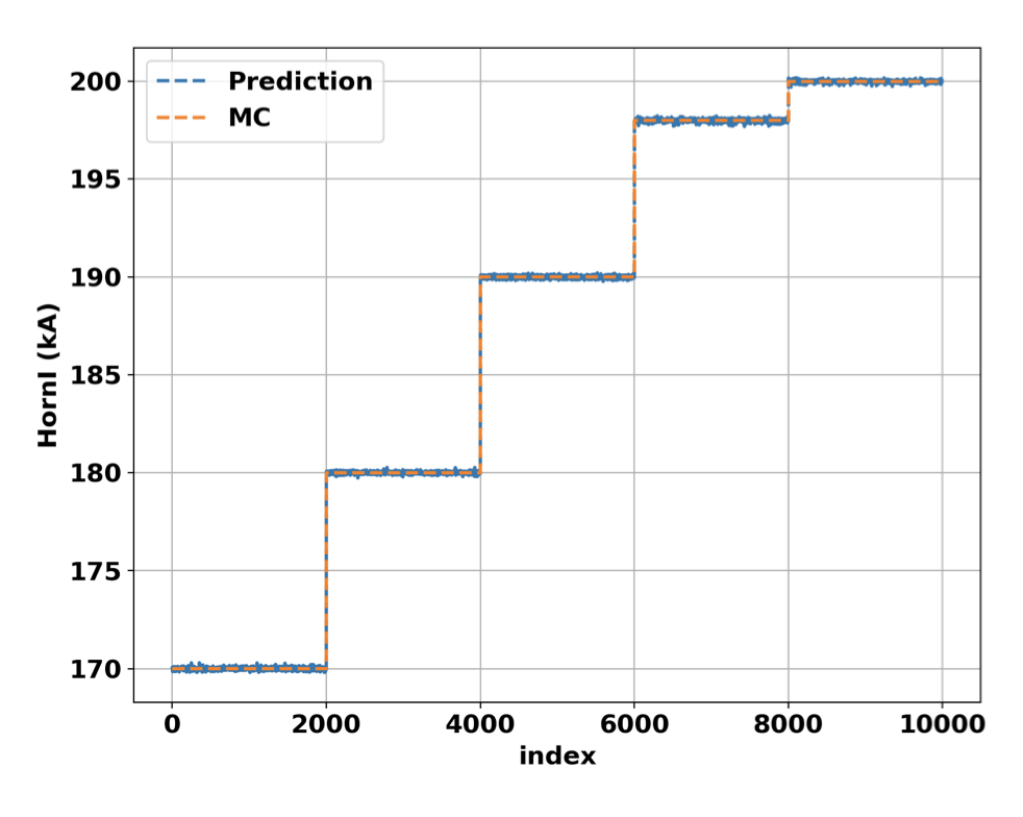}
  \caption{Top: Horizontal beam position prediction, bottom: horn current prediction using a linear regression model.}\label{fig:ML1}
    \end{center}
  \end{figure}
In this technique, we generate a uniformly distributed single simulation data sample on the grid with large statistics for the selected beam variable range. The uniformly distributed sample is then used to generate Gaussian beam profiles for different selected beam parameters. Beam positions, widths etc. can be varied post-processing once we have generated the uniform sample. In this way, computational overhead is greatly reduced. 
A linear regression model, trained using muon monitor pixels as inputs, demonstrated high accuracy in predicting horizontal proton beam position and horn current, as depicted in Fig.~\ref{fig:ML1}.
In this figure, the accuracy of prediction is high because the simulated data does not contain background noises as in real data.
%Based on the uniform beam simulation data, a linear regression model was tested to predict the horizontal proton beam position and the horn current. %To train the linear model, we used muon monitor pixels as inputs. %The model shows high accuracy in predicting beam position (left) and horn current (right) after training as shown in fig.~\ref{fig:ML1}.

\begin{figure}[hp]
  \begin{center}\setlength{\unitlength}{1.0cm}
   \includegraphics[width=\linewidth]{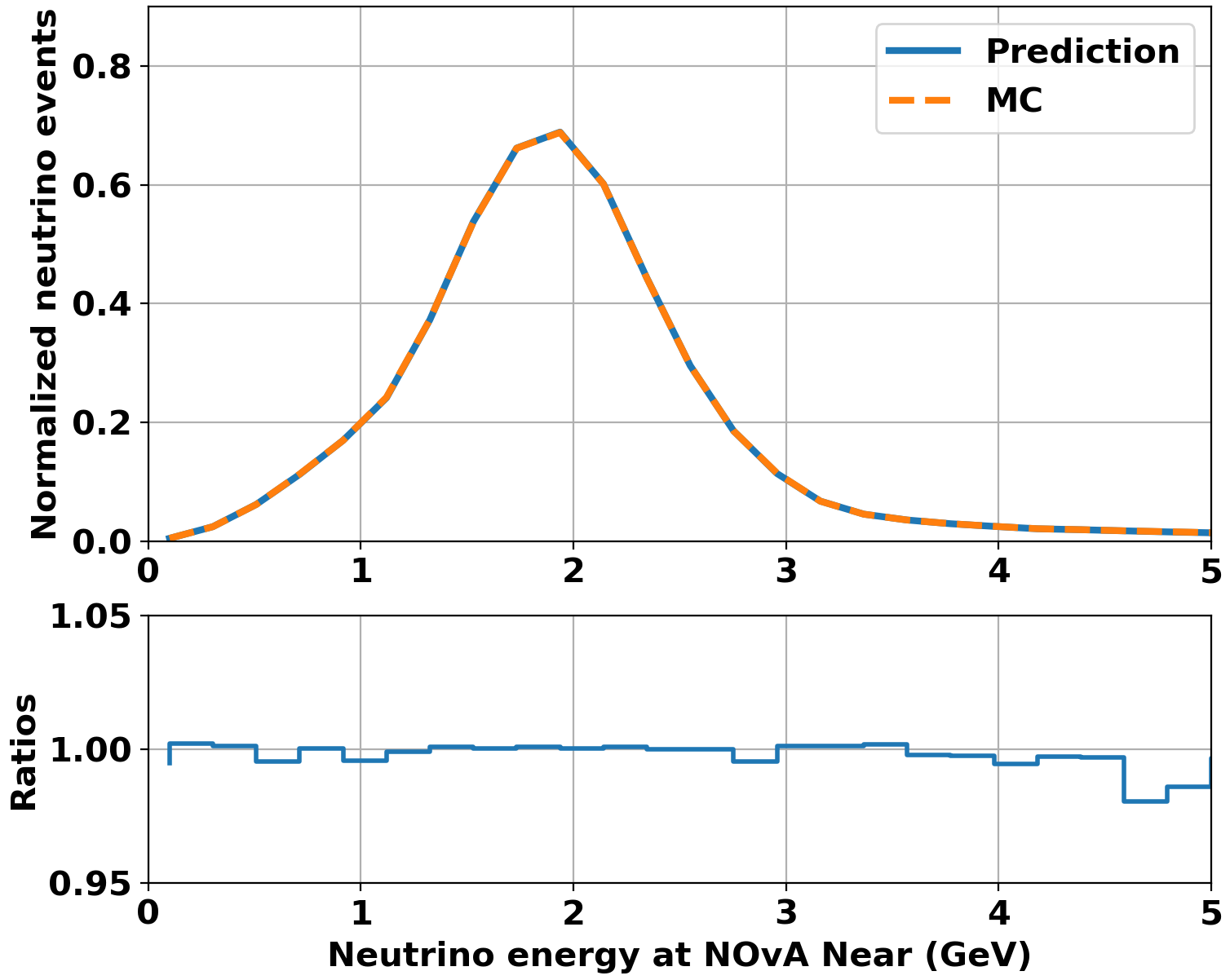}
  \caption{A comparison of predicted (blue) and true (orange) neutrino energy spectra at the NOvA Near detector.}\label{fig:ML}
    \end{center}
  \end{figure}

%The ML algorithm and uniform beam simulations have also been shown to be successful in predicting neutrino flux.
Furthermore, the ML algorithm has been successfully employed to predict neutrino flux at the NOvA Near Detector using simulated muon events from 81 pixels from each muon monitor. 
%We have predicted neutrino flux at the NOvA Near detector using simulated muon events from 81 pixels from each muon monitor. A neural network has been used to make prediction of the NuMI neutrino flux.
Fig.~\ref{fig:ML} shows the predicted and true neutrino energy spectra at the NOvA Near detector where the prediction matches the training sample well. 
%The accuracy of the prediction here is perfect because the simulation is free of background noises. 
Due to the absence of background noises in simulation data, the prediction perfectly matches with the data. 
%The ML predictions provide an additional layer of monitoring of neutrino beam behavior and horn current behavior. %The results show potential for monitoring beam performance, developing trends, and identifying issues during regular beam operations. %The uniform beam simulation has enabled ML models to be used to predict rare incidents and anomalies due to its efficiency and cost-effectiveness in generating large simulation samples.
In conclusion, ML predictions offer an additional layer of monitoring for neutrino beam and horn current behavior, showing potential for monitoring beam performance, identifying trends, and detecting issues during regular beam operations. The efficiency and cost-effectiveness of uniform beam simulation enable ML models to predict rare incidents and anomalies effectively.

\section{Summary and Outlook}\label{sec:summary}
In this paper, we have presented a simulation technique that can be applied to generate as many random beam simulation options desired for beam scan studies. In contrast to generating multiple Gaussian samples with very large statistics, this technique produces large simulation samples with a uniform distribution of beam parameters. 
From a test sample of 1 billion POT with uniformly distributed beam X and Y with g4numi simulation, each generated Gaussian beam with 0.15 cm beam width will contain 540 million weighted interactions. For the same number of interactions, a sample of 250 million POT must be generated every time from a nominal sample. Therefore, instead of generating multiple Gaussian beams with 250 million POT every time, one can generate a uniform simulation with 1 billion POT once and generate as many Gaussian beams from it as desired. 
To generate a single Gaussian beam with 250 million POT with nominal simulation takes more than 2 hrs of wall time on the Fermigrid parallel processing resource at Fermilab computing facility. 
Grid-submitted jobs are subject to varying and lengthy waiting times. 
By using uniform beam simulation with high statistics, one can generate weights for many beam simulation combinations without requiring a large computing resource and a long simulation time. 
In order to demonstrate the application of the uniform simulation technique to NuMI studies, a number of examples have been presented. 

A similar exercise has been performed with g4lbnf simulation, and the studies have been extended to show that the technique can be used to study LBNF beam scans through simulation. 
The results of the uniform simulations are shown to be consistent with the nominal simulation.
A combination of uniform beam simulations and multiple decays has been shown to be applicable to multiple neutrino beamline simulations. 
The high statistics of Gaussian throws from the uniform simulation technique can be leveraged not only for neutrino experiments but also for various simulation requirements related to the upcoming Mu2e experiment at Fermilab or its upgrade, known as Mu2e-II, which is the next generation muon conversion experiment. Due to its efficiency and cost-effectiveness in generating large simulation samples, the utilization of the uniform beam simulation technique has facilitated the use of machine learning models for predicting rare incidents and anomalies. Consequently, this technique will play a crucial role in the application of machine learning algorithms across diverse research contexts.

\begin{acknowledgments}
This work is supported by the Fermi Research Alliance, LLC manages and operates the Fermi National Accelerator Laboratory pursuant to Contract number DE-AC02-07CH11359 with the United States Department of Energy.

This work is partially supported by the U.S. Department of Energy grant DE-SC0019264.
\end{acknowledgments}
	
\bibliographystyle{elsarticle-num}
\bibliography{bibliography} % Produces the bibliography via BibTeX.
%\bibliography{url}
\end{document}